\documentclass[12pt]{article}
\usepackage{amsmath}
\usepackage{graphicx}
\usepackage{enumerate}
\usepackage[round,authoryear]{natbib}
\usepackage{url} 

\newcommand{\blind}{1}

\usepackage{authblk}
\usepackage{hyperref} 
\usepackage{float}
\usepackage{caption}
\usepackage{setspace}
\usepackage{amsfonts}
\usepackage{bm}
\usepackage{cellspace}
\usepackage{booktabs}
\usepackage{multirow} 
\usepackage{rotating}
\hypersetup{colorlinks=true,linkcolor=blue,filecolor=blue,citecolor=blue}
\usepackage{soul}

\newtheorem{theorem}{Theorem}
\newtheorem{corollary}{Corollary}[theorem]
\newtheorem{lemma}{Lemma}
\newtheorem{algorithm}{Algorithm}
\newtheorem{assumption}{A\hspace{-0.3em}}

\newcommand{\indep}{\perp \!\!\! \perp}

\makeatletter
\let\save@mathaccent\mathaccent
\newcommand*\if@single[3]{%
  \setbox0\hbox{${\mathaccent"0362{#1}}^H$}%
  \setbox2\hbox{${\mathaccent"0362{\kern0pt#1}}^H$}%
  \ifdim\ht0=\ht2 #3\else #2\fi
  }
\newcommand*\rel@kern[1]{\kern#1\dimexpr\macc@kerna}
\newcommand*\widebar[1]{\@ifnextchar^{{\wide@bar{#1}{0}}}{\wide@bar{#1}{1}}}
\newcommand*\wide@bar[2]{\if@single{#1}{\wide@bar@{#1}{#2}{1}}{\wide@bar@{#1}{#2}{2}}}
\newcommand*\wide@bar@[3]{%
  \begingroup
  \def\mathaccent##1##2{%
    \let\mathaccent\save@mathaccent
    \if#32 \let\macc@nucleus\first@char \fi
    \setbox\z@\hbox{$\macc@style{\macc@nucleus}_{}$}%
    \setbox\tw@\hbox{$\macc@style{\macc@nucleus}{}_{}$}%
    \dimen@\wd\tw@
    \advance\dimen@-\wd\z@
    \divide\dimen@ 3
    \@tempdima\wd\tw@
    \advance\@tempdima-\scriptspace
    \divide\@tempdima 10
    \advance\dimen@-\@tempdima
    \ifdim\dimen@>\z@ \dimen@0pt\fi
    \rel@kern{0.6}\kern-\dimen@
    \if#31
      \overline{\rel@kern{-0.6}\kern\dimen@\macc@nucleus\rel@kern{0.4}\kern\dimen@}%
      \advance\dimen@0.4\dimexpr\macc@kerna
      \let\final@kern#2%
      \ifdim\dimen@<\z@ \let\final@kern1\fi
      \if\final@kern1 \kern-\dimen@\fi
    \else
      \overline{\rel@kern{-0.6}\kern\dimen@#1}%
    \fi
  }%
  \macc@depth\@ne
  \let\math@bgroup\@empty \let\math@egroup\macc@set@skewchar
  \mathsurround\z@ \frozen@everymath{\mathgroup\macc@group\relax}%
  \macc@set@skewchar\relax
  \let\mathaccentV\macc@nested@a
  \if#31
    \macc@nested@a\relax111{#1}%
  \else
    \def\gobble@till@marker##1\endmarker{}%
    \futurelet\first@char\gobble@till@marker#1\endmarker
    \ifcat\noexpand\first@char A\else
      \def\first@char{}%
    \fi
    \macc@nested@a\relax111{\first@char}%
  \fi
  \endgroup
}
\makeatother

\begin{document}

\def\spacingset#1{\renewcommand{\baselinestretch}%
{#1}\small\normalsize} \spacingset{1}


\if1\blind
{

\renewcommand\footnotemark{}
\renewcommand\footnoterule{}
\renewcommand\Affilfont{\footnotesize}

\title{\bf Robust Causal Inference for EHR-based Studies of Point Exposures with Missingness in Eligibility Criteria}
\author[1]{Luke Benz}
\author[1]{Rajarshi Mukherjee}
\author[1,2,3]{Rui Wang}
\author[4]{David Arterburn}
\author[5]{Heidi Fischer}
\author[6]{Catherine Lee}
\author[7,8]{Susan M. Shortreed}
\author[1]{Sebastien Haneuse*}
\author[9]{Alexander W. Levis*}
\affil[1]{Department of Biostatistics,
Harvard T.H. Chan School of Public Health, Boston, MA, USA}
\affil[2]{Department of Population Medicine, Harvard Pilgrim Health Care Institute, Boston, MA, USA}
\affil[3]{Department of Population Medicine, Harvard Medical School, Boston, MA, USA}
\affil[4]{Kaiser Permanente Washington Health Research Institute, Seattle, WA, USA}
\affil[5]{Department of Research \& Evaluation, Kaiser Permanente Southern California, Pasadena, CA, USA}
\affil[6]{Department of Epidemiology \& Biostatistics, University of California San Francisco, San Francisco, CA, USA}
\affil[7]{Biostatistics Division, Kaiser Permanente Washington Health Research Institute, Seattle, WA, USA}
\affil[8]{Department of Biostatistics, University of Washington School of Public Health, Seattle, WA, USA}
\affil[9]{Department of Biostatistics, Epidemiology \& Informatics,
University of Pennsylvania, Philadelphia, PA, USA}
\thanks{\noindent $^*$ denotes co-last author (AWL and SH).}
  \maketitle
} \fi

\if0\blind
{
  \bigskip
  \bigskip
  \bigskip
  \begin{center}
    {\large \bf Robust Causal Inference for EHR-based Studies of Point Exposures with Missingness in Eligibility Criteria}
\end{center}
  \medskip
} \fi

\vspace{-0.35in}

\begin{abstract}
\noindent Missingness in variables that define study eligibility criteria is a seldom addressed challenge in electronic health record (EHR)-based settings. Patients with incomplete eligibility information are typically excluded from analysis without consideration of the (implicit) assumptions being made, leaving study conclusions subject to potential selection bias. In an effort to ascertain eligibility for more patients, researchers may look back further in time prior to study baseline, inappropriately including individuals who, unbeknownst to the researcher, would fail to meet eligibility at baseline. To the best of our knowledge, however, little work has been done to mitigate these concerns. We propose a robust and efficient estimator of the causal average treatment effect on the treated, defined in the study-eligible population, where eligibility-defining covariates are missing at random. The approach facilitates flexible machine-learning strategies for component nuisance functions while maintaining valid asymptotic inference. This method is applied to EHR data from Kaiser Permanente on 14,809 patients undergoing bariatric surgery, comparing two surgical interventions for long-term weight and glycemic outcomes among severely obese patients with type II diabetes mellitus. Our results confirm a weight-loss advantage for Roux-en-Y gastric bypass over sleeve gastrectomy, but are less suggestive of a diabetes remission advantage than complete-case analyses indicate.

\end{abstract}

\noindent%
{\it \textbf{Keywords}: causal inference, influence functions, missing data, bariatric surgery, diabetes}  

\spacingset{1.9} 

\section{Introduction}
\label{sec:intro}

Observational studies leveraging electronic health record (EHR) databases are seen as important alternatives to randomized clinical trials, particularly when trials may be infeasible due to financial, ethical, or logistical constraints \citep{hudson2017}. In contrast to data collected in randomized trials, however, EHR databases exist to record clinical activity and assist with patient billing, and thus are not collected with any specific research purpose in mind. As such, information that might be routinely collected in a trial or prospective observational study may be unavailable or outdated for some patients in EHR-based studies \citep{haneuse2016a}. 

Well-designed trials and observational studies require that subjects meet a set of study eligibility criteria (inclusion/exclusion criteria) to ensure analysis occurs on the intended population of interest. While measurements necessary to determine inclusion in a research study with primary data collection can be definitively ascertained through appropriate screening prior to enrollment, missingness in variables that define study eligibility criteria poses a serious challenge in EHR-based studies. In practice, subjects with incomplete eligibility data are almost always excluded from analysis \citep{danaei2013observational, McTigue2020}, although this can result in selection bias if subjects with observed eligibility data differ systematically from those whose eligibility status cannot be determined \citep{tompsett2023target, benz2024}.

Superficially, missing eligibility criteria may seem like a standard missing data problem. Unlike other missing data problems, however, the concessions analysts make when faced with missing eligibility, such as looking back further in time to ascertain eligibility, may inadvertently include individuals who, unbeknownst to the analyst, are ineligible at study baseline. Our work is directly motivated by the use of EHR data to understand long-term outcomes following bariatric surgery \citep{arterburn2014bariatric, arterburn2020, coleman2016longterm, coleman2022bariatric, obrien2018microvascular} among patients with type II diabetes mellitus (T2DM). As such, it would be inappropriate to include patients without diabetes in the study.

To date, very few papers have considered the problem of selection bias due to missing eligibility data. In early work, \citet{Pan2014} introduced an approach similar to inverse probability weighting (IPW) for proportional hazards regression, and \citet{heng2014} proposed a sensitivity analysis where subjects with missing eligibility criteria were treated as eligible and later ineligible in two separate analyses, though neither paper had an explicit focus on causal contrasts. More recently, \citet{tompsett2023target} and \citet{austin2023impute} proposed the use of multiple imputation for a single eligibility-defining covariate, with the latter focused on settings where that eligibility covariate was also the primary exposure of interest. \citet{benz2024} introduced inverse probability of eligibility ascertainment weights within a larger IPW framework for common challenges in EHR-based studies.

A notable limitation of the existing literature on missing eligibility criteria is the need to specify all relevant models correctly (e.g., imputation, outcome, treatment, missingness probability) to ensure consistent estimation of causal contrasts \citep{benz2024, austin2023impute}. On the other hand, estimating component models, or ``nuisance functions'', in a flexible/nonparametric manner via machine learning methods often yields slow rates of convergence, and thus invalid statistical inference \citep{kennedy2023semiparametric, chernozhukov2018}. Additionally, standard machine learning models often optimize loss functions that are not tied to specific causal contrasts, such as mean squared prediction error, and thus may result in bias in causal quantities utilizing those predictions \citep{kennedy2023semiparametric, kennedy2016semiparametric}. Tools from semiparametric theory can overcome some of these challenges, leading to estimators that are robust to various degrees of model misspecification, converge to desirable asymptotic distributions which facilitate valid statistical inference, and attain nonparametric efficiency bounds \citep{kennedy2016semiparametric, kennedy2023semiparametric, chernozhukov2018,levis2024robust}.

In this paper, we consider the setting where interest lies in using EHR data to estimate the causal average treatment effect of a point exposure on a given outcome among patients receiving a particular treatment in a targeted population of interest, articulated through a study's eligibility criteria. This is simply the causal average treatment effect on the treated (ATT) among study eligible subjects---as will be made clear, explicit conditioning on eligibility status makes apparent how restricting analysis to subjects with complete information on eligibility-defining covariates can introduce selection bias. Using a novel factorization of the observed data likelihood, we identify this causal estimand and develop various estimators based on semiparametric theory that are robust and asymptotically normal, including one which attains a semiparametric efficiency bound.

The remainder of the paper is organized as follows. Section \ref{sec:motivating_application} provides clinical background on bariatric surgery and elaborates on comparisons between two procedures among a population with obesity and T2DM. A detailed exploratory data analysis that motivates careful consideration of missing eligibility criteria is presented. Section \ref{sec:set_up} formalizes notation, assumptions, and the causal estimand of interest, while Section \ref{sec:estimators} presents a new efficient influence function-based estimator and details its properties. The use of this method in the motivating bariatric surgery study is presented in Section \ref{sec:data_application}, and finally, Section \ref{sec:discussion} concludes with discussion. Detailed proofs and in-depth simulation studies are provided in the Supplementary Material.

\if0\blind{\vspace{-1cm}}\fi
\section{Bariatric Surgery for Patients with Obesity and T2DM}
\if0\blind{\vspace{-0.5cm}}\fi
\label{sec:motivating_application}

\subsection{Background and Clinical Significance}
\label{sec:bariatric_surgery}
Bariatric surgery is an intervention to mitigate obesity and related comorbidities, with typical candidates having a body mass index (BMI) exceeding 35 kg/m$^2$ \citep{eisenberg2022}. The primary bariatric surgery procedures are Roux-en-Y gastric bypass (RYGB) and sleeve gastrectomy (SG), with SG having surpassed RYGB in frequency in the past 15 years \citep{arterburn2014bariatric, Reames2014}. SG is a simpler, less invasive procedure that is offered by more surgeons, and is associated with fewer short-term post-surgical complications \citep{li2021fiveyear}. On the other hand, patients undergoing SG may experience greater long-term re-operation risk \citep{kraljevic2025} and less substantial long-term weight loss \citep{arterburn2020}. Though comparisons between RYGB and SG have been published, evidence regarding their comparative effectiveness among patients with T2DM, particularly for long-term outcomes, is not considered definitive \citep{McTigue2020, jimenez2012long, salminen2018effect}. For example, the American Diabetes Association recommends consideration of bariatric surgery for T2DM patients with sufficiently large BMI as a form of weight and glycemic control, but does not make recommendations about whether SG or RYGB should be preferred \citep{ADA2024}. 

Two important clinical outcomes of interest for T2DM patients following bariatric surgery are relative weight change and remission of diabetes. These outcomes were previously examined in an EHR-based study by \citet{McTigue2020}, who found advantages to RYGB over SG on both outcomes among T2DM patients between 20-79 years of age with BMI $\geq 35$ kg/m$^2$ . However, their analysis excluded at least 10\% of surgical patients due to missing eligibility criteria, the implications of which were not explored.

With this backdrop, we investigate relative weight change and remission of diabetes at 3 years post surgery for a population of 14,809 patients undergoing RYGB or SG at one of three Kaiser Permanente (KP) sites (KP Washington, KP Northern California, KP Southern California) between 2008-2011. Our study utilizes EHR data from DURABLE, an NIH-funded study examining long-term outcomes of bariatric surgery across KP sites \citep{arterburn2020, coleman2016longterm, obrien2018microvascular, coleman2022bariatric, fisher2018association}. Following prior work \citep{obrien2018microvascular, coleman2016longterm, fisher2018association}, study eligibility criteria require that patients have T2DM, a BMI $\geq 35$ kg/m$^2$, and age between 19-79 years. When studying T2DM remission, we additionally require that patients have a DiaRem score of at least 3 points. DiaRem is a metric computed from pre-operative patient characteristics including age, hemoglobin A1c \% (A1c), and medication usage, and relates to the likelihood of experiencing T2DM remission following bariatric surgery, with lower scores indicating greater chance of remission \citep{Still2014}. This additional eligibility restriction places particular emphasis on patients less likely to experience remission (e.g., with greater disease severity), a population where differences between RYGB and SG on T2DM remission rates may be more substantial \citep{McTigue2020}. 

The popularity of SG has grown faster than the availability of long-term evidence for people with diabetes. Our study therefore focuses on the potential long-term benefits that T2DM patients might have missed by choosing SG instead of RYGB. Specifically, our goal is to assess the strength of evidence comparing RYGB and SG on 3-year post-surgical outcomes in light of how missing eligibility data are handled, and the sensitivity of such evidence to the practical decisions researchers make when faced with missing eligibility data. Such evidence can better inform and quantify tradeoffs physicians and patients consider when having shared decision making conversations about bariatric surgery.

Within the past year, recent work \citep{madenci2024, katsoulis2024bariatric} has critiqued prior EHR-based studies of bariatric surgery, including \citet{fisher2018association} which was based on DURABLE data, questioning the strength of evidence generated by these analyses, and suggesting sufficient clinical equipoise to warrant new large-scale randomized clinical trials. A key element in this debate was disagreement over how eligibility criteria should be applied in studies of bariatric surgery. \citet{haneuse2025} reconciled differences between results reported in Fisher et al. and those critiques, emphasizing generalizability of the DURABLE cohort and showing that additional exclusions proposed by critics did not materially affect the evidence supporting bariatric surgery. This recent debate underscores the importance of carefully considering how design choices, and (implicitly) how missing data affect operationalization of those choices, shape the analytic population, and ultimately, the evidence base guiding clinical decision making in bariatric surgery.

\subsection{Ascertainment of Eligibility}
\label{sec:t2dm_ascertainment}

\begin{figure}
    \centering
    \includegraphics[width=\textwidth]{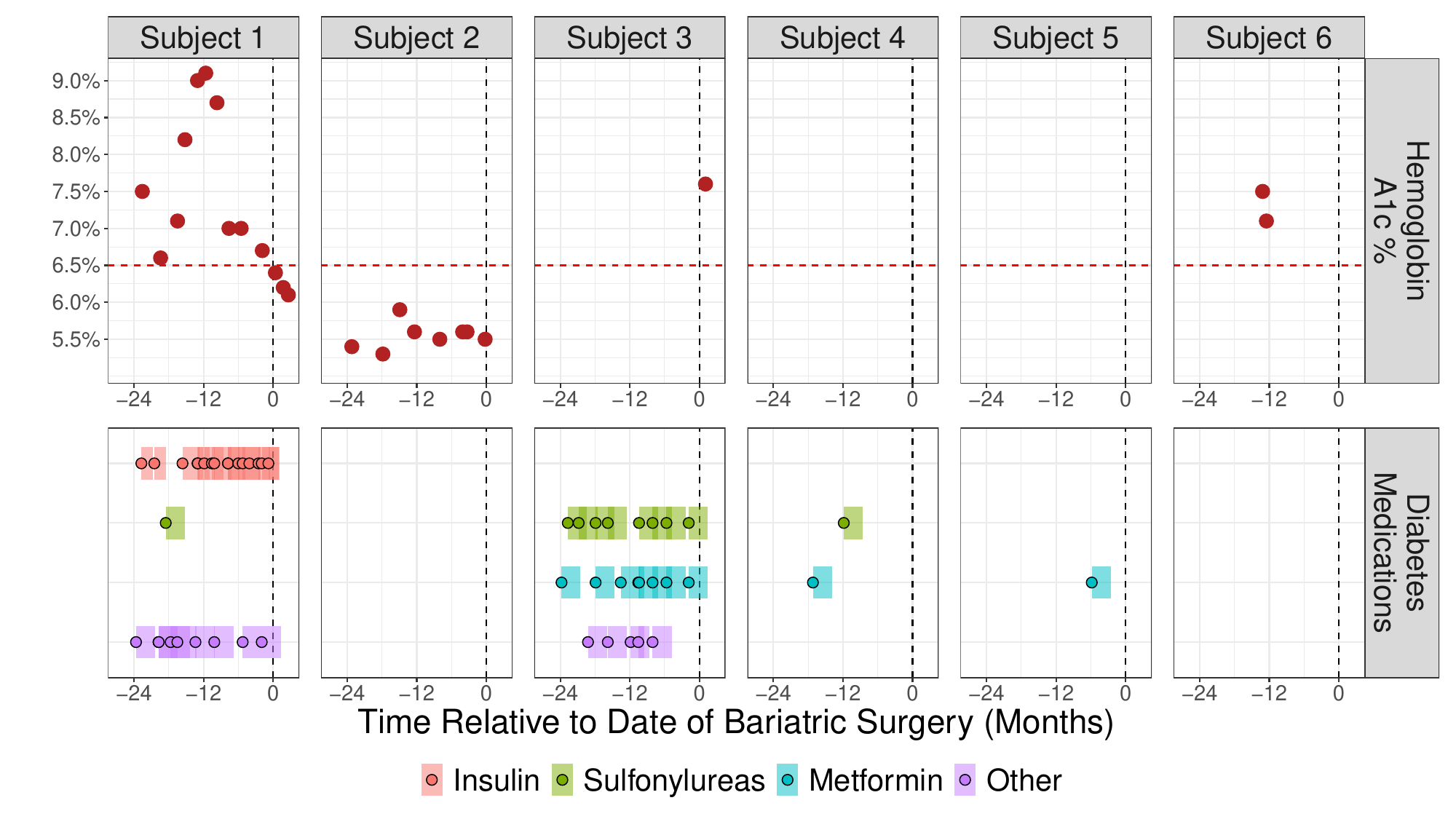}
    \caption{Hemoglobin A1c \% (dark red dots) and diabetic medication usage for six patients undergoing bariatric surgery, information that establishes T2DM status, and thus study eligibility in the 24 months prior to surgery. A1c measurements are shown in relation to a cutoff of 6.5\%, the typical clinical cutoff for T2DM. For medications, points indicate the start of a prescription while shaded bars indicate duration.}
    \label{fig:diabetes_ehr}
\end{figure}

Following previous DURABLE studies \citep{coleman2016longterm}, we define T2DM as a measurement of A1c $\geq 6.5\%$ (fasting blood glucose $\geq 126$ mg/dl), or via prescription for insulin or oral hypoglycemic medication. Metformin as the sole indicator of T2DM (e.g., subject 5 in Figure \ref{fig:diabetes_ehr}) is insufficient to establish diabetic status, and additionally requires an ICD-9 code of 250.x. Remission of T2DM is defined following \citet{coleman2016longterm}, as a measurement of A1c $< 6.5\%$ following a period of at least 90 days without T2DM medication, or fasting blood glucose $< 126$ mg/dl following a period of at least 7 days without T2DM medication. 

In practice however, determining which patients have T2DM at the time of surgery using EHR data is a non-trivial task. To illustrate this complexity, Figure \ref{fig:diabetes_ehr} depicts relevant diabetes information for six surgical patients. Some patients have frequent observations of A1c (e.g., subjects 1 and 2) making it straightforward to classify their diabetes status, with T2DM confirmed by active insulin prescription at the time of surgery in the case of subject 1. In contrast, other subjects have little to no information on A1c prior to surgery (e.g., 3-6). Despite having no pre-surgical A1c measure, subject 3 had an active prescription at the time of surgery of an oral hypoglycemic medication, allowing positive T2DM status to be established. Subjects 4 and 5 have no measures of A1c available prior to surgery, but unlike subject 3, had no active prescriptions at the time of surgery. Finally, subject 6 has multiple A1c measures exceeding 6.5\% but all measures are more than 12 months outdated by the time of surgery. Altogether, there is substantial heterogeneity in the patterns and frequency of observed information in the EHR making ascertainment of T2DM, and thus study eligibility, easy for some patients and difficult or impossible for others.

In determining how far to look back in time to ascertain study eligibility, different papers have used lookback windows of varying lengths \citep{McTigue2020, benz2024, obrien2018microvascular, coleman2016longterm, fisher2018association}, without consideration as to how such decisions might influence study conclusions. To better explore possible tradeoffs associated with this decision, we conduct analysis over a grid of 40 combinations of lookback windows for BMI (1, 3, 6, 12 months) and T2DM ascertainment via lab values (1, 3, 6, 12, 24 months) and medications (active prescription at the date of surgery, or active within 12 months prior to surgery). 

\begin{figure}
    \centering
    \includegraphics[width=0.95\textwidth]{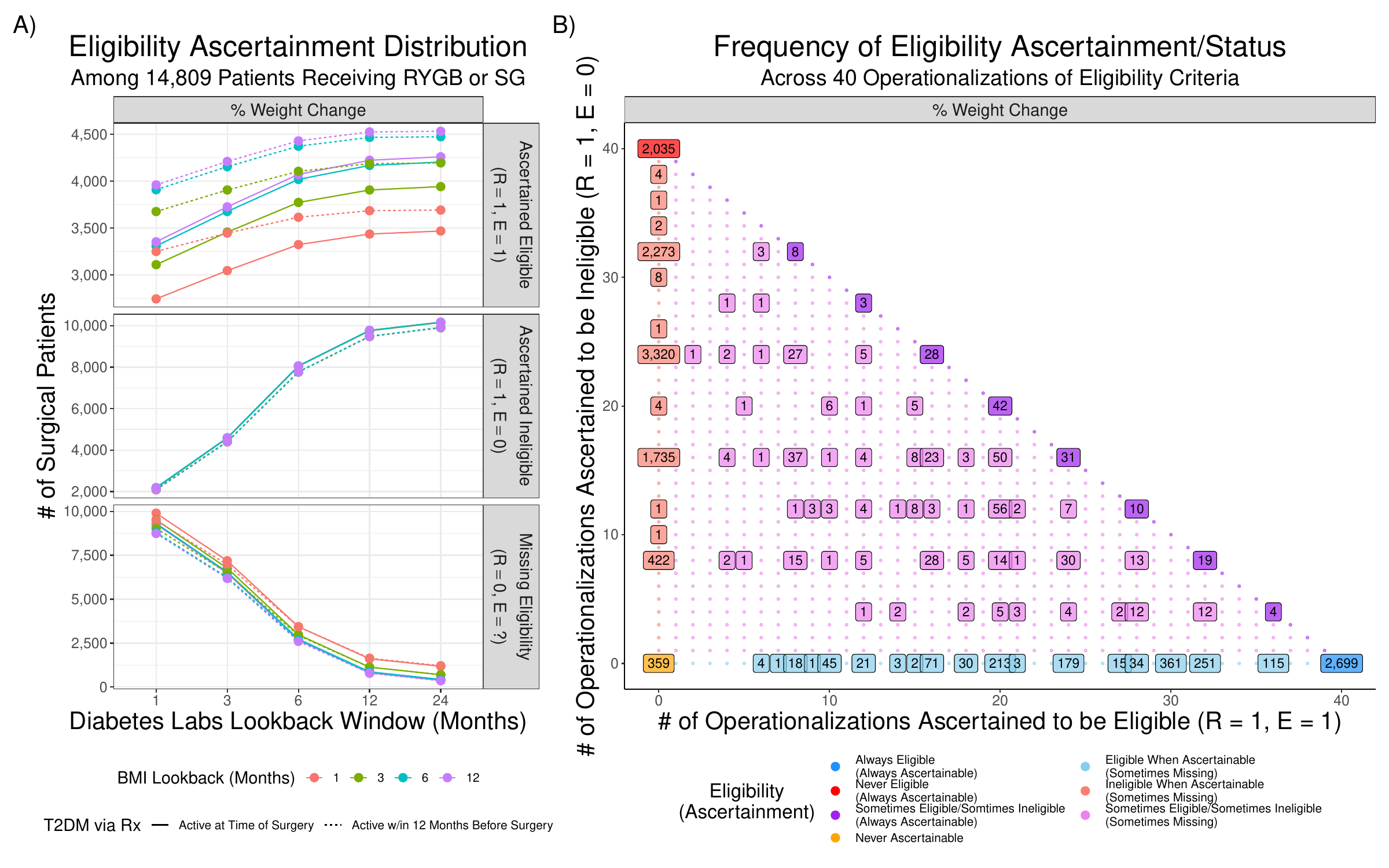}
    \caption{\textbf{A)} Joint distribution of eligibility ascertainment ($R$) and status ($E$) across 40 different possible ways to operationalize the study eligibility criteria. \textbf{B)} Distribution of $(n_{11}, n_{10})$ where $n_{re}$ denotes the number of ways of operationalizing the study eligibility criteria that a subject has $R = r, E = e$. A similar figure for the remission outcome is available in the Supplementary Material.}
    \label{fig:elig_dist}
\end{figure}

Observable information about the joint distribution of indicators for eligibility ascertainment ($R$) and status ($E$) is shown in Figure \ref{fig:elig_dist}. Panel A shows the number of surgical patients for each possible observable pair $(R,E)$ across combinations of lookback windows. In the shortest of lookback window (1 month for BMI and diabetes labs; active T2DM prescription at time of surgery), eligibility status is ascertainable for just 4,912 of 14,809 patients (33\%) and missing for 9,897 patients (67\%). By contrast, eligibility status is deemed missing for just 361 patients (2.4\%) when using the longest lookback windows, which is likely the impetus for certain works adopting longer lookback windows. 

For each patient, we can compute in how many of the 40 combinations they are ascertained to be eligible ($n_{11}$) and ineligible ($n_{10}$). Panel B of Figure \ref{fig:elig_dist} shows the number of patients with each unique combination of $(n_{11}, n_{10})$. When $n_{10} = 0$ or $n_{11} = 0$, subjects are only ever ascertained to be eligible or ineligible, respectively, under combinations where their eligibility status is not missing. Of note, for many subjects both $n_{11} > 0$ and $n_{10} > 0$. That is, several subjects are ascertained to be ineligible under certain strategies for applying inclusion criteria but eligible in others (rather than only eligible/missing or ineligible/missing). As such, simply applying the most relaxed operationalization of the eligibility criteria may inadequately reflect certain patients' eligibility status at surgery.

\subsection{Unique Challenges of Missing Eligibility Data}
\label{sec:unique_chal}

Figures \ref{fig:diabetes_ehr} and \ref{fig:elig_dist} serve to highlight that missing eligibility criteria is a distinct phenomenon from other forms of missing data for a few reasons. To begin with, complete case analyses that discard patients with missing outcomes or confounders do not risk inappropriately including patients in analysis. As is apparent from Figure \ref{fig:elig_dist}, minimizing exclusions due to missing eligibility by looking back further in time may inadvertently include patients in the study population who are not study eligible at the time of surgery (i.e., do not have T2DM). On the other hand, looking back only a short time before surgery to collect the most accurate information yields rates of missingness as substantial as 67\%. While confounders missing with this frequency could plausibly be dropped from analysis and treated as unmeasured, changing the eligibility criteria inherently changes the underlying analysis population and correspondingly, the scientific question of interest. Furthermore, when eligibility is missing with this frequency and longer lookbacks are employed, it can be tricky to determine the intended population to which study conclusions are applicable.

\if0\blind{\vspace{-0.5cm}}\fi
\section{Problem Set-up}
\if0\blind{\vspace{-0.5cm}}\fi
\label{sec:set_up}
\subsection{Notation}
\label{sec:notation}

Let $A \in \{0, 1\}$ denote a binary treatment (e.g., RYGB vs. SG), $Y \in \mathbb{R}$ denote an outcome of interest (e.g., relative weight change, or remission of diabetes) and $\bm L \in \mathbb{R}^k$ denote a set of baseline covariates that are (assumed) fully sufficient for three purposes: (1) to define study eligibility, (2) control for confounding, and (3) predict missingness in eligibility status. We denote eligibility-defining covariates by $\bm L^e$ and let $E = g(\bm L^e)$ denote a binary indicator of eligibility status, where $g(\cdot)$ is some fixed and known eligibility rule. In the context of our motivating application, $\bm L^e$ consists of BMI, age, A1c, DiaRem score (remission outcome only) and usage of insulin or oral hypoglycemic medications, and the eligibility-defining rule is BMI $\geq 35$ kg/m$^2$, $19 \leq \text{age}\leq 79$, and T2DM.

It is frequently the case that components of $\bm L^e$ are missing which precludes ascertainment of eligibility status $E$ (Figure \ref{fig:diabetes_ehr}). We denote the subset of eligibility-defining covariates with any missingness by $\bm L^e_m \in \mathbb{R}^q$, and let $\bm L^* = \bm L \setminus \bm L^e_m$. For example, age is an eligibility-defining covariate but is never missing in our application, and thus is in $\bm L^*$, while remaining eligibility-defining covariates are missing to various degrees and therefore are part of $\bm L^e_m$. Remaining covariates $\bm L^*$ include KP site, race, sex, age, estimated glomerular filtration rate (eGFR), self-reported smoking status, hypertension, dyslipidemia, and calendar year of surgery. Finally, we let $R$ be a binary indicator for each patient for whether or not the set of covariates $\bm L_m^e$ is completely observed, and consider a coarsened version of the observed data, with data units given by $O = (\bm L^*, A, Y, R, R\bm L^e_m)$. Some commentary on this possible discarding of information is offered in the discussion. 

We observe a sample $(O_1, \ldots, O_n)$ of $n$ independent and identically distributed observations from some underlying and unknown data distribution $P$, with corresponding empirical distribution $\mathbb{P}_n$. For any $P$-integrable function $f$ we define $Pf = \mathbb{E}_P[f] = \int f \,dP$ to denote the expectation of $f$ under $P$, and analogously define $\mathbb{P}_nf = \frac{1}{n}\sum_{i = 1}^n f(O_i)$ to express the sample mean of $f$ on data units $(O_1, ..., O_n)$. Lastly, we let $Y(a)$ denote counterfactual outcomes, that is the outcome that would have been observed under treatment $A = a$ \citep{hernan2024}. To simplify the exposition, we assume that there is no missing data outside of $\bm L^e_m$ (particularly in $A$, $Y$ or $\bm L^*$). In the discussion, we offer some considerations for situations where missing data also poses a challenge in outcomes and/or non-eligibility-defining covariates.

\if0\blind{\vspace{-0.5cm}}\fi
\subsection{Average Treatment Effect on the Treated and Study Eligible}
\if0\blind{\vspace{-0.2cm}}\fi
\label{sec:ATTE}
The causal estimand of interest is the average treatment effect on treated individuals meeting inclusion criteria for the study eligible population (ATTE), which we define as 

\begin{equation}\label{eqn:atte}
\theta_\text{ATT}^{\text{elig}} = \mathbb{E}[Y(1) - Y(0)~|~ A = 1, E = 1]    
\end{equation}

\noindent Note, this is the estimand of interest for researchers targeting the ATT after applying carefully constructed inclusion/exclusion criteria. Explicitly conditioning on eligibility status $E = 1$ in Equation~\eqref{eqn:atte} makes clear how the target population (inadvertently) can change when those missing eligibility are discarded. For example, when researchers discard subjects without ascertainable eligibility status $(R = 0)$, they are implicitly targeting the causal parameter $\theta^\text{CC}_\text{ATT} = \mathbb{E}[Y(1) - Y(0)~|~ A = 1, E = 1, R = 1] $. In cases where $\theta^\text{CC}_\text{ATT}  \neq \theta_\text{ATT}^{\text{elig}}$, that is, where the treatment effect among ascertainably eligible treated subjects differs from the treatment effect among all treated subjects meeting the study eligibility criteria, we say there is selection bias with respect to the ATTE.

\if0\blind{\vspace{-0.5cm}}\fi
\subsection{Assumptions}
\label{sec:assumptions}
We articulate two sets of assumptions to simultaneously address the potential for both selection bias and confounding. The first set of assumptions are standard causal inference assumptions and need only hold among the study eligible population. This is particularly salient for other applications comparing treatment to no-treatment, where study eligibility is often designed to exclude patients who are not realistic candidates for treatment. 

\begin{assumption}[Consistency]\label{as:consistency}
  $Y(A) = Y$ when $E = 1$, almost surely
\end{assumption}

\begin{assumption}[Positivity]\label{as:positivity}
  $\exists ~\epsilon > 0$ such that $ \epsilon \leq P(A = 1 ~|~\bm L^*, \bm L^e_m, E = 1) \leq 1 - \epsilon $, almost surely
\end{assumption}

\begin{assumption}[No Unmeasured Confounding]\label{as:NUC}
  $Y(a) \indep A~|~\bm L^*, \bm L^e_m, E = 1$ for $a \in \{0, 1\}$
\end{assumption}

\noindent In the absence of missing eligibility, Assumptions \ref{as:consistency}-\ref{as:NUC} are sufficient to identify $\theta_{\text{ATT}}^{\text{elig}}$, for example via the $g$-formula, $\mathbb{E}_P[Y~|~A = 1, E = 1] - \mathbb{E}_P[\mathbb{E}_P[Y~|~A = 0, \bm L^*, \bm L^e_m, E = 1]~|~A = 1, E = 1]$ \citep{hernan2024}. Given that $\bm L^e_m$, and thus $E$, may be missing, additional assumptions are required.

\begin{assumption}[Eligibility Missing At Random]\label{as:MAR}
  $R \indep (Y, \bm L^e_m)~|~\bm L^*, A$
\end{assumption}

\begin{assumption}[Complete Case Positivity]\label{as:ccpositivity}
$\exists ~\epsilon > 0$ such that $\epsilon \leq P(R = 1 ~|~\bm L^*, A) $, almost surely
\end{assumption}

Assumption \ref{as:MAR} is the core missing data assumption which states that whether or not a patient's eligibility status is observed is jointly independent of outcome and eligibility-defining covariates, conditional on all other completely observed information. While the plausibility of various assumptions is application specific, Assumption \ref{as:MAR} is most likely to be violated when there are either unobserved drivers of missingness or when the actual values of $\bm L^e_m$ influence how closely a patient is followed by their doctor. Alternative MAR assumptions and analysis strategies they might motivate, including the ``removal'' of $Y$ from Assumption \ref{as:MAR}, are examined in the discussion. Of course, analyses that ignore the problem of missing eligibility data altogether implicitly are invoking much stronger missing completely at random (MCAR) assumptions.

Thorough discussion of the validity of both sets of assumptions in the underlying study of bariatric surgery is presented in Section \ref{sec:validity}. Although Assumptions \ref{as:consistency}-\ref{as:ccpositivity} may be reasonably satisfied in our motivating study of bariatric surgery, the broad class of observational studies satisfying these assumptions generally only requires that there are no unmeasured confounders of treatment selection or drivers of missingness. 

In theory, any given covariate could be relevant for only one of the three stated purposes of $\bm L$. As articulated in Assumptions \ref{as:positivity}-\ref{as:NUC}, all of $\bm L$ is taken to be necessary insofar as the control of confounding is concerned. Furthermore, in EHR-based studies, the exact set of covariates which predict missingness is unlikely to be fully understood by researchers, and thus it would be reasonable for analysts to utilize all available information in $\bm L^*$ when trying to account for missingness in $\bm L^e$. This is not to say that Assumption \ref{as:MAR} is implausible in EHR-based studies, but rather to point out there is little downside to expanding the set of variables in $\bm L^*$ to include as many covariates as are readily available, particularly when adopting flexible modeling choices for component nuisance functions. The choice to include a broad set of covariates in $\bm L^*$ also helps mitigate the potential for unmeasured confounding.

It is plausible that in other applications, a more refined partitioning of $\bm L$ which further distinguishes confounders from covariates related only to eligibility status or ascertainment could yield slightly more parsimonious versions of Assumptions \ref{as:positivity}-\ref{as:ccpositivity}. In the Supplementary Material, we offer some exploration of an alternative strategy to further partition $\bm L$, and comment on additional complexities introduced. One example of a covariate serving only a single of $\bm L$'s stated purposes occurs when study eligibility restricts to a single stratum of a categorical eligibility-defining covariate (e.g., positive T2DM). In such cases, that covariate would no longer be a confounder among the study eligible population. Though we choose not to distinguish this further partitioning of $\bm L^e_m$ for ease of notation in subsequent sections, analysts might consider dropping covariates of this form when modeling component nuisance functions outlined in Table \ref{tab:nuisance}.

\subsection{Factorization of the Observed Data Likelihood}\label{sec:factorization}

In the presence of missing data, MAR assumptions allow for decompositions of the observed data likelihood that facilitate identification and estimation strategies for causal contrasts of interest \citep{levis2024robust}. In the present context, Assumptions \ref{as:MAR}-\ref{as:ccpositivity} motivate the following factorization:
\begin{equation}\label{eqn:factorization}
    p(O) = p(\bm L^*)p(A~|~\bm L^*)p(R~|~\bm L^*, A)p(\bm L^e_m~|~\bm L^*, A, R = 1)^Rp(Y~|~\bm L^*, \bm L^e_m, A, R = 1)
\end{equation}

Each component of Equation \eqref{eqn:factorization} is tied closely to nuisance functions which will play a critical role through the remainder of the work. Namely, $\pi(\bm L^*) = P(A = 1~|~\bm L^*)$ is a regression for treatment probability conditional on non-eligibility-defining covariates, and can be thought of as a partial propensity score; $\eta(\bm L^*, A) = P(R = 1~|~\bm L^*, A)$ is a regression for being a complete case, which depends only on fully observed quantities; $\lambda_a(\bm L^e_m;\bm L^*) = p(\bm L^e_m~|~\bm L^*, A = a, R = 1)$ denotes the conditional density of partially observed eligibility-defining covariates given treatment and fully observed covariates, among patients with complete information; and $\mu_a(\bm L^*, \bm L^e_m) = \mathbb{E}_P[Y~|~A = a, \bm L^*, \bm L^e_m, R = 1] = \int y \, dP(y~|~\bm L^*, \bm L^e_m, A=a, R = 1)$ is the mean outcome conditional on all other information, among complete cases. Table \ref{tab:nuisance} provides a summary of how these nuisance functions, as well as others introduced below, are used in various estimation strategies for the ATTE.

While alternative MAR assumptions could motivate alternative ways of factorizing the observed data likelihood, the factorization in Equation \eqref{eqn:factorization} has several appealing properties. Most important is that no component of the likelihood factorization conditions on a covariate stratum with incomplete information. As such, each component corresponds to a nuisance function which is readily estimable from the observed data. Moreover, by construction, components of the factorization are variationally independent, meaning any valid choice of conditional models for each of $\pi, \eta, \lambda$ and $\mu$ can be used to construct a valid joint density \citep{evans_2020, levis2024robust}. Critically, this allows analysts the flexibility to choose any modeling strategy for each component nuisance function separately, including both standard parametric models as well as modern machine learning techniques.

\begin{table}\scriptsize
    \centering
    \begin{tabular}{c@{~}c@{}c@{}c@{}c@{~}c@{~}c@{~}c}
        \toprule
        \textbf{Function} & \textbf{Definition} & \makebox[1.2cm]{\textbf{Section Introduced}} & \textbf{Mechanism(s)} & $\widehat\theta_{\text{CC}}$ & $\widehat\theta_{\text{IWOR}}$  & $\widehat\theta_{\text{IF}}$   & $\widehat\theta_{\text{EIF}}$ \\
        \midrule
        $\pi(\bm L^*)$ & $P(A = 1~|~\bm L^*)$ & \ref{sec:factorization} & Treatment & & & & \\[4pt]
        $\eta(\bm L^*, A)$ & $P(R = 1~|~\bm L^*, A)$ & \ref{sec:factorization} &  Ascertainment & & \checkmark & \checkmark & 
        \checkmark \\[4pt]
        $\lambda_a(\bm L^e_m; \bm L^*)$ & $p(\bm L^e_m~|~\bm L^*, A = a, R = 1)$ & \ref{sec:factorization} &  Imputation \\[4pt]
        $\mu_a(\bm L^*, \bm L^e_m)$ & $\mathbb{E}[Y~|~\bm L^*, \bm L^e_m, A = a, R = 1]$ & \ref{sec:factorization} &  Outcome  &  \checkmark & \checkmark & \checkmark & \checkmark \\[4pt]
        $u(\bm L^*, \bm L^e_m)$ & $P(A = 1~|~\bm L^*, \bm L^e_m, R = 1)$ & \ref{sec:eif_estimator} &  Treatment & & & \checkmark & \checkmark \\[4pt]
        $\varepsilon_a(\bm L^*, Y)$ & $P(E = 1~|~\bm L^*, Y, A = a, R = 1)$ &  \ref{sec:eif_estimator} & Imputation & & & & \checkmark \\[4pt]
        $\xi(\bm L^*, Y)$ & $\mathbb{E}\bigr[E\mu_0(\bm L^*, \bm L^e_m)~|~\bm L^*, Y, A = 1, R = 1]$ & \ref{sec:eif_estimator} &  Imputation/Outcome & & & & \checkmark \\[4pt] 
        $\gamma(\bm L^*, Y)$ & $\mathbb{E}\bigr[E\frac{u(\bm L^*, \bm L^e_m)}{1 - u(\bm L^*, \bm L^e_m)}~\bigr|~\bm L^*, Y, A = 0, R = 1\bigr]$ & \ref{sec:eif_estimator} &  Imputation/Treatment & & & & \checkmark \\[4pt]
        $\chi(\bm L^*, Y)$ & $\mathbb{E}\bigr[E\frac{u(\bm L^*, \bm L^e_m)}{1 - u(\bm L^*, \bm L^e_m)}\mu_0(\bm L^*, \bm L^e_m)~\bigr|~\bm L^*, Y, A = 0, R = 1\bigr]$ & \ref{sec:eif_estimator} &  Imputation/Treatment/Outcome & & & & \checkmark \\[4pt]
        $\nu(\bm L^*)$ & $\mathbb{E}\bigr[E\bigr(Y - \mu_0(\bm L^*, \bm L^e_m)\bigr)~\bigr|~\bm L^*, A = 1, R = 1\bigr]$ & \ref{sec:alternatives} & Imputation/Outcome & & &  \checkmark & \\[4pt]
        $\omega_a(\bm L^*)$ & $P(E = 1~|~\bm L^*, A = a, R = 1)$ & \ref{sec:alternatives} &  Imputation & & & \checkmark & \\[4pt] 
        \bottomrule
    \end{tabular}
    \caption{Definition of nuisance functions and their uses in various estimators of the ATTE}
    \label{tab:nuisance}
\end{table}

\subsection{Identification of the ATTE}
\label{sec:identification}

Under Assumptions \ref{as:consistency}-\ref{as:ccpositivity}, identification of $\theta_{\text{ATT}}^\text{elig}$ is possible, as summarized in Theorem \ref{thm:identification}.

\begin{theorem}\label{thm:identification}
Under Assumptions \ref{as:consistency}-\ref{as:ccpositivity}, $\theta_{\text{ATT}}^\text{elig}$ is identified by the functional $\theta(P) = \frac{\beta(P)}{\alpha(P)}$ where $\beta(P) = \mathbb{E}_P\Bigr[\frac{ARE}{\eta(\bm L^*, 1)}\bigr(Y - \mu_0(\bm L^*, \bm L^e_m)\bigr)\Bigr]$ and $\alpha(P) = \mathbb{E}_P\Bigr[\frac{ARE}{\eta(\bm L^*, 1)}\Bigr]$. 
\end{theorem}

\noindent Briefly, the result is obtained by showing that under Assumptions \ref{as:MAR}-\ref{as:ccpositivity}, $\theta(P)$ recovers the standard $g$-formula for the ATT under causal Assumptions \ref{as:consistency}-\ref{as:NUC} \citep{kennedy2015semiparametric}.

\if0\blind{\vspace{-2cm}}\fi
\section{Robust and Efficient Estimation of the ATTE}\label{sec:estimators}
\subsection{An Efficient Influence Function-Based Estimator of the ATTE}\label{sec:eif_estimator}
Given the form of $\theta(P)$, a natural choice of estimator for $\widehat\theta$ is a plug-in estimator of the form $\widehat\beta/\widehat\alpha$, which in turn motivates the development of robust and efficient one-step estimators for $\alpha(P)$ and $\beta(P)$ \citep{kennedy2023semiparametric, levis2024IV}. Towards that, we introduce the efficient influence functions of $\alpha(P)$ and $\beta(P)$ in Theorem \ref{thm:EIFs}.

\begin{theorem}\label{thm:EIFs}
    The efficient influence functions of $\alpha(P)$ and $\beta(P)$ at $P$ in the semiparametric model induced by Assumption \ref{as:MAR} (on the distribution of the coarsened observed data $O$) are given by
    \begin{equation}\label{eqn:alpha_EIF}
    \dot\alpha_P^*(O) = A\biggr(1 - \frac{R}{\eta(\bm L^*, 1)}\biggr)\varepsilon_1(\bm L^*, Y) + \frac{ARE}{\eta(\bm L^*, 1)} - \alpha(P) 
    \end{equation}
    \begin{equation}\label{eqn:beta_EIF}
    \begin{aligned}
    \dot\beta_P^*(O) &= \frac{AR}{\eta(\bm L^*, 1)}\biggr[\Bigr(E - \varepsilon_1(\bm L^*, Y)\Bigr)Y - 
   \Bigr(E\mu_0(\bm L^*, \bm L^e_m) - \xi(\bm L^*, Y)\Bigr)\biggr]  + A\Bigr( \varepsilon_1(\bm L^*, Y)Y - \xi(\bm L^*, Y)\Bigr)  \\
   & - \frac{(1-A)R}{\eta(\bm L^*, 1)}\biggr[E\frac{ u(\bm L^*, \bm L^e_m)}{1- u(\bm L^*, \bm L^e_m)}\Bigr(Y -  \mu_0(\bm L^*, \bm L^e_m)\Bigr) -\Bigr( \gamma(\bm L^*, Y)Y - \chi(\bm L^*, Y)\Bigr)\biggr] \\
& -(1-A)\frac{ \eta(\bm L^*, 0)}{ \eta(\bm L^*, 1)} \Bigr(\gamma(\bm L^*, Y)Y - \chi(\bm L^*, Y)\Bigr) - \beta(P)
    \end{aligned}
    \end{equation}
\end{theorem}

Both $\dot\alpha_P^*$ and $\dot\beta_P^*$ introduce additional nuisance functions. Most directly interpretable are $u(\bm L^*, \bm L^e_m) = P(A = 1~|~\bm L^*, \bm L^e_m, R = 1)$, the propensity score of treatment among complete cases, and $\varepsilon_a(\bm L^*, Y) = P(E = 1~|~\bm L^*, Y, A = a, R = 1)$, the probability of being eligible for the study conditional on all fully observed covariates, also among complete cases. Three additional nuisance functions appear in $\dot\beta_P^*$, which we refer to as nested nuisance functions given the appearance of simpler, previously defined nuisance functions, pre-multiplied by eligibility status $E$, within the outermost expectation of these more complex nuisance functions: $\xi(\bm L^*, Y) = \mathbb{E}_P\bigr[E\mu_0(\bm L^*, \bm L^e_m)~|~\bm L^*, Y, A = 1, R = 1], \gamma(\bm L^*, Y) = \mathbb{E}_P\bigr[E\frac{u(\bm L^*, \bm L^e_m)}{1 - u(\bm L^*, \bm L^e_m)}~\bigr|~\bm L^*, Y, A = 0, R = 1\bigr]$, and $\chi(\bm L^*, Y) =  \mathbb{E}_P\bigr[E\frac{u(\bm L^*, \bm L^e_m)}{1 - u(\bm L^*, \bm L^e_m)}\mu_0(\bm L^*, \bm L^e_m)~\bigr|~\bm L^*, Y, A = 0, R = 1\bigr]$. One way to interpret these nested nuisance functions is that they attempt to model treatment (e.g., via $u$) and/or outcome (e.g., via $\mu_0$) mechanisms by using marginalization to account for the fact that eligibility status remains unknown, and thus random, without complete knowledge of eligibility-defining covariates $\bm L^e$.

Influence functions are by definition mean zero, so it is helpful to work with uncentered influence functions, $\dot\alpha_P(O) = \dot\alpha^*_P(O) + \alpha(P)$ and $\dot\beta_P(O) = \dot\beta_P^*(O) + \beta(P)$. With these, we propose the estimator $\widehat\theta_{\text{EIF}}$ by taking the ratio of one-step estimators for  $\beta(P)$ and $\alpha(P)$, which each correspond to the sample mean of an estimated uncentered influence function:

\begin{equation}\label{eqn:theta_EIF}
\widehat\theta_{\text{EIF}} =  \frac{\mathbb{P}_n[\dot\beta_{\widehat P}(O)]}{\mathbb{P}_n[\dot\alpha_{\widehat P}(O)]}
\end{equation}

In Equation~\eqref{eqn:theta_EIF}, the notation $\widehat P$ is introduced to indicate that component nuisance functions in Equations~\eqref{eqn:alpha_EIF} and~\eqref{eqn:beta_EIF} are replaced by their corresponding estimated quantities. In simpler studies where there is no missing eligibility data and all subjects included in the study population are known to be eligible, $\widehat\theta_\text{EIF}$ simplifies to the standard one-step (doubly-robust) estimator for the ATT \citep{kennedy2015semiparametric}. We demonstrate this correspondence in the Supplementary Material.

\subsection{Theoretical Properties of \texorpdfstring{$\widehat\theta_{\text{EIF}}$}{Theta EIF}}\label{sec:properties_estimator}
Using one-step estimators for $\beta(P)$ and $\alpha(P)$, rather than $\theta^*_P(O)$ directly, is a matter of practical convenience in implementation; asymptotically the two approaches are equivalent \citep{levis2024IV}. In any case, $\theta^*_P(O)$ is useful in characterizing the asymptotic behavior of $\widehat\theta_\text{EIF}$.

\begin{corollary}\label{corr:theta_EIF}
The efficient influence function of $\theta(P)$ at $P$ in a semiparametric model induced by Assumption \ref{as:MAR} is given by $\dot\theta_P^*(O) = \frac{1}{\alpha(P)}\Bigr(\dot\beta_P(O) - \theta(P)\dot\alpha_P(O)\Bigr)$ 
\end{corollary}

\noindent Corollary \ref{corr:theta_EIF} follows from versions of the product rule and the chain rule for influence functions \citep{kennedy2023semiparametric}. Another pair of important quantities necessary to characterize the asymptotic behavior of $\widehat\theta_\text{EIF}$ are remainder terms $R_\alpha(\widebar P, P)$ and $R_\beta(\widebar P, P)$ from von Mises expansions \citep{kennedy2023semiparametric} for $\alpha(P)$ and $\beta(P)$, defined in the following lemma.

\begin{lemma}\label{lemma:vonMises}
$\alpha(P)$ and $\beta(P)$ satisfy the von Mises expansions 
$$
\begin{aligned}
\alpha(\widebar P) - \alpha(P) &= - \int \dot\alpha_{\widebar P}^*(o)dP(o) + R_\alpha(\widebar P, P) \\
\beta(\widebar P) - \beta(P) &= - \int \dot\beta_{\widebar P}^*(o)dP(o) + R_\beta(\widebar P, P) 
\end{aligned}
$$

\noindent where the remainder terms (omitting inputs for brevity) are as follows:

$$
\begin{aligned}
R_\alpha(\widebar P, P)  &= \mathbb{E}_P\biggr[A(\widebar\varepsilon_1 - \varepsilon_1)\biggr(1 - \frac{\eta_1}{\widebar\eta_1}\biggr)\biggr]
\end{aligned}
$$
$$
\begin{aligned}
R_\beta(\widebar P, P) &= \mathbb{E}_P\biggr[A\biggr(1 - \frac{\eta_1}{\widebar\eta_1}\biggr)\Bigr(Y(\widebar\varepsilon_1 - \varepsilon_1) -  (\widebar\xi - \xi )\Bigr) + \frac{RE}{\widebar\eta_1}(\mu_0 - \widebar\mu_0)\biggr(\frac{u(1 - \widebar u) - \widebar u(1 - u)}{1-\widebar u}\biggr ) \\
&~~~~~~~~~+ (1-A)\frac{(\eta_0 - \widebar\eta_0)}{\widebar\eta_1} \Bigr(Y(\widebar\gamma -   \gamma)  -(\widebar\chi - \chi)\Bigr)\biggr]
\end{aligned}
$$
\end{lemma} 

\noindent The rate of convergence for $\widehat\theta_\text{EIF}$ is closely tied to the second-order remainder terms from the von Mises expansions in Lemma \ref{lemma:vonMises}, and thus depends on products of errors in nuisance function estimation. Leveraging Lemma \ref{lemma:vonMises}, we summarize the asymptotic behavior of $\widehat\theta_\text{EIF}$ in the following theorem, where $\|f\|^2 = \mathbb{E}_P[f(O)^2]$ denotes the squared $L_2(P)$ norm.

\begin{theorem}\label{thm:aysmptotics}
If $\|\dot\alpha_{\widehat P} - \dot\alpha_P\| = o_P(1)$, $\|\dot\beta_{\widehat P} - \dot\beta_P\| = o_P(1)$, $\alpha(P) > 0$, and $P\Bigr[\bigr|\mathbb{P}_n\bigr(\dot\alpha_{\widehat P}(O)\bigr)\bigr| \geq \epsilon\Bigr] \to 1$ as $n \to \infty$ for some $\epsilon > 0$, then

$$
\widehat\theta_\mathrm{EIF} - \theta(P) = \mathbb{P}_n[\dot\theta^*_P(O)] + O_P\Bigr(|R_\alpha(\widehat P, P)| + |R_\beta(\widehat P, P)| \Bigr) + o_P(n^{-1/2})
$$

\noindent Moreover, if $|R_\alpha(\widehat P, P)| + |R_\beta(\widehat P, P)| = o_P(n^{-1/2})$ then $\sqrt{n}\bigr(\widehat\theta_\mathrm{EIF} - \theta(P)\bigr) \overset{d}{\to} \mathcal{N}\bigr(0, \mathrm{Var}_P[\dot\theta^*_P(O)]\bigr)$, whereby $\widehat{\theta}_{\mathrm{EIF}}$ attains the semiparametric efficiency bound induced by Assumption \ref{as:MAR}.
\end{theorem}

\begin{corollary}\label{corr:remainder} Under the conditions of Theorem \ref{thm:aysmptotics} and assuming that $P(\delta \leq 1 - \widehat u \leq 1-\delta) = 1$ for some $\delta > 0$, $P(\widehat\eta_1 > \epsilon) = 1$ for some $\epsilon > 0$, and $|Y| \leq M < \infty$ almost surely,
$$
R_\alpha(\widehat P, P) + R_\beta(\widehat P, P) = O_P\biggr(\|\widehat \mu_0 - \mu_0 \| \|\widehat u - u\| + \|\widehat \eta_1 - \eta_1 \|\Bigr\{ \|\widehat\varepsilon_1 - \varepsilon_1\|  + \|\widehat \xi - \xi\| \Bigr\} +  \|\widehat \eta_0 - \eta_0 \|\Bigr\{ \|\widehat\gamma - \gamma\|  + \|\widehat \chi - \chi\| \Bigr\}\biggr)
$$
    
\end{corollary}

\noindent Theorem \ref{thm:aysmptotics} demonstrates that $\widehat\theta_\text{EIF}$ can attain the parametric $\sqrt{n}$ rate even when component nuisance functions are estimated at slower rates, provided the product-rate conditions of Corollary \ref{corr:remainder} hold. As illustrated by Corollary \ref{corr:remainder}, the rate of convergence for $\widehat\theta_\text{EIF}$ is directly tied to error rates in estimation of component nuisance functions. When these product errors are $o_P(n^{-1/2})$---for example, when all nuisance function error rates are $o_P(n^{-1/4})$---then $\widehat\theta_\text{EIF}$ is not only $\sqrt{n}-$consistent and asymptotically normal, but also attains the semiparametric efficiency bound induced by Assumption \ref{as:MAR}. In particular, careful examination of Corollary \ref{corr:remainder} illustrates that the remainder term $|R_\alpha| + |R_\beta|$ will be $o_P(n^{-1/2})$ if (i) the product of errors for the standard ATT nuisance functions $(\mu_0, u)$, and (ii) the sum of product errors for missing data related nuisance functions ($\eta_0$, $\eta_1$, $\varepsilon_1, \xi, \gamma, \chi$) are each $o_P(n^{-1/2})$. The required rate conditions (e.g., $o_P(n^{-1/4})$ for each nuisance function) are achievable under various structural assumptions, such as sparsity, smoothness, or additivity. Under such conditions, asymptotically valid Wald-style $(1 - q)$-level confidence intervals are readily available, $\widehat\theta_\text{EIF} \pm z_{1-q/2}\sqrt{\frac{1}{n}\mathbb{P}_n[\dot\theta^{*}_{\widehat P}(O)^2 ]}$, where $z_q$ denotes the $q^\text{th}$ quantile of the standard normal distribution.

\if0\blind{\vspace{-0.5cm}}\fi
\subsection{Computation of \texorpdfstring{$\widehat\theta_{\text{EIF}}$}{Theta EIF}}\label{sec:estimation}
\if0\blind{\vspace{-0.5cm}}\fi

Routines for estimation of influence function-based estimators like $\widehat\theta_{\text{EIF}}$ typically involve sample splitting or cross-fitting, particularly when using nonparametric and/or machine learning techniques to estimate component nuisance functions \citep{chernozhukov2018}. In sample splitting,  $(O_1, ..., O_n)$ is randomly split into two disjoint samples $D_0, D_1$, with $D_0$ used to estimate component nuisance functions as well as any hyperparameters for relevant machine learning models. Models for nuisance functions trained on $D_0$ are subsequently applied to observations from the held out dataset $D_1$. Finally, the roles of $D_0$ and $D_1$ are reversed, and estimation of $\widehat\theta_{\text{EIF}}$ can proceed as in Equation \eqref{eqn:theta_EIF} by averaging influence function contributions for each subject. Given that $u$ and $\mu_0$ appear in the respective targets for $\xi, \gamma,$ and $\chi$, estimation of these nested nuisance functions requires predictions of $\widehat\mu_0$ and $\widehat u$ on training set $D_0$ to even train models for $\widehat\xi, \widehat\gamma,$ and $\widehat\chi$. Despite this complication, estimation of $\xi, \gamma,$ and $\chi$ can proceed within traditional sample splitting routines. To illustrate this, we outline the entire procedure for computing $\widehat\theta_{\text{EIF}}$ in a flexible manner in Algorithm \ref{alg:theta_EIF}. We use $k = 2$ splits for illustrative purposes but analogous procedures work for $k > 2$.

\begin{algorithm}[Computation of $\widehat\theta_{\text{EIF}}$]\label{alg:theta_EIF}
\begin{spacing}{1}
Let $D_0$ and $D_1$ be two disjoint, independent splits of $(O_1,...O_n)$ of sizes $n_0$ and $n_1$, respectively, with $n = n_0 + n_1$. The analytical procedure to compute the estimator $\widehat\theta_{\text{EIF}}$ of $\theta(P)$ is given as follows:

\begin{enumerate}
    \item Construct estimators $\widehat\eta$, $\widehat u$, $\widehat\varepsilon_1$, and $\widehat\mu_0$ on $n_0$ samples from $D_0$
    \item For $O_1, O_2, ..., O_{n_0} \in D_0$, compute $\bigr\{E\widehat\mu_0(\bm L^*_i, \bm L^e_{m_i}), E\frac{\widehat u(\bm L^*_i, \bm L^e_{m_i})}{1-\widehat u(\bm L^*_i, \bm L^e_{m_i})}, E\widehat\mu_0(\bm L^*_i, \bm L^e_{m_i})\frac{\widehat u(\bm L^*_i, \bm L^e_{m_i})}{1-\widehat u(\bm L^*_i, \bm L^e_{m_i})}\bigr\}_{i = 1}^{n_0}$, and use these quantities to construct estimators $\widehat\xi, \widehat \gamma$, and $\widehat\chi$,  on $n_0$ samples from $D_0$.
    \item For $O_{n_0 + 1}, O_{n_0 + 2}, ..., O_{n} \in D_1$, construct plug-in estimates of uncentered influence functions $\{\dot\alpha_{\widehat P_0}(O_i)\}_{i = n_0+1}^n$ and $\{\dot\beta_{\widehat P_0}(O_i)\}_{i = n_0+1}^n$ using the nuisance estimators constructed in steps 1 and 2.
    \item Repeat steps 1-3, swapping the roles of $D_0$ and $D_1$, to create sets $\{\dot\alpha_{\widehat P}(O_i)\}_{i = 1}^n = \{\dot\alpha_{\widehat P_0}(O_i)\}_{i = n_0+1}^n \cup \{\dot\alpha_{\widehat P_1}(O_i)\}_{i = 1}^{n_0}$ and $\{\dot\beta_{\widehat P}(O_i)\}_{i = 1}^n = \{\dot\beta_{\widehat P_0}(O_i)\}_{i = n_0+1}^n \cup \{\dot\beta_{\widehat P_1}(O_i)\}_{i = 1}^{n_0}$ (e.g., retaining influence function contributions from corresponding out-of-sample nuisance function predictions).
    \item Compute $\widehat\alpha = n^{-1}\sum_{i = 1}^n\dot\alpha_{\widehat P}(O_i)$ and $\widehat\beta = n^{-1}\sum_{i = 1}^n\dot\beta_{\widehat P}(O_i)$. Report $\widehat\theta_{\text{EIF}}$ as in Eq. \eqref{eqn:theta_EIF}.
    \item Compute $\Bigr\{\dot\theta^{*}_{\widehat P}(O_i) = \frac{1}{\widehat \alpha}\Bigr(\dot\beta_{\widehat P}(O_i) - \frac{\widehat\beta}{\widehat\alpha}\dot\alpha_{\widehat P}(O_i)\Bigr)\Bigr\}_{i = 1}^n$, and estimate $\widehat{\text{Var}}[\widehat\theta_{\text{EIF}}] = n^{-1}\mathbb{P}_n[\dot\theta^*_{\widehat P}(O)^2 ]$
\end{enumerate}
\end{spacing}
\end{algorithm}

\vspace{-0.95cm}
\noindent We note that while neither outcome model, $\mu$, nor propensity model, $u$, conditions on $E = 1$, all contributions of these nuisance functions in $\dot\beta_P(O)$ are pre-multiplied by eligibility indicator $E$, and thus only eligible subjects' values for these nuisances contribute to $\widehat\theta_\text{EIF}$. Because both models already condition on $R = 1$, analysts might additionally restrict the conditioning set of these models to include $E = 1$ and better model the contributions of these nuisance functions for eligible patients if treatment and outcome mechanisms differ drastically for eligibility status.

\if0\blind{\vspace{-0.5cm}}\fi
\subsection{Alternatives to \texorpdfstring{$\widehat\theta_{\text{EIF}}$}{Theta EIF}}\label{sec:alternatives}
\if0\blind{\vspace{-0.3cm}}\fi

In completely nonparametric models, there is only a single influence function for a pathwise differentiable statistical functional, and thus the influence function is the efficient influence function. That is not the case in our problem, however, since Assumption \ref{as:MAR} restricts the tangent space of possible nonparametric models. We derived the efficient influence functions $\dot\alpha_P^*$ and $\dot\beta_P^*$  by first finding nonparametric influence functions $\dot\alpha_P'^{*}$ and $\dot\beta_P'^{*}$ for $\alpha(P)$ and $\beta(P)$, and subsequently projecting them onto the tangent space (Supplementary Material Section 3.3). The uncentered versions of these influence functions, $\dot\alpha'_P$ and $\dot\beta'_P$, are given by
$$
\dot\alpha'_P(O) =  A\biggr(1 - \frac{R}{\eta(\bm L^*, 1)}\biggr)\omega_1(\bm L^*) + \frac{ARE}{\eta(\bm L^*, 1)} 
$$
$$
\dot\beta'_P(O) = A\biggr(1 - \frac{R}{\eta(\bm L^*, 1)}\biggr)\nu(\bm L^*) + \frac{RE}{\eta(\bm L^*, 1)}\Bigr(Y - \mu_0(\bm L^*, \bm L^e_m)\Bigr)\biggr[A - (1-A)\frac{ u(\bm L^*, \bm L^e_m)}{1 - u(\bm L^*, \bm L^e_m)}\biggr]
$$

\noindent $\dot\alpha'_P$ and $\dot\beta'_P$ contain fewer nested nuisance functions and thus motivate an estimator which is simpler to compute, using an analogous procedure to Algorithm \ref{alg:theta_EIF} and Equation \eqref{eqn:theta_EIF}, and replacing $\dot\alpha_P$ and $\dot\beta_P$ with $\dot\alpha_P'$ and $\dot\beta_P'$, respectively. 

$$
\widehat\theta_\text{IF} = \frac{\mathbb{P}_n[\dot\beta'_{\widehat P}(O) ]}{\mathbb{P}_n[\dot\alpha'_{\widehat P}(O)]} 
$$
 
The estimator $\widehat\theta_\text{IF}$ introduces two new nuisance functions, $\omega_a(\bm L^*) = P(E = 1~|~\bm L^*, A = a, R = 1)$, and $\nu(\bm L^*) = \mathbb{E}_P\bigr[E\bigr(Y - \mu_0(\bm L^*, \bm L^e_m)\bigr)~\bigr|~\bm L^*, A = 1, R = 1\bigr]$. See that $\omega$ is very similar to $\varepsilon$ in that it models eligibility among complete cases given completely observed information, only it does not use outcomes $Y$ when modeling this probability. Though $\nu$ is a nested nuisance function, other nested nuisance functions $\xi, \gamma$ and $\chi$ do not need to be estimated for $\widehat\theta_\text{IF}$. While one would expect $\widehat\theta_\text{IF}$ to be less efficient than $\widehat\theta_\text{EIF}$, $\widehat\theta_\text{IF}$ is easier to implement, and could plausibly result in roughly the same (or less) small sample bias without much of a loss (if any) in terms of efficiency. This trade-off is one we explore and offer guidance on through the simulation study in Supplementary Appendix S4.

Additionally, we consider two alternative estimators of $\theta(P)$ which are easier computationally than $\widehat\theta_\text{EIF}$ or $\widehat\theta_\text{IF}$. As a baseline, we consider $\widehat\theta_{\text{CC}}$, the complete case $g$-formula estimator for the ATT, where $\mathcal{C}$ denotes the set of eligible complete cases undergoing RYGB:
$$
\widehat \theta_\text{CC} = \frac{\mathbb{P}_n\bigr[ARE\bigr(Y - \widehat\mu_0(\bm L^*, \bm L^e_m)\bigr)\bigr]}{\mathbb{P}_n[ARE]} = \frac{1}{|\mathcal{C}|}\sum_{i \in \mathcal{C}}\Bigr(Y_i - \widehat \mu_0(\bm L_i^*, \bm L_{m_i}^e)\Bigr)
$$

\noindent  This strategy is representative of the complete-case analyses that are common in practice, and would be valid if $\theta^\text{CC}_\text{ATT} = \theta^\text{elig}_\text{ATT}$. 

Examination of the identification result in Theorem \ref{thm:identification} additionally motivates the following inverse weighted outcome regression (IWOR) estimator, $\widehat\theta_\text{IWOR}$.

$$
\widehat \theta_\text{IWOR} = \frac{\mathbb{P}_n\bigr[\frac{ARE}{\widehat\eta(\bm L^*, 1)}(Y - \widehat\mu_0(\bm L^*, \bm L^e_m)\bigr)\bigr]}{\mathbb{P}_n\bigr[\frac{ARE}{\widehat\eta(\bm L^*, 1)}\bigr]} 
$$

\noindent Unlike $\widehat\theta_\text{CC}$, which ignores the possibility of selection bias, $\widehat \theta_\text{IWOR}$ accounts for the possibility of selection bias and requires only a single extra nuisance function, $\eta$.

Finally, it is worth noting that there is an equivalent form of $\widehat\theta_{\text{EIF}}$, which we denote $\widetilde\theta_\text{EIF}$, that uses the nuisance functions $\pi$ and $\lambda$ instead of $u$, and thus is more directly tied to the likelihood factorization in Equation \eqref{eqn:factorization}. Given that $\lambda$ is a conditional density for $\bm L^e_m$, which itself may be multi-dimensional, ratios of the form $\frac{\lambda_1(\bm L^e_m; \bm L^*)}{\lambda_0(\bm L^e_m; \bm L^*)}$ (as is required by $\widetilde\theta_{\text{EIF}}$) would be difficult to estimate via nonparametric methods. For completeness, we provide additional details on $\widetilde\theta_\text{EIF}$ in the Supplementary Material,  and illustrate the connection between the conditional density ratio, $\frac{\lambda_1(\bm L^e_m; \bm L^*)}{\lambda_0(\bm L^e_m; \bm L^*)}$, and the propensity score ratio, $\frac{u(\bm L^*, \bm L^e_m)}{1 - u(\bm L^*, \bm L^e_m)}$.

\if0\blind{\vspace{-0.5cm}}\fi
\section{Comparison of Bariatric Surgical Procedures}
\if0\blind{\vspace{-0.5cm}}\fi
\label{sec:data_application}
\subsection{Methodological Details}
\if0\blind{\vspace{-0.5cm}}\fi
\label{sec:application_details}

Finally, we return to the bariatric surgery study introduced and discussed in Section \ref{sec:bariatric_surgery}. For both the relative weight change and T2DM remission outcomes, we present results for each estimator proposed in Section \ref{sec:estimators} across all 40 operationalizations of the study eligibility criteria (Figure \ref{fig:elig_dist}). Both influence-function based estimators, $\widehat\theta_\text{EIF}$ and $\widehat\theta_\text{IF}$, were computed using \texttt{SuperLearner} to estimate relevant nuisance functions \citep{superlearner}. The complete-case estimator, $\widehat\theta_\text{CC}$, was computed using a linear model for $\mu$ with pre-specified interactions between surgery type and select $\bm L^*$ covariates. Finally, the inverse-weighted outcome regression estimator, $\widehat\theta_\text{IWOR}$, used the same $\mu$ specification as $\widehat\theta_\text{CC}$, and used a main-effects logistic regression for $\eta$. 95\% confidence intervals for both $\widehat\theta_\text{CC}$ and $\widehat\theta_\text{IWOR}$ were computed via bootstrapping using a normal approximation. Additional methodological details are available in the Supplementary Material.

\if0\blind{\vspace{-0.5cm}}\fi
\subsection{Validity of Assumptions}
\if0\blind{\vspace{-0.3cm}}\fi
\label{sec:validity}
Careful assessment of Assumptions \ref{as:consistency}-\ref{as:ccpositivity} in the context of our bariatric surgery study is critical towards generating high-quality evidence to support clinical practice. First, we note that practical positivity violations would most directly influence $\widehat\theta_\text{EIF}$ through estimation of the complete data propensity score, $u$, along with nested nuisance functions which rely on ratios $\frac{u}{1-u}$. Examination of Equation \eqref{eqn:beta_EIF} reveals that $u$ contributes to $\widehat\theta_\text{EIF}$ only through eligible complete cases undergoing SG. Among such patients, the 1\% and 99\% quantiles of $\widehat u$ were 0.351 and 0.936, respectively, suggesting that Assumption \ref{as:positivity} holds in the present context. Positivity violations are of greater concern in comparisons between surgical and non-surgical patients, given the rare nature of initiating surgery relative to the pool of potentially eligible candidates. In such contexts, trimming of $\widehat u$ might be reasonable \citep{hernan2024}.

Previous studies \citep{Harrington2024, fisher2018association, coleman2022bariatric, obrien2018microvascular} using DURABLE data have examined E-Values \citep{haneuse2019evalue} as a way to quantify the strength of unmeasured confounding that would be required to overturn observed results. Each of these studies has found the strength of association that would be required by unmeasured factors (beyond the dozen plus already adjusted for) to overturn results would be far greater than any known across the literature. We present E-Values for both the point estimate and confidence interval for $\widehat\theta_\text{EIF}$ in Supplementary Materials Table S4. All E-values for the point estimate exceed 2, meaning an unmeasured confounder would need to be associated with both surgical procedure and 3-year weight change by a risk ratio of at least 2-fold each, above and beyond the measured covariates $\bm L^*$ and $\bm L^e_m$, to fully explain away the observed treatment effect. This magnitude of confounding exceeds that observed for any of our measured covariates and is not thought to characterize unmeasured factors in other DURABLE studies, suggesting the strength of any remaining unmeasured confounding is unlikely to bias results in a clinically significant manner.

 \begin{figure}
    \centering
    \includegraphics[width=\textwidth]{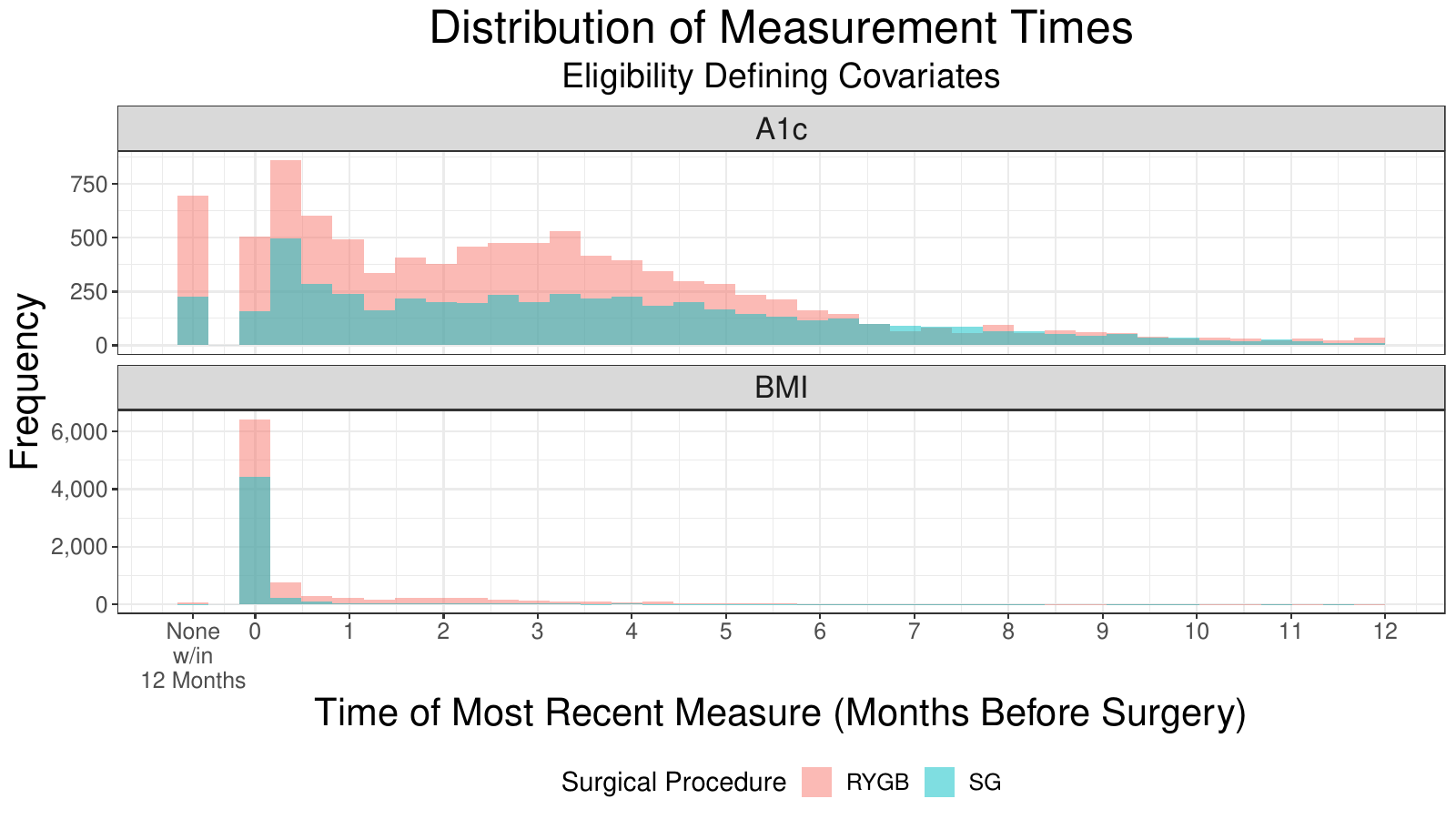}
    \caption{Distributions of time relative to the date of surgery at which the most recent BMI or A1c measure was collected}
    \label{fig:measure_times}
\end{figure}

Bariatric surgery is commonly preceded by a 3-6 month period in which a patient has greater levels of interaction with the health care system in an effort to monitor and encourage possible preoperative lifestyle changes, including weight loss and smoking cessation \citep{madenci2024}. Thus, it seems plausible that whether or not measures of BMI and A1c or medication usage are collected during this preoperative period would depend only on information recorded in the EHR. By contrast, when patients do not have BMI or A1c measures available in the 1-2 years prior to surgery, it might speak to a lower propensity to engage in health-seeking behavior. Altogether, it is certainly plausible that Assumption \ref{as:MAR} is most likely to hold within 3-6 months of surgery and may be less likely to hold when looking 1-2 years prior to surgery to ascertain eligibility. This is reflected in Figure \ref{fig:measure_times}, which displays the distribution of the time relative to the date of surgery at which the most recent BMI and A1c values were recorded in the EHR. In particular, 83.8\% of patients have a BMI measurement recorded in the month prior to surgery, and the 3 month mark is where the A1c distribution begins to fall off. Still, 19.2\% of patients do not have an A1c measurement recorded in the 6 months prior to surgery, compared to just 1.5\% for BMI, suggesting that T2DM status may be the chief driver of possible MAR violations beyond 3-6 months.

\begin{figure}
    \centering
    \includegraphics[width=\textwidth]{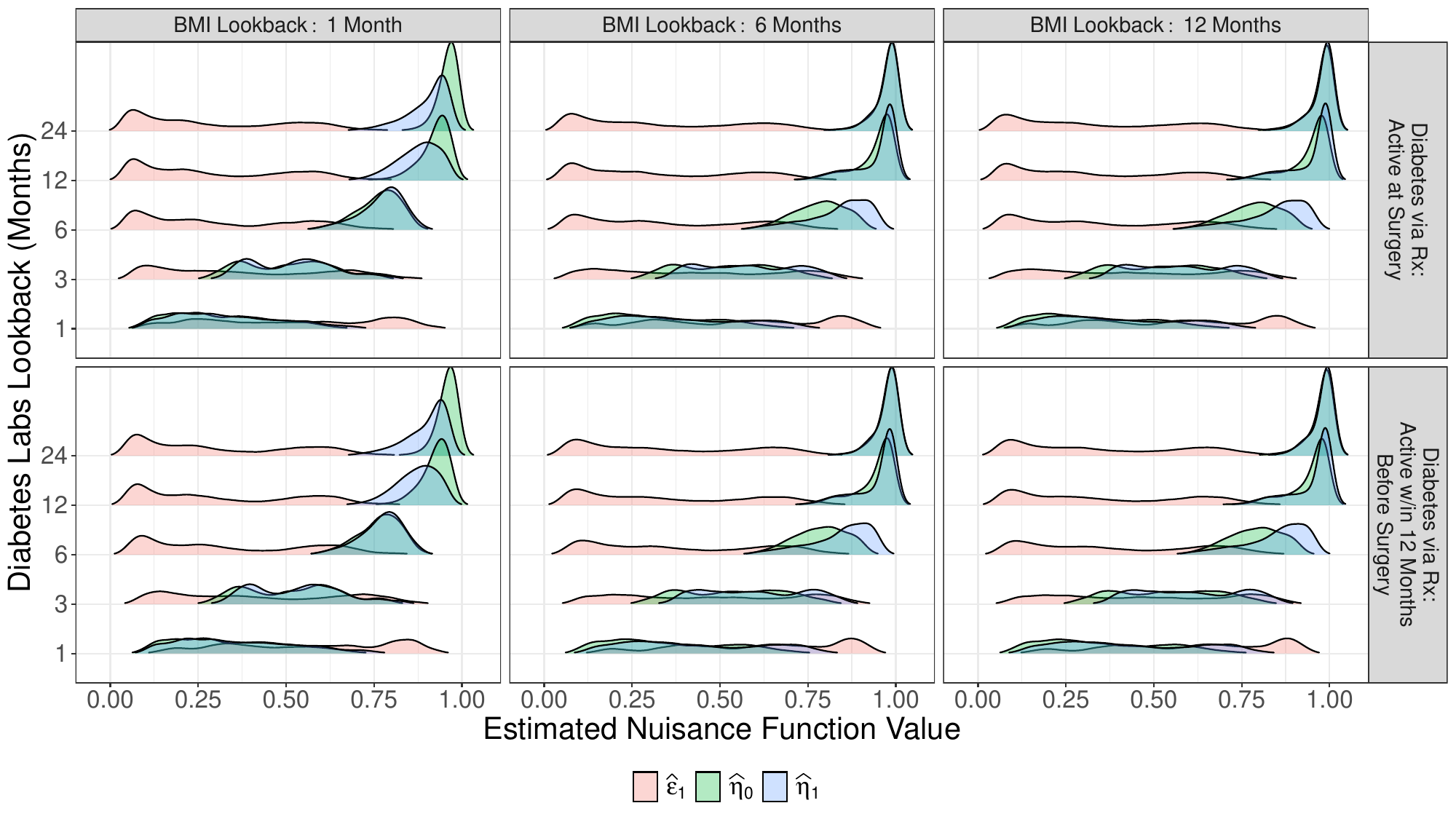}
    \caption{Distributions of select nuisance function estimates from $\widehat\theta_\text{EIF}$ related to ascertainment ($\widehat\eta$) and eligibility ($\widehat\varepsilon$) for relative weight change outcome. An analogous figure for T2DM remission is available in the Supplementary Material. Results for BMI lookback of 3 months are similar to those of 1 month BMI lookback, and omitted for space considerations.}
    \label{fig:nuisance_fx}
\end{figure}

To get a sense of how eligibility-related nuisance function estimates varied by lookback window length, and to gain additional insight into Assumption \ref{as:MAR}, Figure \ref{fig:nuisance_fx} presents the distribution of estimates of complete case probability $\widehat\eta$ and eligibility probability $\widehat\varepsilon$ used in computing $\widehat\theta_\text{EIF}$. Average values of $\varepsilon_1$ ranged from 0.613 in the most stringent settings to 0.300 in the most relaxed, indicating that a higher proportion of subjects with complete information were judged to be eligible in lookback windows with narrower scope. While this observation cannot directly assess the validity of Assumption \ref{as:MAR} across operationalizations, it speaks to the higher quality of information available closer to the time of surgery, and thus in our view, greater plausibility that the MAR assumption holds. In the Supplementary Material, we provide a formal sensitivity analysis framework to query the impact that MNAR violations of Assumption \ref{as:MAR} might have on the substantive clinical conclusions presented in Section \ref{sec:application_results}. This sensitivity model perturbs the ascertainment probability only through the eligibility-defining covariates $\bm L^e_m$, holding $\bm L^*$ and $A$ fixed. As such, it addresses departures from Assumption \ref{as:MAR} in which ascertainment depends on $\bm L^e_m$ given $\bm L^*$ and $A$; it does not address missingness that depends on the outcome $Y$ itself, given $\bm L^*$, $\bm L^e_m$, and $A$.

The distribution of complete case probabilities in Figure \ref{fig:nuisance_fx} displayed greater sensitivity to the lookback length for diabetes labs than for BMI. Similarity between the rows of Figure \ref{fig:nuisance_fx} suggests that ascertainment of T2DM was more affected by availability of lab measures than by medications. The covariate-adjusted average difference in observing complete eligibility information between RYGB and SG, $\mathbb{P}_n[\widehat \eta_1(\bm L^*) - \widehat \eta_0(\bm L^*)]$, ranged from -6.1\% to 7.6\%. Differential missingness by procedure was most prominent when the length of time for BMI and T2DM lookback windows differed. Altogether, Figures \ref{fig:measure_times} and \ref{fig:nuisance_fx} lead us to have greatest confidence in results from more stringent operationalizations of the eligibility criteria. Additional guidance in the form of a diagnostic checklist for assessing assumptions and evaluating possible stability issues for $\widehat\theta_\text{EIF}$ is available in the Supplementary Material.

\if0\blind{\vspace{-0.5cm}}\fi
\subsection{Study Results}
\label{sec:application_results}

\begin{figure}[H]
    \centering
    \includegraphics[width=\textwidth]{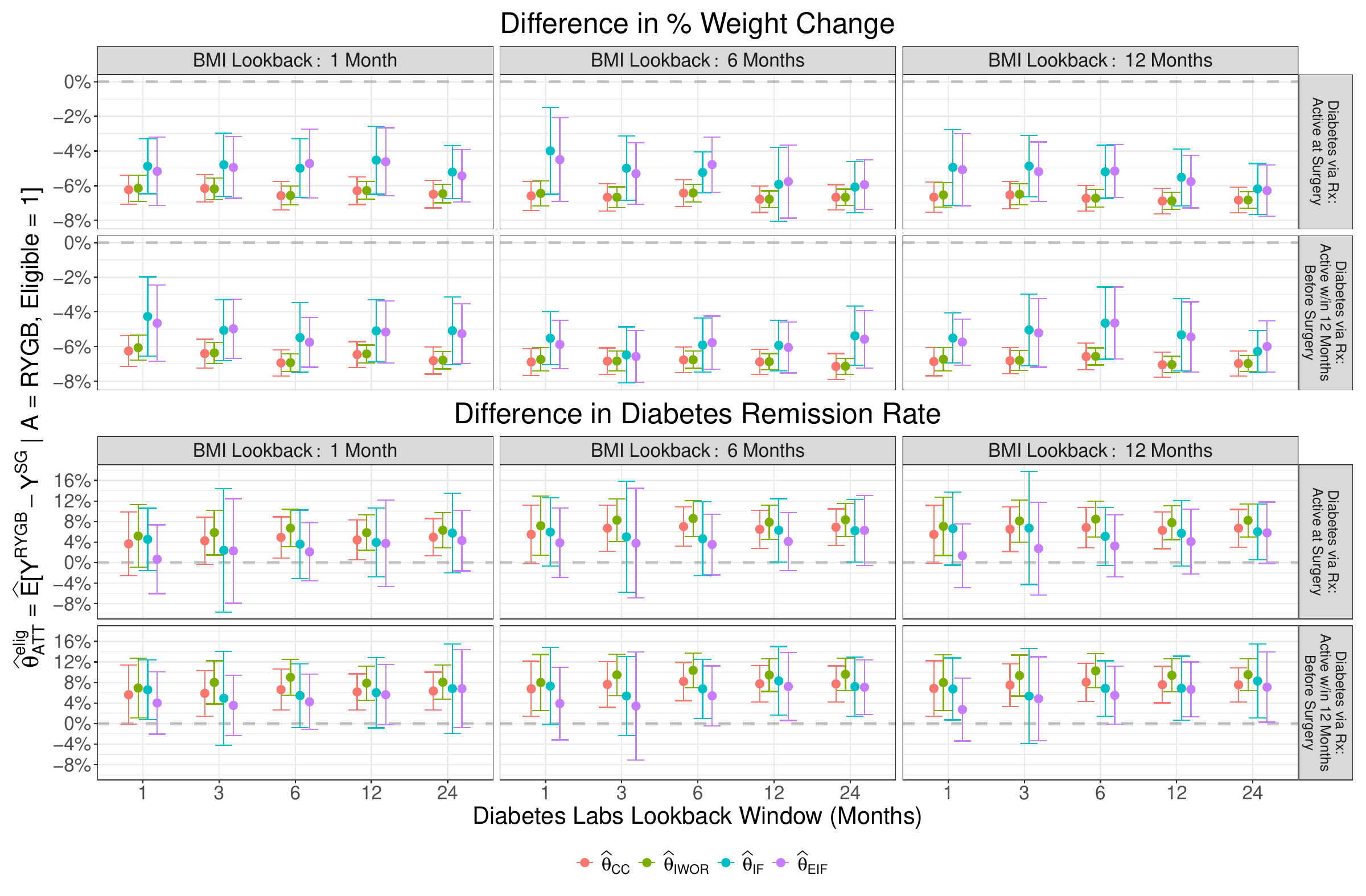}
    \caption{Point estimates and 95\% confidence intervals for four estimators of the average treatment effect of bariatric surgery type on eligible RYGB patients. Estimates are presented for difference in \% weight change and diabetes remission rate 3 years post-surgery. Results for BMI lookback of 3 months are similar to those of 1 month BMI lookback, and omitted for space considerations.}
    \label{fig:results}
\end{figure}

Point estimates and 95\% confidence intervals for $\widehat\theta_\text{CC}, \widehat\theta_\text{IWOR}, \widehat\theta_\text{IF}$, and $\widehat\theta_\text{EIF}$ across application of eligibility for both outcomes are presented in Figure \ref{fig:results}. Point estimates for $\widehat\theta_\text{CC}$ ranged from  -7.1\% to -6.2\% for 3-year post surgical weight change and 3.7\% to 8.2\% for T2DM remission, indicating that eligible subjects undergoing RYGB (with complete information) experienced greater weight loss and rates of remission than had they undergone SG instead. 

$\widehat\theta_\text{IWOR}$ had the same range for weight change estimates as $\widehat\theta_\text{CC}$, but indicated greater differences in T2DM remission (range 5.2\% to 10.3\%). Point estimates for $\widehat\theta_\text{EIF}$ were attenuated towards the null by an average of 19.5\% for relative weight change compared to $\widehat\theta_\text{CC}$ (range -6.6\% to -4.5\%), and 34.1\% for T2DM remission (range 0.6\% to 7.4\%). Estimated standard errors were on average 5.3\% larger for $\widehat \theta_\text{IF}$ than corresponding standard error estimates for $\widehat\theta_\text{EIF}$. Point estimates from influence-function based estimators were most similar to those of $\widehat\theta_\text{CC}$ and $\widehat\theta_\text{IWOR}$ in longer lab lookbacks, with differences exceeding 6.5\%  in a 1-month T2DM labs lookback window. While $\widehat\theta_\text{EIF}$ still shows significant weight loss benefits to RYGB relative to SG, there is less evidence to suggest an RYGB advantage for remission of T2DM, at least among patients with a DiaRem score of $\geq 3$.

Taken holistically, our results confirm the weight loss advantage for RYGB over SG. They also suggest that patients undergoing RYGB would not have experienced a substantially lower T2DM remission rate had they received SG instead. As such, clinicians may consider remission of diabetes to be less significant of a factor in determining which surgical procedure to recommend to prospective patients, at least relative to factors like durability of weight loss and risk of complications associated with surgery. However, clinicians and patients may also consider other factors such as the durability of diabetes remission in making their decisions \citep{McTigue2020}.

Finally, we note that in Figure \ref{fig:results}, 95\% confidence intervals for $\widehat\theta_\text{EIF}$ and $\widehat\theta_\text{IF}$ do not tighten monotonically as the operationalization of study eligibility becomes less stringent. There seems to be a complex interplay between increasing the number of subjects with complete information and the complexity of modeling certain nuisance functions. While increasing the ``$R = 1$'' fraction corresponded to a greater number of patients deemed eligible (Figure \ref{fig:elig_dist}), more stringent lookbacks yielded a smaller fraction of patients with complete eligibility information (Figure \ref{fig:nuisance_fx}). In particular, nested nuisances ($\gamma, \xi$, $\chi$, $\nu$) model compound outcomes that are pre-multiplied by $E$, and thus may simultaneously have to contend with greater degrees of zero-inflation and extreme values (Figure S4 of the Supplementary Material).

In the Supplementary Material, we conduct a tipping point sensitivity analysis to assess the potential impact of violations of Assumption~\eqref{as:MAR}. Briefly, for the confidence interval for RYGB's 3-year weight loss advantage to cross 0, patients with the most severe obesity (BMI $\geq 60$~kg/m$^2$) would need to have roughly $1/20^{\text{th}}$ the odds of ascertainable eligibility status compared to otherwise identical patients with a BMI of 35~kg/m$^2$. By contrast, for the T2DM remission outcome, much smaller values of the sensitivity parameters are enough to make the difference in remission rates favor SG. In particular, the point estimate favors SG under a sensitivity parameter corresponding to an odds ratio of 0.97 per 1~kg/m$^2$ increase in BMI above 35~kg/m$^2$ for having ascertainable eligibility status (about half the odds at a BMI of 60 vs.\ 35~kg/m$^2$, compared with $1/20^{\text{th}}$ for weight loss). These results reinforce the idea that T2DM remission may be a secondary consideration when choosing between SG and RYGB, relative to weight loss and surgical safety.

\if0\blind{\vspace{-0.7cm}}\fi
\section{Discussion}
\if0\blind{\vspace{-0.3cm}}\fi
\label{sec:discussion}

Although there is substantial observational evidence regarding long-term outcomes following bariatric surgery, debate continues with some arguing that sufficient clinical equipoise exists to warrant large-scale randomized clinical trials. Among the various aspects of this debate is how to best design trials with clear eligibility criteria that are relevant to clinical practice. While previous observational studies have tended to adopt pragmatic strategies when dealing with missing data in eligibility criteria (e.g., exclusion and lookback windows), this is the first paper examining long-term outcomes following bariatric surgery using real-world EHR data coupled with formalized methods in causal inference that acknowledge missingness in eligibility criteria. The results reported in Section \ref{sec:application_results} align with previous studies \citep{McTigue2020, arterburn2020}, strengthening the evidence base for current clinical guidelines and adding support to the argument that there is insufficient clinical equipoise to justify new large-scale randomized trials.

When designing and analyzing observational studies from EHR databases, the potential for confounding bias is often considered paramount, with missing data frequently viewed as a secondary consideration. While analysts might consider methods for dealing with missing data in certain covariates or outcomes after finalizing their analysis population, it is rare that they account for the possibility that discarding subjects without ascertainable eligibility when building an analysis population can introduce selection bias. Towards addressing this form of selection bias, we introduced a pair of estimators, $\widehat\theta_\text{EIF}$ and $\widehat\theta_\text{IF}$ which are robust to various degrees of model misspecification, attain $\sqrt{n}$-convergence to normal distributions even under flexible machine learning modeling strategies, and in the case of $\widehat\theta_\text{EIF}$, can achieve semiparametric efficiency bounds. 

In prior work, the multiple imputation strategy of \citet{tompsett2023target} and \citet{austin2023impute} requires modeling $\lambda(\bm L^e_m; \bm L^*)$ directly. Such a task does not readily facilitate nonparametric modeling strategies, particularly when $\bm L^e_m$ is multidimensional. Of greater practical concern however is how this strategy would be implemented in EHR-based studies where 67\% of subjects are missing eligibility (Figure \ref{fig:elig_dist}). \citet{benz2024} used IPW based on $\eta(\bm L^*, A)$, which is a similar idea to $\widehat\theta_\text{IWOR}$. Their work focused on sequential target trial emulations (an observational study design for time-to-event endpoints) and additional challenges inherent to that study design, including time-varying eligibility. Given that the problem of missing eligibility data is likely to impact most EHR-based study designs, extensions of our work to designs where study eligibility can vary over time are a valuable direction for future research. In particular, such extensions can help carry ideas from this work into settings studying time-varying exposures or dynamic treatment regimes.

Critical to addressing the possibility of selection bias due to missing eligibility data is the MAR assumption for $\bm L^e_m$ (Assumption \ref{as:MAR}). Within our MAR statement we also assumed that conditionally on fully observed covariates and treatment, eligibility ascertainment was independent of the outcome. The set of studies that work with Assumption \ref{as:MAR} is quite broad, including \citet{benz2024}, \citet{tompsett2023target}, and \citet{austin2023impute}. Relaxation of this assumption to remove $Y$ from the joint independence is possible but would ultimately entail integration of several nuisance functions over the entire conditional distribution of $\bm L^e_m$,  similar to the strategy used by \citet{levis2024robust} in the case of missing confounders. Given the presence of multiple nested nuisance functions, this solution would likely pose significant computational challenges, to the point where it might inhibit the practical use of the resulting method. 

For the purposes of developing $\widehat\theta_\text{EIF}$ and $\widehat\theta_\text{IF}$, we worked with a coarsened version of the data, treating $\bm L^e_m$ as fully observed or completely missing. This avoids the complexities associated with non-monotone missingness. Assessing the plausibility of a version of Assumption \ref{as:MAR} considering $2^q - 1$ possible missingness patterns may often be very challenging. More practically, the best existing methods for non-monotone missingness under MAR \citep{sun2018nonmonotone} rely on parametric models which condition on unobservable covariate strata \citep{levis2024robust}. In practice, analysts may consider treating $R$ as an indicator for whether $E$ is determined given the available information in $\bm L^e_m$ to reduce information loss, particularly if the portions of $\bm L^e_m$ which are missing are not confounders.

The focus of this work was on missingness in eligibility-defining covariates $\bm L^e$, and thus implicit to this work was the assumption that there was no missingness in other variables. In practice, that is unlikely in complex EHR-based studies, where missing data can affect ascertainment of confounders, treatment, and outcomes, perhaps simultaneously. In cases where outcomes $Y$ or non-eligibility related covariates $\bm L^*$ are missing with less frequency than $\bm L^e$ (Figure \ref{fig:elig_dist}), imputation may be a reasonable approach. We adopted this approach in our study and comment on specifics in the Supplementary Material. If the primary source of missing data came from $\bm L^*$, the path of least resistance might be to move the missing components of $\bm L^*$ into $\bm L^e_m$, but such an approach is likely to introduce non-monotone missingness when multiple components of $\bm L^*$ are missing, not to mention reducing the likelihood that Assumption \ref{as:MAR} holds.

When missing outcomes are of concern, analysts may consider $\widehat\theta_\text{IF}$ instead of $\widehat\theta_\text{EIF}$, as the former does not use $Y$ in the conditioning set of any component nuisance function. One area of future work is monotone missingness in $Y$ and $\bm L^e$, for example when $\bm L^e$ is needed to define $Y$, as for weight change post surgery.  Under this scenario, it seems possible to develop similar influence function-based estimation strategies, albeit with additional assumptions.

\if1\blind
{
\section*{Disclosure Statement}
The authors report there are no competing interests to declare.

\section*{Data Availability Statement}
The DURABLE EHR data contain protected health information and cannot be shared publicly.

\section*{Code Availability}
All code for analysis and simulations is made available on GitHub at \url{https://github.com/lbenz730/semiparametric_missing_elig}.

\section*{Acknowledgments}
This work was supported by NIH Grants R01 DK128150-01 (LB, RM, RW, DA, CL, HF, SMS, SH) and F31 DK141237-01 (LB).
\clearpage
\bibliographystyle{plainnat}
} \fi

\if0\blind
{
\bibliographystyle{plainnat}
} \fi

\spacingset{0.8} 
{\small\setlength{\bibsep}{3pt}
\bibliography{references}}
\end{document}


\def\spacingset#1{\renewcommand{\baselinestretch}%
{#1}\small\normalsize} \spacingset{1}


\if1\blind
{

\renewcommand\footnotemark{}
\renewcommand\footnoterule{}
\renewcommand\Affilfont{\footnotesize}

\title{\bf Supplementary Material for ``Robust Causal Inference for EHR-based Studies of Point Exposures with Missingness in Eligibility Criteria''}
\author[1]{Luke Benz}
\author[1]{Rajarshi Mukherjee}
\author[1,2,3]{Rui Wang}
\author[4]{David Arterburn}
\author[5]{Heidi Fischer}
\author[6]{Catherine Lee}
\author[7,8]{Susan M. Shortreed}
\author[1]{Sebastien Haneuse*}
\author[9]{Alexander W. Levis*}
\affil[1]{Department of Biostatistics,
Harvard T.H. Chan School of Public Health, Boston, MA, USA}
\affil[2]{Department of Population Medicine, Harvard Pilgrim Health Care Institute, Boston, MA, USA}
\affil[3]{Department of Population Medicine, Harvard Medical School, Boston, MA, USA}
\affil[4]{Kaiser Permanente Washington Health Research Institute, Seattle, WA, USA}
\affil[5]{Department of Research \& Evaluation, Kaiser Permanente Southern California, Pasadena, CA, USA}
\affil[6]{Department of Epidemiology \& Biostatistics, University of California San Francisco, San Francisco, CA, USA}
\affil[7]{Biostatistics Division, Kaiser Permanente Washington Health Research Institute, Seattle, WA, USA}
\affil[8]{Department of Biostatistics, University of Washington School of Public Health, Seattle, WA, USA}
\affil[9]{Department of Biostatistics, Epidemiology \& Informatics,
University of Pennsylvania, Philadelphia, PA, USA}
\thanks{\noindent $^*$ denotes co-last author (AWL and SH).}
  \maketitle
} \fi

\if0\blind
{
  \bigskip
  \bigskip
  \bigskip
  \begin{center}
    {\large\bf Supplementary Material for ``Robust Causal Inference for EHR-based Studies of Point Exposures with Missingness in Eligibility Criteria''}
\end{center}
  \medskip
} \fi

\makeatletter
\renewcommand \thesection{S\@arabic\c@section}
\renewcommand\thetable{S\@arabic\c@table}
\renewcommand \thefigure{S\@arabic\c@figure}
\renewcommand \theequation{S\@arabic\c@equation}
\renewcommand \thelemma{S\@arabic\c@lemma}
\renewcommand \thetheorem{S\@arabic\c@theorem}
\makeatother
\setcounter{figure}{0}
\setcounter{table}{0}
\setcounter{equation}{0}

\vspace{-1.5cm}
{\footnotesize\tableofcontents}

\spacingset{1} 

\section{Identification of \texorpdfstring{$\theta_{\text{ATT}}^{\text{elig}}$}{ATTE} (Proof of Theorem 1)}\label{suppSec:identification}
In this section, we will prove Theorem 1, that under Assumptions 1-5, $\theta_{\text{ATT}}^{\text{elig}} = \mathbb{E}_P[Y(1) - Y(0)~|~A = 1, E = 1]$ is identified by the functional 

$$
\theta(P) = \frac{\beta(P)}{\alpha(P)} = \frac{\mathbb{E}_P\biggr[\frac{ARE}{\eta(\bm L^*, 1)}\Bigr(Y - \mu_0(\bm L^*, \bm L^e_m)\Bigr)\biggr]}{\mathbb{E}_P\Bigr[\frac{ARE}{\eta(\bm L^*, 1)}\Bigr]}
$$

\begin{lemma}\label{lemma:bvt}
Let $A, B, C$ be random variables, with $C \in \{0, 1\}$. Then 
$$
\mathbb{E}[A~|~B, C = 1] = \frac{\mathbb{E}[AC~|~B]}{P(C = 1~|~B)}
$$
\end{lemma}
\noindent This result follows immediately from the law of total expectation by observing that 

$$
\mathbb{E}[AC~|~B] = \mathbb{E}[A~|~B, C = 1]P(C = 1~|~B) + 0\times P(C = 0~|~B)
$$

\noindent Using this lemma, and Assumptions 1-5, we have that

$$
\begin{aligned}
P(E = 1~|~A = 1) &= \mathbb{E}_P[\mathbb{E}_P(E~|~\bm L^*, A = 1)~|~ A = 1)] \\
&= \mathbb{E}_P[\mathbb{E}_P(E~|~\bm L^*, A = 1, R = 1)~|~ A = 1)]~~~(\text{A4, A5}) \\
&= \mathbb{E}_P\biggr[\mathbb{E}_P\biggr(\frac{RE}{\eta(\bm L^*, 1)}~\Bigr|~\bm L^*, A = 1\biggr)~\Bigr|~ A = 1\biggr]~~~(\text{Lemma \ref{lemma:bvt}, Defn. of $\eta$}) \\
&= \mathbb{E}_P\biggr[\frac{RE}{\eta(\bm L^*, 1)}~\Bigr|~ A = 1\biggr]\\
&= \frac{\mathbb{E}_P\biggr[\frac{ARE}{\eta(\bm L^*, 1)}\biggr]}{P(A = 1)}~~~(\text{Lemma \ref{lemma:bvt}})\\
&= \frac{\alpha(P)}{P(A = 1)}\\
\end{aligned}
$$

\noindent Next, we have

$$
\begin{aligned}
\mathbb{E}_P[E Y~|~A = 1] &= 
\mathbb{E}_P[\mathbb{E}_P(E Y~|~\bm L^*, \bm L^e_m, A = 1)~|~ A = 1)] \\
&= \mathbb{E}_P[E\mathbb{E}_P(Y~|~\bm L^*, \bm L^e_m, A = 1)~|~ A = 1)]~~~\text{($E = g(\bm L^e)$, fixed function of $\bm L^e$)} \\
&= \mathbb{E}_P[E\mathbb{E}_P(Y~|~\bm L^*, \bm L^e_m, A = 1, R = 1)~|~ A = 1)]~~~\text{(A4, A5)} \\
&= \mathbb{E}_P\biggr[E\mathbb{E}_P\biggr(\frac{RY}{P(R = 1~|~\bm L^*, \bm L^e_m, A = 1)}~\Bigr|~\bm L^*, \bm L^e_m, A = 1\biggr)~\Bigr|~ A = 1\biggr]~~~(\text{Lemma \ref{lemma:bvt}}) \\
&= \mathbb{E}_P\biggr[\frac{E}{\eta(\bm L^*, 1)}\mathbb{E}_P[RY~|~\bm L^*, \bm L^e_m, A = 1]~\Bigr|~A = 1 \biggr] ~~~(\text{A4, A5}) \\
&= \mathbb{E}_P\biggr[\frac{RE}{\eta(\bm L^*, 1)}Y~\Bigr|~ A = 1\biggr]\\
&= \frac{\mathbb{E}_P\Bigr[\frac{ARE}{\eta(\bm L^*, 1)}Y\Bigr]}{P(A = 1)}~~~(\text{Lemma \ref{lemma:bvt}}) \\
\end{aligned}
$$
$$
\begin{aligned}
\mathbb{E}_P[E\mu_0(\bm L^*, \bm L^e_m)~|~A = 1] &= 
\mathbb{E}_P\biggr[\frac{\mathbb{E}_P(R~|~\bm L^*, A = 1, \bm L^e_m)}{\eta(\bm L^*, 1)} E \mu_0(\bm L^*, \bm L^e_m)~\Bigr|~A = 1\biggr]~~~\text{(A4, A5)} \\
&= \mathbb{E}_P\biggr[\mathbb{E}_P\biggr(\frac{R}{\eta(\bm L^*, 1)} E \mu_0(\bm L^*, \bm L^e_m)~\Bigr|~\bm L^*, \bm L^e_m, A = 1\biggr)~\Bigr|~A = 1\biggr] \\
&= \mathbb{E}_P\biggr[\frac{RE}{\eta(\bm L^*, 1)}  \mu_0(\bm L^*, \bm L^e_m)~\Bigr|~A = 1\biggr] \\
&= \frac{\mathbb{E}_P\Bigr[\frac{ARE}{\eta(\bm L^*, 1)}  \mu_0(\bm L^*, \bm L^e_m)~\Bigr]}{P(A = 1)}
\end{aligned}
$$
\\~\\
\noindent Applying Assumptions 1-3, the causal inference assumptions among study eligible subjects, we have
$$
\begin{aligned}
\mathbb{E}_P[Y(1)~|~A = 1, E = 1] &= \mathbb{E}_P[Y~|~A = 1, E = 1] = \frac{\mathbb{E}_P[E  Y~|~A = 1]}{P(E = 1~|~A = 1)}
\end{aligned}
$$

$$
\begin{aligned}
\mathbb{E}_P[Y(0)~|~A = 1, E = 1 ] &= \mathbb{E}_P[\mathbb{E}_P(Y(0)~|~A = 1, E = 1, \bm L^*, \bm L^e_m)~|~A = 1, E = 1 ] \\ 
&= \mathbb{E}_P[\mathbb{E}_P(Y(0)~|~A = 0, E = 1, \bm L^*, \bm L^e_m)~|~A = 1, E = 1] ~~~\text{(A2, A3)} \\ 
&= \mathbb{E}_P[\mathbb{E}_P(Y~|~A = 0, E = 1, \bm L^*, \bm L^e_m)~|~A = 1, E = 1] ~~~\text{(A1)} \\ 
&= \mathbb{E}_P[\mathbb{E}_P(Y~|~A = 0, E, \bm L^*, \bm L^e_m)~|~A = 1, E = 1] \\
&= \mathbb{E}_P[\mathbb{E}_P(Y~|~A = 0, \bm L^*, \bm L^e_m)~|~A = 1, E = 1]~~~\text{($E = g(\bm L^e)$ fixed)} \\
&= \mathbb{E}_P[\mu_0(\bm L^*, \bm L^e_m)~|~A = 1, E = 1]~~~\text{(A4)} \\
&= \frac{\mathbb{E}_P[E  \mu_0(\bm L^*, \bm L^e_m)~|~A = 1]}{P(E = 1~|~A = 1)}~~~\text{(Lemma \ref{lemma:bvt})} 
\end{aligned}
$$

\noindent Combining pieces, we obtain the desired result, namely
$$
\begin{aligned}
\theta_{\text{ATT}}^{\text{elig}} &= \mathbb{E}_P[Y(1) - Y(0)~|~A = 1, E = 1] \\
&= \frac{\mathbb{E}_P\bigr[E\bigr(Y-  \mu_0(\bm L^*, \bm L^e_m)\bigr)~|~A = 1\bigr]}{P(E = 1~|~A = 1)} \\
&= \frac{\mathbb{E}_P\biggr[\frac{ARE}{\eta(\bm L^*, 1)}\Bigr(Y -  \mu_0(\bm L^*, \bm L^e_m)\Bigr)~\biggr]\Bigr/P(A = 1)}{\alpha(P)/P(A = 1)} \\
&= \frac{\beta(P)}{\alpha(P)} = \theta(P)
\end{aligned}
$$

\section{Alternative Representations of \texorpdfstring{$\widehat{\theta}_\text{EIF}$}{Theta Hat EIF} and \texorpdfstring{$\widehat{\theta}_\text{IF}$}{Theta Hat IF}}
\subsection{Reparameterization of \texorpdfstring{$\dot\beta_P(O)$}{Beta EIF} and \texorpdfstring{$\dot\beta_P'(O)$}{Beta IF}}\label{supSec:reparam}
The likelihood factorization in Equation (2) of the main paper was closely tied to four nuisance functions, ($\pi$, $\eta$, $\lambda$, $\mu$). Perhaps surprisingly, $\pi$ and $\lambda$ do not appear in any estimator for $\theta(P)$. In this section, we show how both nuisance functions actually are relevant to estimation of $\theta(P)$, and why we chose an alternative nuisance function parametrization which did not require estimation of either $\pi$ or $\lambda$. 

Using Bayes' theorem, we see the following relationship between $\lambda$, $\eta$, $\pi$, and the complete case propensity score $u$.

$$
\begin{aligned}
       \frac{\lambda_1(\bm L^e_m;\bm L^*)}{\lambda_0(\bm L^e_m;\bm L^*)} &= \frac{p(\bm L^e_m~|~ \bm L^*, A = 1, R = 1)}{p(\bm L^e_m~|~ \bm L^*, A = 0, R = 1)} \\
        &= \frac{p(\bm L^e_m,  \bm L^*, A = 1, R = 1)/p(\bm L^*, A = 1, R = 1)}{p(\bm L^e_m, \bm L^*, A = 0, R = 1)/p(\bm L^*, A = 0, R = 1)} \\
        &= \frac{p(A = 1~|~\bm L^e_m,  \bm L^*, R = 1)/p(\bm L^*, A = 1, R = 1)}{p(A = 0~|~ \bm L^e_m, \bm L^*,R = 1)/p(\bm L^*, A = 0, R = 1)} \\
        &= \frac{u(\bm L^*, \bm L^e_m)}{1 - u(\bm L^*, \bm L^e_m)} \frac{[1-\pi(\bm L^*)]\eta(\bm L^*, 0)}{\pi(\bm L^*)\eta(\bm L^*, 1)}
\end{aligned}
$$

\noindent This identity yields alternative parametrizations of $\dot\beta_P(O)$ and $\dot\beta'_P(O)$, respectively, as follows:

$$
\begin{aligned}
\dot\beta_P(O) &= \frac{AR}{\eta(\bm L^*, 1)}\biggr[\Bigr(E - \varepsilon_1(\bm L^*, Y)\Bigr)Y - 
   \Bigr(E\mu_0(\bm L^*, \bm L^e_m) - \xi(\bm L^*, Y)\Bigr)\biggr]  + A\Bigr( \varepsilon_1(\bm L^*, Y)Y - \xi(\bm L^*, Y)\Bigr)  \\
   & - \frac{(1-A)R}{\eta(\bm L^*, 0)}\frac{\pi(\bm L^*)}{1-\pi(\bm L^*)}\biggr[E\frac{\lambda_1(\bm L^e_m;\bm L^*)}{\lambda_0(\bm L^e_m;\bm L^*)}\Bigr(Y -  \mu_0(\bm L^*, \bm L^e_m)\Bigr) -\Bigr( \gamma'(\bm L^*, Y)Y - \chi'(\bm L^*, Y)\Bigr)\biggr] \\
& -(1-A)\frac{\pi(\bm L^*)}{1-\pi(\bm L^*)}\Bigr(\gamma'(\bm L^*, Y)Y - \chi'(\bm L^*, Y)\Bigr) \\
\dot\beta'_P(O) &= A\biggr(1 - \frac{R}{\eta(\bm L^*, 1)}\biggr)\nu(\bm L^*) + RE\Bigr(Y - \mu_0(\bm L^*, \bm L^e_m)\Bigr)\biggr[\frac{A}{\eta(\bm L^*, 1)} - \frac{(1-A)}{\eta(\bm L^*, 0)}\frac{\pi(\bm L^*)}{1 - \pi(\bm L^*)}\frac{\lambda_1(\bm L^e_m;\bm L^*)}{\lambda_0(\bm L^e_m;\bm L^*)}\biggr]
\end{aligned}
$$

\noindent where 
$$
\gamma'(\bm L^*, Y) =\mathbb{E}\Bigr[E\frac{\lambda_1(\bm L^e_m;\bm L^*)}{\lambda_0(\bm L^e_m; \bm L^*)}~\Bigr|~\bm L^*, A = 0, R = 1, Y\Bigr]
$$
\noindent and  
$$
\chi'(\bm L^*, Y) = \mathbb{E}\Bigr[E\frac{\lambda_1(\bm L^e_m;\bm L^*)}{\lambda_0(\bm L^e_m; \bm L^*)}\mu_0(\bm L^*, \bm L^e_m)~\Bigr|~\bm L^*, A = 0, R = 1, Y\Bigr]
$$ are versions of $\gamma$ and $\chi$ induced by the switch from propensity score odds $\frac{u(\bm L^*, \bm L^e_m)}{1-u(\bm L^*, \bm L^e_m)}$ to density ratio $\frac{\lambda_1(\bm L^e_m;\bm L^*)}{\lambda_0(\bm L^e_m;\bm L^*)}$. 

In fact, the above versions of $\dot\beta_P(O)$ and $\dot\beta'_P(O)$ were the original expressions we derived, and follow much more directly from the likelihood factorization in Equation (2). The presence of density ratio $\frac{\lambda_1(\bm L^e_m;\bm L^*)}{\lambda_0(\bm L^e_m;\bm L^*)}$ makes nonparametric estimation more challenging given that both $\bm L^e_m$ and $\bm L^*$ can be multidimensional and in general, conditional density estimation is a very challenging statistical problem. While other techniques for density ratio estimation exist \citep{sugiyama2012density}, our reparameterization follows the same technique as \citet{diaz2023}.

When estimating $\theta(P)$ using a one-step estimator for $\beta(P)$ based on this alternative representation of $\dot\beta_P(O)$ or $\dot\beta'_P(O)$, we denote such estimators by $\dot\beta_{\widetilde P}(O)$ and $\dot\beta'_{\widetilde P}(O)$, respectively, and corresponding estimators for $\theta(P)$ by $\widetilde\theta_\text{EIF}$ and $\widetilde\theta_\text{IF}$. A choice to use $\widetilde\theta_\text{EIF}$ and $\widetilde\theta_\text{IF}$ rather than $\widehat\theta_\text{EIF}$ or $\widehat\theta_\text{IF}$ might be more appropriate if analysts had understanding of the exact process giving rise to eligibility defining covariates $\bm L^e$. For completeness, we provide summary of the asymptotics of these alternative estimators in Section \ref{suppSec:theory_extra}.

\subsection{Correspondence Between \texorpdfstring{$\widehat{\theta}_\text{EIF}$}{Theta Hat EIF} and One-Step Estimation of the ATT}
In the absence of missing eligibility data, $R = 1$ for all subjects. Suppose further that all $n$ remaining subjects are eligible for the study $(E = 1)$. Such might be the case if analysis is restricted to $n$ eligible subjects in an EHR database with $N > n$ subjects. This is analogous to the typical way cohorts are built in observational studies for point exposures if missing eligibility criteria is not an issue. We will show informally that when $(R = 1, E = 1)$ for all $n$ subjects on which we observe data units $(O_1, ..., O_n)$, $\widehat\theta_\text{EIF}$ simplifies to the usual one-step estimator for the average treatment effect on the treated (ATT). Under this scenario
\begin{itemize}
    \item $R = 1, E = 1$ for all subjects
    \item $\eta(\bm L^*, A) = 1$ for $A \in \{0, 1\}$ (All $n$ patients are complete cases)
    \item $\varepsilon_a(\bm L^*, Y) = 1$ for $a \in \{0, 1\}$ (All $n$ patients are study eligible)
\end{itemize}

$$
\begin{aligned}
  \mathbb{P}_n[\dot\beta_P(O)] = \mathbb{P}_n&\Biggr\{\frac{AR}{\widehat\eta(\bm L^*, 1)}\biggr[\Bigr(E - \widehat\varepsilon_1(\bm L^*, Y)\Bigr)Y - 
   \Bigr(E\widehat\mu_0(\bm L^*, \bm L^e_m) - \widehat\xi(\bm L^*, Y)\Bigr)\biggr] \\
   &~+ A\Bigr( \widehat\varepsilon_1(\bm L^*, Y)Y - \widehat\xi(\bm L^*, Y)\Bigr)  \\
   &~- \frac{(1-A)R}{\eta(\bm L^*, 1)}\biggr[E\frac{ \widehat u(\bm L^*, \bm L^e_m)}{1- \widehat u(\bm L^*, \bm L^e_m)}\Bigr(Y -  \mu_0(\bm L^*, \bm L^e_m)\Bigr) -\Bigr( \gamma(\bm L^*, Y)Y - \chi(\bm L^*, Y)\Bigr)\biggr] \\
& -(1-A)\frac{ \eta(\bm L^*, 0)}{ \eta(\bm L^*, 1)} \Bigr(\gamma(\bm L^*, Y)Y - \chi(\bm L^*, Y)\Bigr)\Biggr\}\\
=\mathbb{P}_n&\Biggr\{-A 
   \Bigr(\widehat\mu_0(\bm L^*, \bm L^e_m) - \widehat\xi(\bm L^*, Y)\Bigr) + A\Bigr(Y - \widehat\xi(\bm L^*, Y)\Bigr)  \\
   &~-(1-A)\biggr[\frac{ \widehat u(\bm L^*, \bm L^e_m)}{1- \widehat u(\bm L^*, \bm L^e_m)}\Bigr(Y -  \mu_0(\bm L^*, \bm L^e_m)\Bigr) -\Bigr( \gamma(\bm L^*, Y)Y - \chi(\bm L^*, Y)\Bigr)\biggr] \\
& -(1-A)\Bigr(\gamma(\bm L^*, Y)Y - \chi(\bm L^*, Y)\Bigr)\Biggr\} \\
=\mathbb{P}_n&\Biggr\{A \Bigr(Y-\widehat\mu_0(\bm L^*, \bm L^e_m) \Bigr)-(1-A)\biggr[\frac{ \widehat u(\bm L^*, \bm L^e_m)}{1- \widehat u(\bm L^*, \bm L^e_m)}\Bigr(Y -  \mu_0(\bm L^*, \bm L^e_m)\Bigr)\biggr]\Biggr\} \\
=\mathbb{P}_n&\Biggr\{\Bigr(Y-\widehat\mu_0(\bm L^*, \bm L^e_m) \Bigr)\biggr[A -(1-A)\biggr(\frac{\widehat  u(\bm L^*, \bm L^e_m)}{1- \widehat u(\bm L^*, \bm L^e_m)}\biggr)\biggr]\Biggr\} \\
\end{aligned}
$$

\noindent Similarly,

$$
\begin{aligned}
      \mathbb{P}_n[\dot\alpha_P(O)] = \mathbb{P}_n&\Biggr\{\ A\biggr(1 - \frac{R}{\widehat\eta(\bm L^*, 1)}\biggr)\widehat\varepsilon_1(\bm L^*, Y) + \frac{ARE}{\widehat\eta(\bm L^*, 1)}\Biggr\} = \mathbb{P}_n[A]
\end{aligned}
$$

\noindent Thus 
$$
\widehat\theta_\text{EIF} = \frac{ \mathbb{P}_n[\dot\beta_P(O)] }{ \mathbb{P}_n[\dot\alpha_P(O)] }=  \mathbb{P}_n\Biggr\{\frac{1}{\mathbb{P}_n[A]}\Bigr(Y-\widehat\mu_0(\bm L^*, \bm L^e_m) \Bigr)\biggr[A -(1-A)\biggr(\frac{\widehat  u(\bm L^*, \bm L^e_m)}{1- \widehat u(\bm L^*, \bm L^e_m)}\biggr)\biggr]\Biggr\}
$$

\noindent which is nothing more than the one-step estimator (eg., doubly-robust) for the ATT \citep{kennedy2015semiparametric, mercatanti2014debit, moodie2018doubly}. A very similar argument yields the same result for $\widehat\theta_\text{IF}$.

\section{Proofs of Theoretical Properties}\label{suppSec:theory}
\subsection{von Mises Expansion of \texorpdfstring{$\alpha(P)$}{Alpha} and \texorpdfstring{$\beta(P)$}{Beta} with \texorpdfstring{$\dot\alpha_P^*(O)$}{Alpha EIF} and \texorpdfstring{$\dot\beta_P^*(O)$}{Beta EIF} (Proof of Lemma 1)}\label{suppSec:IF_a}
Before proving Lemma 1, we will note that 

$$
\begin{aligned}
\alpha(P) &= \mathbb{E}_{P}\biggr[\frac{ARE}{\eta(\bm L^*, 1)}\biggr] \\
&=   \mathbb{E}_{P}\biggr[\frac{AR}{\eta(\bm L^*, 1)}\mathbb{E}_P[E~|~\bm L^*, A = 1, R = 1, Y]\biggr]\\
&= \mathbb{E}_{P}\biggr[\frac{AR}{\eta(\bm L^*, 1)}\varepsilon_1(\bm L^*, Y)\biggr] \\
&= \mathbb{E}_P[A\varepsilon_1(\bm L^*, Y)]
\end{aligned}
$$

\noindent where the last line follows by applying iterated expectation conditioning on $\bm L^*, A, Y$ and than applying Assumption 4 (MAR) to note that $\mathbb{E}_P[R~|~\bm L^*, A = 1, Y] = \mathbb{E}_P[R~|~\bm L^*, A = 1] = \eta(\bm L^*, 1)$. Similarly, we note that 

$$
\begin{aligned}
\beta(P) &= \mathbb{E}_P\biggr[\frac{ARE}{\eta(\bm L^*, 1)}\Bigr(Y - \mu_0(\bm L^*, \bm L^e_m)\Bigr)\biggr] \\
    &= \mathbb{E}_P\biggr[\frac{AR}{\eta(\bm L^*, 1)}\mathbb{E}_P\biggr[E\Bigr(Y - \mu_0(\bm L^*, \bm L^e_m)\Bigr)\Bigr| \bm L^*, A = 1, R = 1, Y\biggr]\biggr] \\
    &= \mathbb{E}_P\biggr[\frac{AR}{\eta(\bm L^*, 1)}\Bigr(Y\varepsilon_1(\bm L^*, Y) - \xi(\bm L^*, Y)\Bigr)\biggr]~~~({\text{Defn. of $\varepsilon_a, \xi$}}) \\
    &= \mathbb{E}_P\Bigr[A\Bigr(Y\varepsilon_1(\bm L^*, Y) - \xi(\bm L^*, Y)\Bigr)\Bigr]
\end{aligned}
$$

\noindent where again the last line follows by applying iterated expectation conditioning on $\bm L^*, A, Y$ and than applying Assumption 4 (MAR) to note that $\mathbb{E}_P[R~|~\bm L^*, A = 1, Y] = \mathbb{E}_P[R~|~\bm L^*, A = 1] = \eta(\bm L^*, 1)$. Now to prove Lemma 1:

\footnotesize
$$
\begin{aligned}
R_\alpha(\widebar P, P) &= \alpha(\widebar P) - \alpha(P) +  \int \dot\alpha^*_{\widebar P}(o)dP(o) \\
&= \alpha(\widebar P) - \alpha(P) +  \mathbb{E}_P\biggr[A\biggr(1 - \frac{R}{\widebar \eta(\bm L^*, 1)}\biggr)\widebar\varepsilon_1(\bm L^*, Y) + \frac{ARE}{\widebar \eta(\bm L^*, 1)} - \alpha(\widebar P) \biggr] \\
&= \mathbb{E}_P\biggr[A\biggr(1 - \frac{R}{\widebar \eta(\bm L^*, 1)}\biggr)\widebar\varepsilon_1(\bm L^*, Y) + \frac{ARE}{\widebar \eta(\bm L^*, 1)} - A\varepsilon_1(\bm L^*, Y)\biggr] \\
&= \mathbb{E}_P\biggr[A\biggr(1 - \frac{R}{\widebar \eta(\bm L^*, 1)}\biggr)\widebar\varepsilon_1(\bm L^*, Y) + \frac{AR\varepsilon_1(\bm L^*, Y)}{\widebar \eta(\bm L^*, 1)} - A\varepsilon_1(\bm L^*, Y)\biggr] ~~~\text{(It. Exp. on $\bm L^*, A, R, Y$)} \\
&= \mathbb{E}_P\biggr[A\biggr(1 - \frac{\eta(\bm L^*, 1)}{\widebar\eta(\bm L^*, 1)}\ \biggr)\widebar\varepsilon_1(\bm L^*)  + A \frac{\eta(\bm L^*, 1)\varepsilon_1(\bm L^*)}{\widebar\eta(\bm L^*, 1)} - A\varepsilon_1(\bm L^*) \biggr]~~~\text{(It. Exp. on $\bm L^*, A, Y$ + A4)} \\
&= \mathbb{E}_P\biggr[A(\widebar\varepsilon_1 - \varepsilon_1)\biggr(1 - \frac{\eta_1}{\widebar\eta_1}\biggr) \biggr]
\end{aligned}
$$

$$
\begin{aligned}
R_\beta(\widebar P, P)  &= \beta(\widebar P) - \beta(P) +  \int \dot\beta^*_{\widebar P}(o)dP(o) \\
&= \beta(\widebar P) - \beta(P) +\\ &~~~~\mathbb{E}_P\biggr[\frac{AR}{\widebar\eta(\bm L^*, 1)}\biggr\{\Bigr(E - \widebar\varepsilon_1(\bm L^*, Y)\Bigr)Y - \Bigr(E\widebar\mu_0(\bm L^*, \bm L^e_m) - \widebar\xi(\bm L^*, Y)\Bigr)\biggr\}  + A\Bigr(\widebar\varepsilon_1(\bm L^*, Y) Y - \widebar\xi(\bm L^*, Y)\Bigr)  \\
&~~~~~~~~- \frac{(1-A)R}{\widebar\eta(\bm L^*, 1)}\biggr\{E\frac{\widebar u(\bm L^*, \bm L^e_m)}{1 -\widebar u(\bm L^*, \bm L^e_m)}\Bigr(Y - \widebar\mu_0(\bm L^*, \bm L^e_m)\Bigr) -\Bigr(\widebar{\gamma}(\bm L^*, Y)Y - \widebar{\chi}(\bm L^*, Y)\Bigr)\biggr\} \\
&~~~~~~~~-(1-A)\frac{\widebar\eta(\bm L^*, 0)}{\widebar\eta(\bm L^*, 1)} \Bigr(\widebar{\gamma}(\bm L^*, Y)Y - \widebar{\chi}(\bm L^*, Y)\Bigr) - \beta(\widebar P)\biggr]\\
&= \mathbb{E}_P\biggr[\frac{AR}{\widebar\eta(\bm L^*, 1)}\biggr\{\Bigr(E - \widebar\varepsilon_1(\bm L^*, Y)\Bigr)Y - \Bigr(E\widebar\mu_0(\bm L^*, \bm L^e_m) - \widebar\xi(\bm L^*, Y)\Bigr)\biggr\}  + A\Bigr(\widebar\varepsilon_1(\bm L^*, Y) Y - \widebar\xi(\bm L^*, Y)\Bigr)  \\
&~~~~~~~~- \frac{(1-A)R}{\widebar\eta(\bm L^*, 1)}\biggr\{E\frac{\widebar u(\bm L^*, \bm L^e_m)}{1 -\widebar u(\bm L^*, \bm L^e_m)}\Bigr(Y - \widebar\mu_0(\bm L^*, \bm L^e_m)\Bigr) -\Bigr(\widebar{\gamma}(\bm L^*, Y)Y - \widebar{\chi}(\bm L^*, Y)\Bigr)\biggr\} \\
&~~~~~~~~-(1-A)\frac{\widebar\eta(\bm L^*, 0)}{\widebar\eta(\bm L^*, 1)} \Bigr(\widebar{\gamma}(\bm L^*, Y)Y - \widebar{\chi}(\bm L^*, Y)\Bigr) - A\Bigr(\varepsilon_1(\bm L^*, Y) Y - \xi(\bm L^*, Y)\Bigr)\biggr]
\end{aligned}
$$

\normalsize
\noindent Adding and subtracting the term $E\mu_0(\bm L^*, \bm L^e_m)$ on line 1 and applying iterated expectation (conditioning on $\bm L^*, A, Y, \bm L^e_m$ in line 2) yields
\footnotesize
$$
\begin{aligned}
R_\beta(\widebar P, P) &= \mathbb{E}_P\biggr[\frac{AR}{\widebar\eta(\bm L^*, 1)}\biggr\{\Bigr(E - \widebar\varepsilon_1(\bm L^*, Y)\Bigr)Y - \Bigr(E\widebar\mu_0(\bm L^*, \bm L^e_m) {\color{blue} + E\mu_0(\bm L^*, \bm L^e_m) - E\mu_0(\bm L^*, \bm L^e_m)} - \widebar\xi(\bm L^*, Y)\Bigr)\biggr\}\\
&~~~~~~~~- \frac{(1-A)R}{\widebar\eta(\bm L^*, 1)}\biggr\{E\frac{\widebar u(\bm L^*, \bm L^e_m)}{1 -\widebar u(\bm L^*, \bm L^e_m)}\Bigr(\mu_0(\bm L^*, \bm L^e_m) - \widebar\mu_0(\bm L^*, \bm L^e_m)\Bigr) -\Bigr(\widebar{\gamma}(\bm L^*, Y)Y - \widebar{\chi}(\bm L^*, Y)\Bigr)\biggr\} \\
&~~~~~~~~-(1-A)\frac{\widebar\eta(\bm L^*, 0)}{\widebar\eta(\bm L^*, 1)} \Bigr(\widebar{\gamma}(\bm L^*, Y)Y - \widebar{\chi}(\bm L^*, Y)\Bigr)\\ 
&~~~~~~~~- A\Bigr(\varepsilon_1(\bm L^*, Y) Y - \xi(\bm L^*, Y)\Bigr) + A\Bigr(\widebar\varepsilon_1(\bm L^*, Y) Y - \widebar\xi(\bm L^*, Y)\Bigr)\biggr] 
\end{aligned}
$$

\normalsize
\noindent Next, we rearrange and combine terms and begin simplifying. In particular, we apply additional iterated expectations in lines 1 (on $\bm L^*, A, R, Y$) and 3 (on $\bm L^*, A, Y$, followed by Assumption 4) of the below expression, and finally add and subtract $\gamma(\bm L^*, Y)Y$ in the final term below.
\footnotesize
$$
\begin{aligned}
    R_\beta(\widebar P, P) &=  \mathbb{E}_P\biggr[A\biggr(1 - \frac{\eta(\bm L^*, 1)}{\widebar\eta(\bm L^*, 1)}\biggr)\biggr\{Y\Bigr(\widebar\varepsilon_1(\bm L^*, Y) - \varepsilon_1(\bm L^*, Y)\Bigr) -  \Bigr(\widebar\xi(\bm L^*, Y) - \xi(\bm L^*, Y)\Bigr)\biggr\}  \\
&~~~~~~~~+ \frac{RE}{\widebar\eta(\bm L^*, 1)}\Bigr(\mu_0(\bm L^*, \bm L^e_m) - \widebar\mu_0(\bm L^*, \bm L^e_m)\Bigr)\biggr\{ A - (1-A) \frac{\widebar u(\bm L^*, \bm L^e_m)}{1-\widebar u(\bm L^*, \bm L^e_m)}\biggr \} \\
&~~~~~~~~+(1-A)\frac{\Bigr(\eta(\bm L^*, 0) - \widebar\eta(\bm L^*, 0)\Bigr)}{\widebar\eta(\bm L^*, 1)}  \Bigr(\widebar{\gamma}(\bm L^*, Y)Y   {\color{blue} + \gamma(\bm L^*, Y)Y - \gamma(\bm L^*, Y)Y} -\widebar{\chi}(\bm L^*, Y)\Bigr) 
\biggr]
\end{aligned}
$$

\normalsize
\noindent Applying iterated expectation once more (to line two, on $\bm L^*, A, R, \bm L^e_m$), we arrive at the desired result.
\footnotesize
$$
\begin{aligned}
    R_\beta(\widebar P, P) &=  \mathbb{E}_P\biggr[A\biggr(1 - \frac{\eta(\bm L^*, 1)}{\widebar\eta(\bm L^*, 1)}\biggr)\biggr\{Y\Bigr(\widebar\varepsilon_1(\bm L^*, Y) - \varepsilon_1(\bm L^*, Y)\Bigr) -  \Bigr(\widebar\xi(\bm L^*, Y) - \xi(\bm L^*, Y)\Bigr)\biggr\}  \\
&~~~~~~~~+ \frac{RE}{\widebar\eta(\bm L^*, 1)}\Bigr(\mu_0(\bm L^*, \bm L^e_m) - \widebar\mu_0(\bm L^*, \bm L^e_m)\Bigr)\biggr\{\frac{u(\bm L^*, \bm L^e_m)\bigr(1-\widebar u(\bm L^*, \bm L^e_m)\bigr) - \widebar u(\bm L^*, \bm L^e_m)\bigr(1- u(\bm L^*, \bm L^e_m)\bigr)}{1-\widebar u(\bm L^*, \bm L^e_m)}\biggr \} \\
&~~~~~~~~+(1-A)\frac{\Bigr(\eta(\bm L^*, 0) - \widebar\eta(\bm L^*, 0)\Bigr)}{\widebar\eta(\bm L^*, 1)}  \biggr\{Y\Bigr(\widebar{\gamma}(\bm L^*, Y) - \gamma(\bm L^*, Y)\Bigr) -\Bigr(\widebar{\chi}(\bm L^*, Y) - \chi(\bm L^*, Y)\Bigr)\biggr\}\biggr] \\
&= \mathbb{E}_P\biggr[A\biggr(1 - \frac{\eta_1}{\widebar\eta_1}\biggr)\Bigr(Y(\widebar\varepsilon_1 - \varepsilon_1) -  (\widebar\xi - \xi )\Bigr) + \frac{RE}{\widebar\eta_1}(\mu_0 - \widebar\mu_0)\biggr(\frac{u(1 - \widebar u) - \widebar u(1 - u)}{1-\widebar u}\biggr ) \\
&~~~~~~~~~+ (1-A)\frac{(\eta_0 - \widebar\eta_0)}{\widebar\eta_1} \Bigr(Y(\widebar\gamma -   \gamma)  -(\widebar\chi - \chi)\Bigr)\biggr]
\end{aligned}
$$

\normalsize
This finishes the proof of Lemma 1, by showing that the claimed von Mises expansions are satisfied. An immediate consequence of Lemma 1---invoking Lemma 2 in \citet{kennedy2023}---is that $\dot\alpha^*_{P}(O)$ and $\dot\beta^*_{P}(O)$ are influence functions at $P$ in a semiparametric model induced by Assumption 4. As mentioned in the main paper, Assumption 4 restricts the tangent space of the model, and as such demonstrating that the claimed von Mises expansions are satisfied is not sufficient to prove that $\dot\alpha^*_{P}(O)$ and $\dot\beta^*_{P}(O)$ the \textbf{efficient} influence functions. In order to prove Theorem 2, that $\dot\alpha^*_{P}(O)$ and $\dot\beta^*_{P}(O)$ are indeed the efficient influence functions in the induced semiparametric model, we will first show that $\dot\alpha_P'^{*}(O)$ and $\dot\beta_P'^{*}(O)$ satisfy a different set of von Mises expansions for $\alpha(P)$ and $\beta(P)$, respectively, and thus are themselves influence functions (Section \ref{suppSec:IF_b}). Then, we will go through the process of projecting $\dot\alpha_P'^{*}(O)$ and $\dot\beta_P'^{*}(O)$ onto the tangent space of the model, and show that projection yields $\dot\alpha_P^{*}(O)$ and $\dot\beta_P^{*}(O)$, thereby proving $\dot\alpha_P^{*}(O)$ and $\dot\beta_P^{*}(O)$ are indeed the efficient influence functions (Section \ref{suppSec:projection_IF})

\subsection{von Mises Expansion of \texorpdfstring{$\alpha(P)$}{Alpha} and \texorpdfstring{$\beta(P)$}{Beta} with \texorpdfstring{$\dot\alpha_P'^{*
}(O)$}{Alpha IF} and \texorpdfstring{$\dot\beta_P'^{*}(O)$}{Beta IF}}\label{suppSec:IF_b}

\begin{lemma}\label{lemma:vonMises_IF}
    $\alpha(P)$ and $\beta(P)$ satisfy the von Mises expansions 
$$
\begin{aligned}
\alpha(\widebar P) - \alpha(P) &= - \int \dot\alpha'^{*}_{\widebar P}(o)dP(o) + R'_\alpha(\widebar P, P) \\
\beta(\widebar P) - \beta(P) &= - \int \dot\beta'^{*}_{\widebar P}(o)dP(o) + R'_\beta(\widebar P, P) 
\end{aligned}
$$

\noindent where the remainder terms (omitting inputs for brevity) are as follows:

$$
\begin{aligned}
R'_\alpha(\widebar P, P)  &= \mathbb{E}_P\biggr[A(\widebar\omega_1 - \omega_1)\biggr(1 - \frac{\eta_1}{\widebar\eta_1}\biggr)\biggr] \\
R'_\beta(\widebar P, P) &= \mathbb{E}_P\biggr[A\biggr(1 - \frac{\eta_1}{\widebar\eta_1}\biggr)(\widebar \nu - \nu)   + \frac{RE}{\widebar\eta_1}(\mu_0 - \widebar\mu_0)\biggr(\frac{u(1 - \widebar u) - \widebar u(1 - u)}{1-\widebar u}\biggr )\biggr]
\end{aligned}
$$
\end{lemma} 

The proof of Lemma \ref{lemma:vonMises_IF} is very similar to that of Lemma 1, from the previous section. Before proving Lemma \ref{lemma:vonMises_IF}, we note that 

$$
\begin{aligned}
\alpha(P) &= \mathbb{E}_{P}\biggr[\frac{ARE}{\eta(\bm L^*, 1)}\biggr] \\
&=   \mathbb{E}_{P}\biggr[\frac{AR}{\eta(\bm L^*, 1)}\mathbb{E}_P[E~|~\bm L^*, A = 1, R = 1]\biggr]\\
&= \mathbb{E}_{P}\biggr[\frac{AR}{\eta(\bm L^*, 1)}\omega_1(\bm L^*)\biggr] \\
&= \mathbb{E}_P[A\omega_1(\bm L^*)]
\end{aligned}
$$

\noindent where the last line follows by applying iterated expectation conditioning on $\bm L^*, A$. Additionally, we note that 

$$
\begin{aligned}
    \beta(P) &= \mathbb{E}_P\biggr[\frac{ARE}{\eta(\bm L^*, 1)}\Bigr(Y - \mu_0(\bm L^*, \bm L^e_m)\Bigr)\biggr] \\
    &= \mathbb{E}_P\biggr[\frac{AR}{\eta(\bm L^*, 1)}\mathbb{E}_P\biggr[E\Bigr(Y - \mu_0(\bm L^*, \bm L^e_m)\Bigr)\Bigr| \bm L^*, A = 1, R = 1\biggr]\biggr] \\
    &= \mathbb{E}_P\biggr[\frac{AR}{\eta(\bm L^*, 1)}\nu(\bm L^*)\biggr]~~~({\text{Defn. of $\nu$}}) \\
    &= \mathbb{E}_P[A\nu(\bm L^*)]~~~(\text{It. Exp. on $\bm L^*, A$})
\end{aligned}
$$

\noindent Now to prove Lemma \ref{lemma:vonMises_IF}

\footnotesize
$$
\begin{aligned}
R'_\alpha(\widebar P, P) &= \alpha(\widebar P) - \alpha(P) +  \int \dot\alpha'^{*}_{\widebar P}(o)dP(o) \\
&= \alpha(\widebar P) - \alpha(P) +  \mathbb{E}_P\biggr[A\biggr(1 - \frac{R}{\widebar \eta(\bm L^*, 1)}\biggr)\widebar\omega_1(\bm L^*) + \frac{ARE}{\widebar \eta(\bm L^*, 1)} - \alpha(\widebar P) \biggr] \\
&= \mathbb{E}_P\biggr[A\biggr(1 - \frac{R}{\widebar \eta(\bm L^*, 1)}\biggr)\widebar\omega_1(\bm L^*) + \frac{ARE}{\widebar \eta(\bm L^*, 1)} - A\omega_1(\bm L^*)\biggr] \\
&= \mathbb{E}_P\biggr[A\biggr(1 - \frac{R}{\widebar \eta(\bm L^*, 1)}\biggr)\widebar\omega_1(\bm L^*) + \frac{AR\omega_1(\bm L^*)}{\widebar \eta(\bm L^*, 1)} - A\omega_1(\bm L^*)\biggr] ~~~\text{(It. Exp. on $\bm L^*, A, R$)} \\
&= \mathbb{E}_P\biggr[A\biggr(1 - \frac{\eta(\bm L^*, 1)}{\widebar\eta(\bm L^*, 1)} \biggr)\widebar\omega_1(\bm L^*)  + A \frac{\eta(\bm L^*, 1)\omega_1(\bm L^*)}{\widebar\eta(\bm L^*, 1)} - A\omega_1(\bm L^*) \biggr]~~~\text{(It. Exp. on $\bm L^*, A$)} \\
&= \mathbb{E}_P\biggr[A(\widebar\omega_1 - \omega_1)\biggr(1 - \frac{\eta_1}{\widebar\eta_1}\biggr) \biggr]
\end{aligned}
$$
\normalsize
\noindent Next we have that 

\footnotesize
$$
\begin{aligned}
R'_\beta(\widebar P, P) &= \beta(\widebar P) - \beta(P)  + \int \dot\beta_{\widebar P}'^{*}(o)dP(o)\\
 &=\beta(\widebar P) - \beta(P) + \mathbb{E}_P\biggr[A\biggr(1 - \frac{R}{\widebar\eta(\bm L^*, 1)}\biggr)\widebar\nu(\bm L^*)
+\frac{ARE}{\widebar\eta(\bm L^*, 1)}\Bigr(Y - \widebar\mu_0(\bm L^*, \bm L^e_m)\Bigr)\\
&~~~~~~~~~~~~~~~~~~~~~~~~~~~~~-\frac{(1-A)RE}{\widebar\eta(\bm L^*, 1)}\frac{\widebar u(\bm L^*, \bm L^e_m)}{1 - \widebar u(\bm L^*, \bm L^e_m)}\Bigr(Y - \widebar\mu_0(\bm L^*, \bm L^e_m)\Bigr)
-A\nu(\bm L^*)\biggr] \\
&= \mathbb{E}_P\biggr[A\biggr(1 - \frac{R}{\widebar\eta(\bm L^*, 1)}\biggr)\widebar\nu(\bm L^*) - A\nu(\bm L^*) +\frac{ARE}{\widebar\eta(\bm L^*, 1)}\Bigr(Y - \widebar\mu_0(\bm L^*, \bm L^e_m) {\color{blue}+\mu_0(\bm L^*, \bm L^e_m) - \mu_0(\bm L^*, \bm L^e_m)}\Bigr)\\
&~~~~~~~~~-\frac{(1-A)RE}{\widebar\eta(\bm L^*, 1)}\frac{\widebar u(\bm L^*, \bm L^e_m)}{1-\widebar u(\bm L^*, \bm L^e_m)}\Bigr(Y - \widebar\mu_0(\bm L^*, \bm L^e_m)\Bigr)
\biggr] \\
\end{aligned}
$$
\normalsize

\noindent Now rearranging terms, and applying iterated expectations, we have 

\footnotesize
$$
\begin{aligned}
R'_\beta(\widebar P, P) &= \mathbb{E}_P\biggr[A\biggr(1 - \frac{\eta(\bm L^*, 1)}{\widebar\eta(\bm L^*, 1)}\biggr)\Bigr(\widebar\nu(\bm L^*) -\nu(\bm L^*)\Bigr)~~~\text{(It. Exp. on $\bm L^*, A, R$, then again on $\bm L^*, A$)}\\
&~~~~~~~~~+\frac{ARE}{\widebar\eta(\bm L^*, 1)}\biggr(\mu_0(\bm L^*, \bm L^e_m) - \widebar\mu_0(\bm L^*, \bm L^e_m) )\biggr)\\
&~~~~~~~~~-\frac{(1-A)RE}{\widebar\eta(\bm L^*, 1)}\frac{\widebar u(\bm L^*, \bm L^e_m)}{1-\widebar u(\bm L^*, \bm L^e_m)}\biggr(\mu_0(\bm L^*, \bm L^e_m) - \widebar\mu_0(\bm L^*, \bm L^e_m)\biggr)
\biggr]~~~\text{(It. Exp. on $\bm L^*, A, R, \bm L^e_m)$} \\
&= \mathbb{E}_P\biggr[A\biggr(1 - \frac{\eta(\bm L^*, 1)}{\widebar\eta(\bm L^*, 1)}\biggr)\Bigr(\widebar\nu(\bm L^*) -\nu(\bm L^*)\Bigr)
+\frac{RE}{\widebar\eta(\bm L^*, 1)}\Bigr(\mu_0(\bm L^*, \bm L^e_m) - \widebar\mu_0(\bm L^*, \bm L^e_m)\Bigr)\biggr\{ A - (1-A) \frac{\widebar u(\bm L^*, \bm L^e_m)}{1-\widebar u(\bm L^*, \bm L^e_m)}\biggr \}\biggr] \\
&= \mathbb{E}_P\biggr[A\biggr(1 - \frac{\eta(\bm L^*, 1)}{\widebar\eta(\bm L^*, 1)}\biggr)\Bigr(\widebar\nu(\bm L^*) -\nu(\bm L^*)\Bigr)\\
&~~~~~~~~+\frac{RE}{\widebar\eta(\bm L^*, 1)}\Bigr(\mu_0(\bm L^*, \bm L^e_m) - \widebar\mu_0(\bm L^*, \bm L^e_m)\Bigr)\biggr\{ u(\bm L^*, \bm L^e_m) - \Bigr(1 - u(\bm L^*, \bm L^e_m)\Bigr)\frac{\widebar u(\bm L^*, \bm L^e_m)}{1-\widebar u(\bm L^*, \bm L^e_m)}\biggr \}\biggr]
\end{aligned}
$$
\normalsize
\noindent where the final line follows from iterated expectations on  $\bm L^*, \bm L^e_m, R$. Omitting conditioning terms we see that this error term is of second order

$$
\mathbb{E}_P\biggr[A\biggr(1 - \frac{\eta_1}{\widebar \eta_1}\biggr)(\widebar \nu - \nu) + \frac{RE(\mu_0 - \widebar\mu_0)}{\widebar \eta_1}\biggr\{\frac{u(1-\widebar u) - \widebar u (1 - u)}{1-\widebar u}\biggr\}\biggr]
$$

This proves Lemma \ref{lemma:vonMises_IF} and shows that both $\dot\alpha_P'^{*}(O)$ and $\dot\beta_P'^{*}(O)$ are valid influence functions for $\alpha(P)$ and $\beta(P)$ at $P$, respectively, as were $\dot\alpha^*_{P}(O)$ and $\dot\beta^*_{P}(O)$. In Section \ref{suppSec:projection_IF}, we will go through the process of projecting $\dot\alpha_P'^{*}(O)$ and $\dot\beta_P'^{*}(O)$ onto the tangent space of the model, and show that projection yields $\dot\alpha_P^{*}(O)$ and $\dot\beta_P^{*}(O)$, thereby proving $\dot\alpha_P^{*}(O)$ and $\dot\beta_P^{*}(O)$ are indeed the efficient influence functions of $\alpha(P)$ and $\beta(P)$.

\subsection{Derivation of \texorpdfstring{$\dot\alpha_P^*(O)$}{Alpha EIF} and \texorpdfstring{$\dot\beta_P^*(O)$}{Beta EIF} From \texorpdfstring{$\dot\alpha_P(O)$}{Alpha IF} and \texorpdfstring{$\dot\beta_P(O)$}{Beta IF} (Proof of Theorem 2)}\label{suppSec:projection_IF}

Assumption 4 ($R \indep (Y, \bm L^e_m)~|~\bm L^*, A$) is equivalent to the conjunction of two conditions, which play different roles. Throughout, $Y$ is observed for all subjects and $\bm L^e_m$ is either completely observed ($R = 1$) or completely missing ($R = 0$). The first condition, $R \indep Y~|~\bm L^*, A$, is a restriction on the law of the observed data $O$: it implies that the joint distribution of $(R, Y)$ given $(\bm L^*, A)$ factorizes as $p(R~|~\bm L^*, A)\,p(Y~|~\bm L^*, A)$. This restriction shrinks the tangent space, so that influence functions for $\alpha(P)$ and $\beta(P)$ are no longer unique; this is why $\dot\alpha_P'^{*}$ and $\dot\beta_P'^{*}$ are valid but not efficient. The second condition, $R \indep \bm L^e_m~|~\bm L^*, A, Y$, places no restriction on the observed data law, since $\bm L^e_m$ is unobserved whenever $R = 0$, and so it does not change the tangent space. Identification of $\theta^{\text{elig}}_{\text{ATT}}$ uses Assumption 4 as a whole: together, the two conditions equate complete-case quantities such as $\mu_0$, $\varepsilon_1$ and the nested nuisance functions with their full-population counterparts (Section S1). The efficiency bound in Theorem 2, however, is induced by the first condition alone. We let $\Lambda_P$ denote the tangent space of the model, which by Lemma 24 of \citet{rotnitzky2020} decomposes into orthogonal subspaces as follows:

$$
\Lambda_P = \Lambda_{\bm L^*, A} \oplus \Lambda_{R|\bm L^*, A} \oplus \Lambda_{Y|\bm L^*, A} \oplus \Lambda_{R\bm L^e_m|\bm L^*, A, R, Y}
$$

\noindent where in the above 
$$
\Lambda_{W|V} = \bigr\{f(w, v) \in L_2^0(P)~|~ \mathbb{E}_P[f~|~V] = 0\bigr\}
$$

\noindent Next, we introduce the projection operator $\Pi$. For any function $f\in L_2(P)$, 
$$
\Pi(f, \Lambda_{W|V}) = \mathbb{E}_P[f(O)~|~W, V] - \mathbb{E}_P[f(O)~|~V]
$$

\noindent\underline{\textbf{Projection of $\alpha_P'^{*}(O)$ onto $\Lambda_P$}}\\
\noindent First notice that because $A\Bigr(1 - \frac{R}{\eta(\bm L^*, 1)}\Bigr)\omega_1(\bm L^*)$ is mean zero given $\bm L^*, A$, we have that 

$$
\alpha_P'^{*}(O) = \underbrace{A\biggr(1 - \frac{R}{\eta(\bm L^*, 1)}\biggr)\omega_1(\bm L^*)}_{\in \Lambda_{R|\bm L^*, A}} + \frac{ARE}{\eta(\bm L^*, 1)} - \alpha(P)
$$

\noindent Thus, we only need to project the latter two terms on $\Lambda_P$. Furthermore, notice that term $\alpha(P)$ is a constant (given the distribution $P$), it will cancel out in projection onto every orthogonal subspace with conditioning statements, and thus we need only keep track of it when projecting onto orthogonal subspace without conditioning statements.

\footnotesize
$$
\begin{aligned}
\Pi \biggr(\frac{ARE}{\eta(\bm L^*, 1)} - \alpha(P)~\Bigr|~ \Lambda_{\bm L^*, A, R} \biggr) &= \mathbb{E}\biggr[\frac{ARE}{\eta(\bm L^*, 1)} - \alpha(P)~\Bigr|~ \bm L^*, A, R\biggr] = \frac{AR}{\eta(\bm L^*, 1)}\omega_1(\bm L^*) - \alpha(P)   
\end{aligned}
$$

$$
\begin{aligned}
&\Pi \biggr(\frac{ARE}{\eta(\bm L^*, 1)} - \alpha(P)~\Bigr|~\Lambda_{Y|\bm L^*, A} \biggr)\\ &= \mathbb{E}\biggr[\frac{ARE}{\eta(\bm L^*, 1)}~\Bigr|~ \bm L^*, A, Y\biggr] -  \mathbb{E}\biggr[\frac{ARE}{\eta(\bm L^*, 1)}~\Bigr|~\bm L^*, A\biggr] \\
&=  \mathbb{E}\biggr[\frac{AR}{\eta(\bm L^*, 1)}\mathbb{E}[E~|~\bm L^*, A=1, Y, R = 1]~\Bigr|~\bm L^*, A, Y\biggr] - \mathbb{E}\biggr[\frac{AR}{\eta(\bm L^*, 1)}\mathbb{E}[E~|~\bm L^*, A, R = 1]~\Bigr|~\bm L^*, A\biggr] \\
&=  \mathbb{E}\biggr[\frac{AR}{\eta(\bm L^*, 1)}\varepsilon_1(\bm L^*, Y)~\Bigr|~\bm L^*, A, Y\biggr]  - \mathbb{E}\biggr[\frac{AR}{\eta(\bm L^*, 1)}\omega_1(\bm L^*)~\Bigr|~\bm L^*, A\biggr] \\ 
&= A\Bigr(\varepsilon_1(\bm L^*, Y) - \omega_1(\bm L^*)\Bigr)~~~\text{(A4 and defn. of } \eta )
\end{aligned}
$$

$$
\begin{aligned}
\Pi \biggr(\frac{ARE}{\eta(\bm L^*, 1)} - \alpha(P)~\Bigr|~ \Lambda_{\bm L^e_m|\bm L^*, A, R, Y} \biggr) &= \mathbb{E}\biggr[\frac{ARE}{\eta(\bm L^*, 1)} ~\Bigr|~ \bm L^*, A, R, Y, \bm L^e_m\biggr] -  \mathbb{E}\biggr[\frac{ARE}{\eta(\bm L^*, 1)} ~\Bigr|~ \bm L^*, A, R, Y\biggr]  \\
&= \frac{ARE}{\eta(\bm L^*, 1)} - \frac{AR}{\eta(\bm L^*, 1)}\mathbb{E}[E ~|~ \bm L^*, A = 1, R = 1, Y] \\
&= \frac{ARE}{\eta(\bm L^*, 1)} - \frac{AR}{\eta(\bm L^*, 1)}\varepsilon_1(\bm L^*, Y)
\end{aligned}
$$

\normalsize
\noindent Summing terms, we have 

$$
\Pi\Bigr(\alpha_P'^{*}(O), \Lambda_P\Bigr) = A\biggr(1 - \frac{R}{\eta(\bm L^*, 1)}\biggr)\varepsilon_1(\bm L^*, Y) + \frac{ARE}{\eta(\bm L^*, 1)} - \alpha(P) 
$$
\noindent which is exactly $\alpha^*_P(O)$. Thus, $\alpha^*_P(O)$ is the efficient influence function of $\alpha(P)$.\\~\\

\noindent\underline{\textbf{Projection of $\beta_P'^{*}(O)$ onto $\Lambda_P$}}\\
\noindent We begin by presenting $\dot\beta'^{*}_P(O)$ with terms slightly rearranged, which will facilitate the projection of terms onto the requisite orthogonal subspaces.

$$
\begin{aligned}
\dot\beta'^{*}_P(O) &= A\biggr(1 - \frac{R}{\eta(\bm L^*, 1)}\biggr)\nu(\bm L^*) + \frac{RE}{\eta(\bm L^*, 1)}\biggr(Y - \mu_0(\bm L^*, \bm L^e_m)\biggr)\biggr[A - (1-A)\frac{ u(\bm L^*, \bm L^e_m)}{1 - u(\bm L^*, \bm L^e_m)}\biggr] - \beta(P) \\
&=\frac{AR}{\eta(\bm L^*, 1)}\biggr\{E \Bigr(Y - \mu_0(\bm L^*, \bm L^e_m)\Bigr) - \nu(\bm L^*)\biggr\} + \underbrace{A\nu(\bm L^*) - \beta(P)}_{\in \Lambda_{\bm L^*, A}} \\
&~~~~-\frac{(1-A)R}{\eta(\bm L^*, 1)}\frac{ u(\bm L^*, \bm L^e_m)}{1 - u(\bm L^*, \bm L^e_m)}\biggr\{E \Bigr(Y - \mu_0(\bm L^*, \bm L^e_m)\Bigr)  \biggr\}
\end{aligned}
$$

\noindent where $A\nu(\bm L^*) - \beta(P)\in \Lambda_{\bm L^*, A}$ as $\mathbb{E}_P[A\nu(\bm L^*)] = \beta(P)$. Thus we only need to project the components of the first and third term on each portion of the tangent space.

\footnotesize

$$
\begin{aligned}
    \Pi\biggr(\frac{AR}{\eta(\bm L^*, 1)}\biggr\{E \Bigr(Y - \mu_0(\bm L^*, \bm L^e_m) \Bigr) - \nu(\bm L^*)\biggr\}~\Bigr|~ \Lambda_{\bm L^*, A, R} \biggr) &= \mathbb{E}\biggr[\frac{AR}{\eta(\bm L^*, 1)}\biggr\{E \Bigr(Y - \mu_0(\bm L^*, \bm L^e_m) \Bigr) - \nu(\bm L^*)\biggr\}~\Bigr|~ \bm L^*, A, R \biggr] \\
    &= \frac{AR}{\eta(\bm L^*, 1)}\biggr\{\mathbb{E}\biggr[E \Bigr(Y - \mu_0(\bm L^*, \bm L^e_m) \Bigr) \Bigr|~ \bm L^*, A, R \biggr] - \nu(\bm L^*) \biggr\} \\
    &= \frac{AR}{\eta(\bm L^*, 1)}\Bigr(\nu(\bm L^*) - \nu(\bm L^*) \Bigr) \\
    &= 0
\end{aligned}
$$

$$
\begin{aligned}
     &\Pi\biggr(\frac{AR}{\eta(\bm L^*, 1)}\biggr\{E \Bigr(Y - \mu_0(\bm L^*, \bm L^e_m) \Bigr) - \nu(\bm L^*)\biggr\}~\Bigr|~ \Lambda_{\bm L^e_m| \bm L^*, A, R, Y} \biggr) \\
     &= \frac{AR}{\eta(\bm L^*, 1)}\biggr\{E \Bigr(Y - \mu_0(\bm L^*, \bm L^e_m) \Bigr) - \nu(\bm L^*)\biggr\} - \mathbb{E}\biggr[\frac{AR}{\eta(\bm L^*, 1)}\biggr\{E \Bigr(Y - \mu_0(\bm L^*, \bm L^e_m) \Bigr) - \nu(\bm L^*)\biggr\}~\Bigr|~ \bm L^*, A, R, Y \biggr] \\
     &= \frac{AR}{\eta(\bm L^*, 1)}\biggr\{E \Bigr(Y - \mu_0(\bm L^*, \bm L^e_m) \Bigr) - \nu(\bm L^*)\biggr\} - \frac{AR}{\eta(\bm L^*, 1)}\biggr\{Y\mathbb{E}[E~|~\bm L^*, A, R, Y] - \mathbb{E}[E\mu_0(\bm L^*,  \bm L^e_m)~|~\bm L^*, A, R, Y] - \nu(\bm L^*)\biggr\} \\
     &= \frac{AR}{\eta(\bm L^*, 1)}\biggr\{E \Bigr(Y - \mu_0(\bm L^*, \bm L^e_m) \Bigr) - \nu(\bm L^*)\biggr\} - \frac{AR}{\eta(\bm L^*, 1)}\biggr\{\varepsilon_1(\bm L^*, Y)Y - \xi(\bm L^*, Y) -  \nu(\bm L^*)\biggr\} \\
     &= \frac{AR}{\eta(\bm L^*, 1)}\biggr\{\Bigr(E - \varepsilon_1(\bm L^*, Y)\Bigr)Y - \Bigr(E\mu_0(\bm L^*, \bm L^e_m) - \xi(\bm L^*, Y)\Bigr)\biggr\}
\end{aligned}
$$

$$
\begin{aligned}
     &\Pi\biggr(\frac{AR}{\eta(\bm L^*, 1)}\biggr\{E \Bigr(Y - \mu_0(\bm L^*, \bm L^e_m) \Bigr) - \nu(\bm L^*)\biggr\}~\Bigr|~ \Lambda_{ Y| \bm L^*, A} \biggr) \\
     &=  \mathbb{E}\biggr[\frac{AR}{\eta(\bm L^*, 1)}\biggr\{E \Bigr(Y - \mu_0(\bm L^*, \bm L^e_m) \Bigr) - \nu(\bm L^*)\biggr\}~\Bigr|~ \bm L^*, A, Y \biggr] - \mathbb{E}\biggr[\frac{AR}{\eta(\bm L^*, 1)}\biggr\{E \Bigr(Y - \mu_0(\bm L^*, \bm L^e_m) \Bigr) - \nu(\bm L^*)\biggr\}~\Bigr|~ \bm L^*, A \biggr] \\
  &=  \mathbb{E}\biggr[A\biggr\{E \Bigr(Y - \mu_0(\bm L^*, \bm L^e_m) \Bigr) - \nu(\bm L^*)\biggr\}~\Bigr|~ \bm L^*, A, Y, R = 1 \biggr] - \mathbb{E}\biggr[A\biggr\{E \Bigr(Y - \mu_0(\bm L^*, \bm L^e_m) \Bigr) - \nu(\bm L^*)\biggr\}~\Bigr|~ \bm L^*, A, R =1 \biggr]~~~\text{(\ref{lemma:bvt})} \\
  &= A\biggr(Y \mathbb{E}[E~|~\bm L^*, A = 1, Y, R = 1] - \mathbb{E}[E\mu_0(\bm L^*,\bm L^e_m)~|~\bm L^*, A = 1, Y, R = 1] - \nu(\bm L^*)\biggr)\\ &~~~- A\biggr(\mathbb{E}\Bigr[E\bigr(Y -\mu_0(\bm L^*,\bm L^e_m)\bigr)~\Bigr|~\bm L^*, A = 1, R = 1\Bigr] - \nu(\bm L^*)\biggr)\\
  &= A\Bigr(\varepsilon_1(\bm L^*, Y) Y - \xi(\bm L^*, Y) - \nu(\bm L^*)\Bigr) -A\Bigr( \nu(\bm L^*) - \nu(\bm L^*)\Bigr) \\&= A\Bigr(\varepsilon_1(\bm L^*, Y) Y - \xi(\bm L^*, Y) - \nu(\bm L^*)\Bigr)
\end{aligned}
$$
\normalsize
\noindent Now for the final term, $\frac{(1-A)R}{\eta(\bm L^*, 1)}\frac{ u(\bm L^*, \bm L^e_m)}{1 - u(\bm L^*, \bm L^e_m)}\biggr\{E \Bigr(Y - \mu_0(\bm L^*, \bm L^e_m)\Bigr)\biggr\}$
\footnotesize
$$
\begin{aligned}
    &\Pi\biggr(\frac{(1-A)R}{\eta(\bm L^*, 1)}\frac{ u(\bm L^*, \bm L^e_m)}{1 - u(\bm L^*, \bm L^e_m)}\biggr\{E \Bigr(Y - \mu_0(\bm L^*, \bm L^e_m)\Bigr)\biggr\}~\Bigr|~ \Lambda_{\bm L^*, A, R} \biggr)\\
    &= \mathbb{E}\biggr[\frac{(1-A)R}{\eta(\bm L^*, 1)}\frac{ u(\bm L^*, \bm L^e_m)}{1 - u(\bm L^*, \bm L^e_m)}\biggr\{E \Bigr(Y - \mu_0(\bm L^*, \bm L^e_m)\Bigr)\biggr\}~\Bigr|~ \bm L^*, A, R \biggr]\\
    &= \mathbb{E}\biggr[\frac{(1-A)R}{\eta(\bm L^*, 1)}\frac{ u(\bm L^*, \bm L^e_m)}{1 - u(\bm L^*, \bm L^e_m)}E \mathbb{E}[ Y - \mu_0(\bm L^*, \bm L^e_m)~|~\bm L^*, A = 0, R = 1, \bm L^e_m]~\Bigr|~ \bm L^*, A, R \biggr]\\
    &= \mathbb{E}\biggr[\frac{(1-A)R}{\eta(\bm L^*, 1)}\frac{ u(\bm L^*, \bm L^e_m)}{1 - u(\bm L^*, \bm L^e_m)}E\Bigr(\mu_0(\bm L^*, \bm L^e_m) - \mu_0(\bm L^*, \bm L^e_m)\biggr)\Bigr]\\
    &= 0
\end{aligned}
$$

$$
\begin{aligned}
    &\Pi\biggr(\frac{(1-A)R}{\eta(\bm L^*, 1)}\frac{ u(\bm L^*, \bm L^e_m)}{1 - u(\bm L^*, \bm L^e_m)}\biggr\{E \Bigr(Y - \mu_0(\bm L^*, \bm L^e_m) \Bigr)\biggr\}~\Bigr|~ \Lambda_{\bm L^e_m~|~\bm L^*, A, R, Y} \biggr)\\
    &= \frac{(1-A)R}{\eta(\bm L^*, 1)}\frac{ u(\bm L^*, \bm L^e_m)}{1 - u(\bm L^*, \bm L^e_m)}\biggr\{E \Bigr(Y - \mu_0(\bm L^*, \bm L^e_m)\Bigr) \biggr\} -
    \mathbb{E}\biggr[\frac{(1-A)R}{\eta(\bm L^*, 1)}\frac{ u(\bm L^*, \bm L^e_m)}{1 - u(\bm L^*, \bm L^e_m)}\biggr\{E \Bigr(Y - \mu_0(\bm L^*, \bm L^e_m)\Bigr) \biggr\}~\Bigr|~ \bm L^*, A, R, Y\biggr] \\
    &= \frac{(1-A)R}{\eta(\bm L^*, 1)}\frac{ u(\bm L^*, \bm L^e_m)}{1 - u(\bm L^*, \bm L^e_m)}\biggr\{E \Bigr(Y - \mu_0(\bm L^*, \bm L^e_m)\Bigr) \biggr\}\\ &-
    \frac{(1-A)R}{\eta(\bm L^*, 1)} \biggr\{Y \mathbb{E}\biggr[E \frac{ u(\bm L^*, \bm L^e_m)}{1 - u(\bm L^*, \bm L^e_m)}~\Bigr|~ \bm L^*, A, R, Y\biggr]  -  \mathbb{E}\biggr[E  \frac{ u(\bm L^*, \bm L^e_m)}{1 - u(\bm L^*, \bm L^e_m)}\mu_0(\bm L^*, \bm L^e_m)~\Bigr|~ \bm L^*, A, R, Y\biggr]\biggr\}\\
    &= \frac{(1-A)R}{\eta(\bm L^*, 1)}\biggr\{E\frac{ u(\bm L^*, \bm L^e_m)}{1 - u(\bm L^*, \bm L^e_m)}\Bigr(Y - \mu_0(\bm L^*, \bm L^e_m\Bigr) -\Bigr( \gamma(\bm L^*, Y)Y - \chi(\bm L^*, Y)\Bigr)\biggr\}
\end{aligned}
$$

$$
\begin{aligned}
    &\Pi\biggr(\frac{(1-A)R}{\eta(\bm L^*, 1)}\frac{ u(\bm L^*, \bm L^e_m)}{1 - u(\bm L^*, \bm L^e_m)}\biggr\{E \Bigr(Y - \mu_0(\bm L^*, \bm L^e_m) \Bigr)\biggr\}~\Bigr|~ \Lambda_{\bm Y~|~\bm L^*, A} \biggr)\\
    &=  \mathbb{E}\biggr[\frac{(1-A)R}{\eta(\bm L^*, 1)}\frac{ u(\bm L^*, \bm L^e_m)}{1 - u(\bm L^*, \bm L^e_m)}\biggr\{E \Bigr(Y - \mu_0(\bm L^*, \bm L^e_m)\Bigr) \biggr\}~\Bigr|~ \bm L^*, A, Y\biggr]\\ &-
    \mathbb{E}\biggr[\frac{(1-A)R}{\eta(\bm L^*, 1)}\frac{ u(\bm L^*, \bm L^e_m)}{1 - u(\bm L^*, \bm L^e_m)}\biggr\{E \Bigr(Y - \mu_0(\bm L^*, \bm L^e_m)\Bigr) \biggr\}~\Bigr|~ \bm L^*, A\biggr] \\
    &= (1-A)\frac{\eta(\bm L^*, 0)}{\eta(\bm L^*, 1)}\Biggr\{Y \mathbb{E}\biggr[E\frac{u(\bm L^*, \bm L^e_m)}{1-u(\bm L^*, \bm L^e_m)}~|~\bm L^*, A = 0, R=1, Y\biggr] - \mathbb{E}\biggr[E \frac{u(\bm L^*, \bm L^e_m)}{1-u(\bm L^*, \bm L^e_m)}\mu_0(\bm L^*, \bm L^e_m)~|~\bm L^*, A = 0, R=1, Y\biggr]\Biggr\} \\
    &- (1-A)\frac{\eta(\bm L^*, 0)}{\eta(\bm L^*, 1)}\Biggr\{\mathbb{E}\biggr[E Y\frac{u(\bm L^*, \bm L^e_m)}{1-u(\bm L^*, \bm L^e_m)}~|~\bm L^*, A = 0, R=1\biggr] - \mathbb{E}\biggr[E\frac{u(\bm L^*, \bm L^e_m)}{1-u(\bm L^*, \bm L^e_m)}\mu_0(\bm L^*, \bm L^e_m)~|~\bm L^*, A = 0, R=1\biggr]\Biggr\} \\
    &= (1-A)\frac{\eta(\bm L^*, 0)}{\eta(\bm L^*, 1)} \biggr(\gamma(\bm L^*, Y)Y - \chi(\bm L^*, Y)\biggr)
\end{aligned}
$$
\normalsize

\noindent The second/third to last lines follow from Lemma \ref{lemma:bvt}. That the 2nd term disappears in the final line is because

$$
\begin{aligned}
&~\mathbb{E}\biggr[E Y\frac{u(\bm L^*, \bm L^e_m)}{1-u(\bm L^*, \bm L^e_m)}~|~\bm L^*, A = 0, R=1\biggr] - \mathbb{E}\biggr[E\frac{u(\bm L^*, \bm L^e_m)}{1-u(\bm L^*, \bm L^e_m)}\mu_0(\bm L^*, \bm L^e_m)~|~\bm L^*, A = 0, R=1\biggr] \\
&= \mathbb{E}\biggr[E \Bigr(Y-\mu_0(\bm L^*, \bm L^e_m)\Bigr)\frac{u(\bm L^*, \bm L^e_m)}{1-u(\bm L^*, \bm L^e_m)}~\Bigr|~\bm L^*, A = 0, R=1\biggr]\\
&= \mathbb{E}\biggr[\mathbb{E}\biggr[E \Bigr(Y-\mu_0(\bm L^*, \bm L^e_m)\Bigr)\frac{u(\bm L^*, \bm L^e_m)}{1-u(\bm L^*, \bm L^e_m)}~|~\bm L^*, A = 0, R=1, \bm L^e_m\biggr]~|~\bm L^*, A = 0, R=1\biggr] \\
&= \mathbb{E}\biggr[E \Bigr(\mu_0(\bm L^*, \bm L^e_m)-\mu_0(\bm L^*, \bm L^e_m)\Bigr)\frac{u(\bm L^*, \bm L^e_m)}{1-u(\bm L^*, \bm L^e_m)}~|~\bm L^*, A = 0, R=1\biggr] \\
&= 0
\end{aligned}
$$

\noindent Putting the pieces together we have that 

$$
\begin{aligned}
\Pi\bigr(\beta_P'^{*}(O), \Lambda_P\bigr) &=  \frac{AR}{\eta(\bm L^*, 1)}\biggr[\Bigr(E - \varepsilon_1(\bm L^*, Y)\Bigr)Y -\Bigr(E\mu_0(\bm L^*, \bm L^e_m) - \xi(\bm L^*, Y)\Bigr)\biggr] \\
& + A\Bigr( \varepsilon_1(\bm L^*, Y)Y - \xi(\bm L^*, Y)\Bigr)  \\
& - \frac{(1-A)R}{\eta(\bm L^*, 1)}\biggr[E\frac{ u(\bm L^*, \bm L^e_m)}{1- u(\bm L^*, \bm L^e_m)}\Bigr(Y -  \mu_0(\bm L^*, \bm L^e_m)\Bigr) -\Bigr( \gamma(\bm L^*, Y)Y - \chi(\bm L^*, Y)\Bigr)\biggr] \\
& -(1-A)\frac{ \eta(\bm L^*, 0)}{ \eta(\bm L^*, 1)} \Bigr(\gamma(\bm L^*, Y)Y - \chi(\bm L^*, Y)\Bigr) - \beta(P)
\end{aligned}
$$

\noindent which is exactly $\beta^*_P(O)$. Thus, $\beta^*_P(O)$ is the efficient influence function of $\beta(P)$, concluding the proof of Theorem 2.

\subsection{Asymptotic Behavior of \texorpdfstring{$\widehat{\theta}_\text{EIF}$}{Theta Hat EIF} (Proof of Theorem 3)}\label{suppSec:theory_asymptotics}
\noindent\underline{\textbf{Proof of Theorem 3}} \\
\noindent To prove Theorem 3, we follow arguments used for ratio parameters by (Kennedy et al., 2023) (Theorem 3) \citep{kennedy2023} and Levis et al. (Theorem 1) \citep{levis2024IV}. For ease of notation, we use the shorthand $P[f] = \mathbb{E}_{P}[f(O)]$ for the mean of any function $f$ under distribution $P$. We then have that

$$
\begin{aligned}
\widehat\theta_\text{EIF} - \theta(P) &= \frac{\mathbb{P}_n[\dot\beta_{\widehat P}(O)]}{\mathbb{P}_n[\dot\alpha_{\widehat P}(O)]} - \frac{P[\dot\beta_P(O)]}{P[\dot\alpha_P(O)]} \\
&= \frac{1}{\mathbb{P}_n[\dot\alpha_{\widehat P}(O)]}\biggr\{ \mathbb{P}_n[\dot\beta_{\widehat P}(O)]-  P[\dot\beta_P(O)] - \theta(P)\Bigr(\mathbb{P}_n[\dot\alpha_{\widehat P}(O)]-  P[\dot\alpha_P(O)] \Bigr)\biggr\}
\end{aligned}
$$

\noindent Now applying the results of the von Mises expansions in Lemma \ref{lemma:vonMises_IF},  we have the following decompositions: 

$$
\begin{aligned}
\mathbb{P}_n[\dot\beta_{\widehat P}(O)]-  P[\dot\beta_P(O)] &= (\mathbb{P}_n - P)[\dot\beta_P(O)] + (\mathbb{P}_n - P)[\dot\beta_{\widehat P}(O) - \dot\beta_P(O)] + R_\beta(\widehat P, P) \\
\mathbb{P}_n[\dot\alpha_{\widehat P}(O)]-  P[\dot\alpha_P(O)] &= (\mathbb{P}_n - P)[\dot\alpha_P(O)] + (\mathbb{P}_n - P)[\dot\alpha_{\widehat P}(O) - \dot\alpha_P(O)] + R_\alpha(\widehat P, P)
\end{aligned}
$$

\noindent Because our estimation procedure for $\widehat\theta_\text{EIF}$ leverages sample splitting (Algorithm 1), Lemma 2 of (Kennedy, 2020) guarantees that $(\mathbb{P}_n - P)[\dot\beta_{\widehat P}(O) - \dot\beta_P(O)]$ and $(\mathbb{P}_n - P)[\dot\alpha_{\widehat P}(O) - \dot\alpha_P(O)]$ are both $o_P(n^{-1/2})$ \citep{kennedy_2020}. Plugging in these decompositions, and using the fact that $\mathbb{P}_n[\dot\alpha_{\widehat P}(O)]$ is bounded away from zero, we have that 

$$
\widehat\theta_\text{EIF} - \theta(P) = \frac{1}{\mathbb{P}_n[\dot\alpha_{\widehat P}(O)]}(\mathbb{P}_n - P)\bigr(\dot\beta_P(O) - \theta(P)\dot\alpha_P(O)\bigr) + O_P(|R_\alpha| + |R_\beta|) + o_P(n^{-1/2})
$$

\noindent Next, we observe that 

\begin{equation}\label{eqn:alpha_decomp}
\begin{aligned}
 \mathbb{P}_n[\dot\alpha_{\widehat P}(O)] - \alpha(P) &= \mathbb{P}_n[\dot\alpha_{\widehat P}(O)] - P[\dot\alpha_P(O)] \\
 &= (\mathbb{P}_n - P)\bigr(\dot\alpha_P(O)\bigr) + (\mathbb{P}_n - P)\bigr(\dot\alpha_{\widehat P}(O) - \dot\alpha_P(O)\bigr) + P\bigr(\dot\alpha_{\widehat P}(O) - \dot\alpha_P(O)\bigr)\\
 &= O_P(n^{-1/2}) + O_P(n^{-1/2}) + o_P(1) \\
&= o_P(1)
\end{aligned}
\end{equation}

\noindent where the third equality follows from the central limit theorem (term 1), Lemma 2 of (Kennedy, 2020) \citep{kennedy_2020} (term 2), and that for term 3, $\Bigr|P\bigr(\dot\alpha_{\widehat P}(O) - \dot\alpha_P(O)\bigr)\Bigr| \lesssim  \| \dot\alpha_{\widehat P}(O) - \dot\alpha_P(O) \| = o_P(1)$ by assumption. Recalling from Corollary 2.1 that $\dot\theta^*_P(O) = \frac{1}{\alpha(P)}\bigr(\dot\beta_P(O) - \theta(P)\dot\alpha_P(O)\bigr)$, observe that

$$
\begin{aligned}
&~~~~\frac{1}{\mathbb{P}_n[\dot\alpha_{\widehat P}(O)]}(\mathbb{P}_n - P)\bigr(\dot\beta_P(O) - \theta(P)\dot\alpha_P(O)\bigr) - (\mathbb{P}_n - P)\bigr(\theta_P^*(O)\bigr) \\
&= \biggr(\frac{1}{\mathbb{P}_n[\dot\alpha_{\widehat P}(O)]} - \frac{1}{\alpha(P)}\biggr)(\mathbb{P}_n - P)\bigr(\dot\beta_P(O) - \theta(P)\dot\alpha_P(O)\bigr) \\
&= \frac{\alpha(P) - \mathbb{P}_n[\dot\alpha_{\widehat{P}}(O)]}{\mathbb{P}_n[\dot\alpha_{\widehat{P}}(O)]\alpha(P)}\cdot O_p(n^{-1/2})~~~\text{(Central Limit Theorem)} \\
&=o_p(1)O_p(n^{-1/2}) ~~~\text{(Equation \eqref{eqn:alpha_decomp} and $\mathbb{P}_n[\dot\alpha_{\widehat P}(O)]$ bounded away from zero)}
\\ &= o_P(n^{-1/2})
\end{aligned}
$$

\noindent Combining pieces, we have the desired result, that 

$$
\widehat\theta_\mathrm{EIF} - \theta(P) = \mathbb{P}_n[\dot\theta^*_P(O)] + O_P(|R_\alpha| + |R_\beta|) + o_P(n^{-1/2})
$$

\clearpage

\noindent\underline{\textbf{Proof of Corollary 3.1}}\\
\noindent Since $A \in \{0, 1\}$ is bounded, and $P(\widehat\eta_1 > \epsilon) = 1$ for some $\epsilon > 0$, we have that
$$
\begin{aligned}
|R_\alpha(\widehat P, P)|  &= \Biggr|\mathbb{E}_P\biggr[A(\widehat\varepsilon_1 - \varepsilon_1)\Bigr(1 - \frac{\eta_1}{\widehat\eta_1}\Bigr)\biggr]\Biggr|\\
 &\leq \mathbb{E}_P\biggr[\Bigr|\frac{A}{\widehat \eta_1}(\widehat\varepsilon_1 - \varepsilon_1)( \widehat \eta_1 - \eta_1)\Bigr|\biggr] \\
 &\leq\frac{1}{\epsilon}\|\widehat\varepsilon_1 - \varepsilon_1 \|\|\widehat \eta_1 - \eta_1\|~~~\text{(Cauchy-Schwarz)} \\
 &= O_P(\|\widehat\varepsilon_1 - \varepsilon_1 \|\|\widehat \eta_1 - \eta_1\|)
\end{aligned}
$$

\noindent Similarly, since $|Y| \leq M < \infty$ almost surely, we have that

$$
\begin{aligned}
\Biggr|\mathbb{E}_P\biggr[A\Bigr(1 - \frac{\eta_1}{\widehat\eta_1}\Bigr)\Bigr(Y(\widehat\varepsilon_1 - \varepsilon_1)\Bigr)\biggr]\Biggr| &\leq \mathbb{E}_P\biggr[\Bigr|\frac{AY}{\widehat \eta_1}(\widehat\eta_1 - \eta_1)(\widehat\varepsilon_1 - \varepsilon_1)\Bigr|\biggr] \\
&\leq \frac{1}{\epsilon}\mathbb{E}_P\biggr[|Y|\Bigr|(\widehat\eta_1 - \eta_1)(\widehat\varepsilon_1 - \varepsilon_1)\Bigr|\biggr]~~~(|A| \leq 1,\ \widehat\eta_1 > \epsilon) \\
&\leq \frac{M}{\epsilon} \|\widehat\varepsilon_1 - \varepsilon_1 \|\|\widehat \eta_1 - \eta_1\|~~~\text{($|Y| \leq M$ and Cauchy-Schwarz)} \\
 &= O_P(\|\widehat\varepsilon_1 - \varepsilon_1 \|\|\widehat \eta_1 - \eta_1\|)
\end{aligned}
$$

$$
\begin{aligned}
\Biggr|\mathbb{E}_P\biggr[-A\Bigr(1 - \frac{\eta_1}{\widehat\eta_1}\Bigr)(\widehat\xi - \xi )\biggr]\Biggr| &\leq \mathbb{E}_P\biggr[\frac{1}{\widehat\eta_1}\Bigr|(\widehat\eta_1 - \eta_1)(\widehat\xi - \xi )\Bigr|\biggr] \\
&\leq \frac{1}{\epsilon} \|\widehat\eta_1 - \eta_1 \|\|\widehat\xi - \xi \|~~~\text{(Cauchy-Schwarz)} \\
 &= O_P(\|\widehat\eta_1 - \eta_1 \|\| \widehat\xi - \xi  \|)
\end{aligned}
$$

\noindent Furthermore, since $P[\delta \leq 1 - \widehat u \leq 1 - \delta] = 1$ for some $\delta > 0$, and $R, E \in \{0, 1\}$, we have that 

$$
\begin{aligned}
    \Biggr|\mathbb{E}_P\biggr[\frac{RE}{\widehat\eta_1}(\mu_0 - \widehat\mu_0)\biggr(\frac{u(1 - \widehat u) - \widehat u(1 - u)}{1-\widehat u}\biggr )\biggr]\Biggr| &\leq \mathbb{E}_P\biggr[\Bigr|\frac{RE}{\widehat\eta_1(1-\widehat u)}(\mu_0 - \widehat\mu_0)\Bigr(u(1 - \widehat u) - \widehat u(1 - u)\Bigr)\Bigr|\biggr] \\
    &\leq\frac{1}{\epsilon\delta}\|\widehat \mu_0 - \mu_0\|\|u(1 - \widehat u) - \widehat u(1 - u)\| ~~~\text{(Cauchy-Schwarz)} \\
    &= O_P(\|\widehat \mu_0 - \mu_0\|\|\widehat u - u\|)
\end{aligned}
$$

\noindent Finally

$$
\begin{aligned}
\Biggr|\mathbb{E}_P\biggr[(1-A)\frac{(\eta_0 - \widehat\eta_0)}{\widehat\eta_1} \Bigr(Y(\widehat\gamma -   \gamma)\Bigr) \Biggr| &\leq \mathbb{E}_P\biggr[\Bigr|\frac{(1-A)Y}{\widehat \eta_1}(\widehat\eta_0 - \eta_0)(\widehat \gamma - \gamma)\Bigr|\biggr] \\
&\leq \frac{1}{\epsilon}\mathbb{E}_P\biggr[|Y|\Bigr|(\widehat\eta_0 - \eta_0)(\widehat\gamma - \gamma)\Bigr|\biggr]~~~(|1-A| \leq 1,\ \widehat\eta_1 > \epsilon) \\
&\leq \frac{M}{\epsilon}\|\widehat\eta_0 - \eta_0 \|\|\widehat\gamma - \gamma \|~~~\text{($|Y| \leq M$ and Cauchy-Schwarz)} \\
 &= O_P(\|\widehat\eta_0 - \eta_0 \|\|\widehat\gamma - \gamma \|)
\end{aligned}
$$

$$
\begin{aligned}
\Biggr|\mathbb{E}_P\biggr[-(1-A)\frac{(\eta_0 - \widehat\eta_0)}{\widehat\eta_1}(\widehat\chi - \chi)\Biggr| &\leq \mathbb{E}_P\biggr[\Bigr|\frac{(1-A)}{\widehat \eta_1}(\widehat\eta_0 - \eta_0)(\widehat\chi - \chi)\Bigr|\biggr] \\
&\leq \frac{1}{\epsilon}\|\widehat\eta_0 - \eta_0 \|\|\widehat\chi - \chi \|~~~\text{(Cauchy-Schwarz)} \\
 &= O_P(\|\widehat\eta_0 - \eta_0 \|\|\widehat\chi - \chi \|)
\end{aligned}
$$

\noindent Combining pieces we have that
\footnotesize
$$
\begin{aligned}
R_\beta(\widehat P, P) = O_P\biggr(\|\widehat \mu_0 - \mu_0 \| \|\widehat u - u\| + \|\widehat \eta_1 - \eta_1 \|\Bigr\{ \|\widehat\varepsilon_1 - \varepsilon_1\|  + \|\widehat \xi - \xi\| \Bigr\} +  \|\widehat \eta_0 - \eta_0 \|\Bigr\{ \|\widehat\gamma - \gamma\|  + \|\widehat \chi - \chi\| \Bigr\}\biggr)
\end{aligned}
$$
\normalsize
\noindent and thus 
\footnotesize
$$
R_\alpha(\widehat P, P) + R_\beta(\widehat P, P) = 
O_P\biggr(\|\widehat \mu_0 - \mu_0 \| \|\widehat u - u\| + \|\widehat \eta_1 - \eta_1 \|\Bigr\{ \|\widehat\varepsilon_1 - \varepsilon_1\|  + \|\widehat \xi - \xi\| \Bigr\} +  \|\widehat \eta_0 - \eta_0 \|\Bigr\{ \|\widehat\gamma - \gamma\|  + \|\widehat \chi - \chi\| \Bigr\}\biggr)
$$
\normalsize

\noindent\underline{\textit{Remark on nested nuisance functions.}} In Algorithm 1, the regression targets for $\widehat\xi$, $\widehat\gamma$ and $\widehat\chi$ are built from $\widehat\mu_0$ and $\widehat u$ evaluated on the same training split used to fit them. We do not use nested cross-fitting for these functions, since it is considerably more computationally intensive. Because $\xi$, $\gamma$ and $\chi$ are smooth functionals of $\mu_0$ and $u$ (and $E \in \{0, 1\}$, $\delta \leq 1 - \widehat u \leq 1 - \delta$), the error in each nested function is bounded, up to constants, by the error of its second-stage regression plus $\|\widehat\mu_0 - \mu_0\|$ or $\|\widehat u - u\|$. The bound in Corollary 3.1 therefore holds under the additional assumption that the second-stage regression errors $\|\widehat\xi - \xi_{\widehat\mu_0}\|$, $\|\widehat\gamma - \gamma_{\widehat u}\|$ and $\|\widehat\chi - \chi_{\widehat\mu_0, \widehat u}\|$ are not inflated by the in-sample fit of $\widehat\mu_0$ and $\widehat u$, which is reasonable for regularized learners. Here $\xi_{\widehat\mu_0}(\bm L^*, Y) = \mathbb{E}_P[E\widehat\mu_0(\bm L^*, \bm L^e_m)~|~\bm L^*, Y, A = 1, R = 1]$ is the regression function of the fitted target with $\widehat\mu_0$ held fixed, and $\gamma_{\widehat u}$, $\chi_{\widehat\mu_0, \widehat u}$ are defined analogously. The additional products of $\|\widehat\eta_1 - \eta_1\|$ and $\|\widehat\eta_0 - \eta_0\|$ with $\|\widehat\mu_0 - \mu_0\|$ and $\|\widehat u - u\|$ are $o_P(n^{-1/2})$ under the same $o_P(n^{-1/4})$ conditions as the rest of the corollary.\\

\subsection{Summary of Asymptotic Behavior of  \texorpdfstring{$\widetilde{\theta}_\text{EIF}$}{Theta Tilde EIF}, \texorpdfstring{$\widehat{\theta}_\text{IF}$}{Theta Hat IF}, and \texorpdfstring{$\widetilde{\theta}_\text{IF}$}{Theta Tilde IF} }\label{suppSec:theory_extra}

In this section, we briefly summarize the asymptotic behavior of $\widetilde\theta_\text{EIF}$, as well as $\widehat\theta_\text{IF}$ and $\widetilde\theta_\text{IF}$. We omit formal proofs, but note that justification of results in this section are nearly identical to the proofs of Theorem 3 and Corollary 3.1. \\

\noindent\underline{\textbf{Asymptotic Behavior of $\widetilde\theta_\text{EIF}$}}
\begin{lemma}\label{lemma:widetilde_EIF}
If $\|\dot\alpha_{\widehat P} - \dot\alpha_P\| = o_P(1)$, $\|\dot\beta_{\widetilde P} - \dot\beta_P\| = o_P(1)$, $\alpha(P) > 0$, and $P\Bigr[\bigr|\mathbb{P}_n\bigr(\dot\alpha_{\widehat P}(O)\bigr)\bigr| \geq \epsilon\Bigr] \to 1$ as $n \to \infty$ for some $\epsilon > 0$, then

$$
\widetilde\theta_\mathrm{EIF} - \theta(P) = \mathbb{P}_n[\dot\theta^*_P(O)] + O_P\Bigr(|R_\alpha(\widehat P, P)| + |\widetilde R_\beta(\widehat P, P)|\Bigr) + o_P(n^{-1/2})
$$

\noindent where remainder term, omitting inputs for brevity, is given by 
$$
\begin{aligned}
\widetilde R_\beta(\widebar P, P) &= \mathbb{E}_P\biggr[A\Bigr(1 - \frac{\eta_1}{\widebar \eta_1}\Bigr)\Bigr(Y(\widebar\varepsilon_1 -\varepsilon_1) - (\widebar \xi - \xi) \Bigr) \\
&~~~~~~~~~~+ \frac{RE(\mu_0 - \widebar\mu_0)}{\lambda_1\eta_1\pi + \lambda_0\eta_0(1-\pi)}\biggr\{\lambda_1\pi\Bigr(\frac{\eta_1}{\widebar\eta_1} - 1\Bigr) + (\lambda_1\pi - \widebar\lambda_1\widebar\pi) + \widebar\lambda_1\widebar\pi\Bigr(1 - \frac{\eta_0\lambda_0(1-\pi)}{\widebar\eta_0\widebar\lambda_0(1-\widebar\pi)}\Bigr)\biggr\}\\ 
&~~~~~~~~~~ + (1-A)\frac{\widebar \pi}{1 - \widebar \pi}\cdot\frac{\eta_0 - \widebar\eta_0}{\widebar\eta_0}\biggr\{Y(\widebar \gamma' - \gamma') - (\widebar\chi' - \chi')\biggr\}\biggr]
\end{aligned}
$$

\noindent Moreover, if $|R_\alpha(\widehat P, P)| + |\widetilde R_\beta(\widehat P, P)| = o_P(n^{-1/2})$ then $\sqrt{n}\bigr(\widetilde\theta_\mathrm{EIF} - \theta(P)\bigr) \overset{d}{\to} \mathcal{N}\bigr(0, \mathrm{Var}_P[\dot\theta^*_P(O)]\bigr)$, attaining the semiparametric efficiency bound induced by Assumption 4.
\end{lemma}

\begin{corollary}\label{corr:remainder} Under the conditions of Lemma \ref{lemma:widetilde_EIF} and assuming that $P(\delta \leq 1 - \widehat \pi \leq 1-\delta) = 1$ for some $\delta > 0$, $P(\widehat\eta_a > \epsilon) = 1$ for some $\epsilon > 0, a \in \{0,1\}$, $P(\widehat\lambda_0 > c) = 1$ for some $c > 0$ and $|Y| \leq M < \infty$ almost surely

$$
\begin{aligned}
R_\alpha(\widehat P, P) + \widetilde R_\beta(\widehat P, P) &= O_P\biggr(\|\widehat\eta_1 - \eta_1\|\Bigr\{\|\widehat\varepsilon_1 - \varepsilon_1\| + \|\widehat \xi - \xi\|\Bigr\} + \|\widehat\eta_0 - \eta_0\|\Bigr\{\|\widehat\gamma' - \gamma'\| + \|\widehat \chi' - \chi'\|\Bigr\} \\ &+ \|\widehat\mu_0 - \mu_0\|  \Bigr\{\|\widehat\eta_1 - \eta_1\| + \|\widehat\lambda_1 - \lambda_1\| + \|\widehat \pi - \pi \| + \|\widehat\eta_0 - \eta_0\| + \|\widehat\lambda_0 - \lambda_0 \| \Bigr\}\biggr)
\end{aligned}
$$ 
\end{corollary}

\noindent\underline{\textbf{Asymptotic Behavior of $\widehat\theta_\text{IF}$}}\\~\\
Analogous to Corollary 2.1, we note that an influence function for $\theta(P)$ follows directly from $\dot\alpha_P'(O)$ and $\dot\beta_P'(O)$ by $\dot\theta_P'^{*}(O) = \frac{1}{\alpha(P)}\bigr(\dot\beta'_P(O) - \theta(P)\dot\alpha'_P(O)\bigr)$.

\begin{lemma}\label{lemma:widehat_IF}
If $\|\dot\alpha'_{\widehat P} - \dot\alpha'_P\| = o_P(1)$, $\|\dot\beta'_{\widehat P} - \dot\beta'_P\| = o_P(1)$, $\alpha(P) > 0$, and $P\Bigr[\bigr|\mathbb{P}_n\bigr(\dot\alpha'_{\widehat P}(O)\bigr)\bigr| \geq \epsilon\Bigr] \to 1$ as $n \to \infty$ for some $\epsilon > 0$, then

$$
\widehat\theta_\mathrm{IF} - \theta(P) = \mathbb{P}_n[\dot\theta'^{*}_P(O)] + O_P\Bigr(|R'_\alpha(\widehat P, P)| + |R'_\beta(\widehat P, P)|\Bigr) + o_P(n^{-1/2})
$$

\noindent Moreover, if $|R'_\alpha(\widehat P, P)| + |R'_\beta(\widehat P, P)| = o_P(n^{-1/2})$ then $\sqrt{n}\bigr(\widehat\theta_\mathrm{IF} - \theta(P)\bigr) \overset{d}{\to} \mathcal{N}\bigr(0, \mathrm{Var}_P[\dot\theta'^{*}_P(O)]\bigr)$
\end{lemma}

\begin{corollary} Under the conditions of Lemma \ref{lemma:widehat_IF} and Corollary 3.1

$$
\begin{aligned}
R'_\alpha(\widehat P, P) + R'_\beta(\widehat P, P) &= O_P\biggr(\|\widehat\eta_1 - \eta_1\|\Bigr\{\|\widehat\nu - \nu\| + \|\widehat\omega_1 - \omega_1\|\Bigr\} +  \|\widehat\mu_0 - \mu_0\| \| \widehat u - u\|\biggr)
\end{aligned}
$$ 
\end{corollary}

\noindent\underline{\textbf{Asymptotic Behavior of $\widetilde\theta_\text{IF}$}}
\begin{lemma}\label{lemma:widetilde_IF}
If $\|\dot\alpha'_{\widehat P} - \dot\alpha'_P\| = o_P(1)$, $\|\dot\beta'_{\widetilde P} - \dot\beta'_P\| = o_P(1)$, $\alpha(P) > 0$, and $P\Bigr[\bigr|\mathbb{P}_n\bigr(\dot\alpha'_{\widehat P}(O)\bigr)\bigr| \geq \epsilon\Bigr] \to 1$ as $n \to \infty$ for some $\epsilon > 0$, then

$$
\widetilde\theta_\mathrm{IF} - \theta(P) = \mathbb{P}_n[\dot\theta_P'^{*}(O)] + O_P\Bigr(|R'_\alpha(\widehat P, P)| + |\widetilde R'_\beta(\widehat P, P)|\Bigr) + o_P(n^{-1/2})
$$

\noindent where remainder term, omitting inputs for brevity, is given by 
\footnotesize
$$
\begin{aligned}
\widetilde R'_\beta(\widebar P, P) = \mathbb{E}_P\biggr[A\Bigr(1 - \frac{\eta_1}{\widebar \eta_1}\Bigr)(\widebar\nu - \nu) +
\frac{RE(\mu_0 - \widebar\mu_0)}{\lambda_1\eta_1\pi + \lambda_0\eta_0(1-\pi)}\biggr\{\lambda_1\pi\Bigr(\frac{\eta_1}{\widebar\eta_1} - 1\Bigr) + (\lambda_1\pi - \widebar\lambda_1\widebar\pi) + \widebar\lambda_1\widebar\pi\Bigr(1 - \frac{\eta_0\lambda_0(1-\pi)}{\widebar\eta_0\widebar\lambda_0(1-\widebar\pi)}\Bigr)\biggr\}
\biggr]
\end{aligned}
$$
\normalsize

\noindent Moreover, if $|R'_\alpha| + |\widetilde R'_\beta| = o_P(n^{-1/2})$ then $\sqrt{n}\bigr(\widetilde\theta_\mathrm{IF} - \theta(P)\bigr) \overset{d}{\to} \mathcal{N}\bigr(0, \mathrm{Var}_P[\dot\theta_P'^{*}(O)]\bigr)$.
\end{lemma}

\begin{corollary}Under the conditions of Lemma \ref{lemma:widetilde_IF} and Corollary \ref{corr:remainder}
$$
\begin{aligned}
R'_\alpha(\widehat P, P) + \widetilde R'_\beta(\widehat P, P) &= O_P\biggr(\|\widehat\eta_1 - \eta_1\|\Bigr\{\|\widehat\omega_1 - \omega_1\| + \|\widehat \nu - \nu\|\Bigr\} \\ &+ \|\widehat\mu_0 - \mu_0\|  \Bigr\{\|\widehat\eta_1 - \eta_1\| + \|\widehat\lambda_1 - \lambda_1\| + \|\widehat \pi - \pi \| + \|\widehat\eta_0 - \eta_0\| + \|\widehat\lambda_0 - \lambda_0 \| \Bigr\}\biggr)
\end{aligned}
$$ 
\end{corollary}~\\

There are a few comments worth pointing out regarding these asymptotic results. To begin with $\widehat\theta_\text{EIF}$ and $\widetilde\theta_\text{EIF}$ have the same asymptotic distribution, and likewise with $\widehat\theta_\text{IF}$ and $\widetilde\theta_\text{IF}$. Of course, the asymptotic results rely on slightly different assumptions regarding respective von Mises remainder terms. Moreover, performance may not be the same in finite samples. Given the appearance of $\lambda$ in remainder terms for both $\widetilde\theta_\text{EIF}$ and $\widetilde\theta_\text{IF}$, these estimators may be unrealistic in practice unless $\bm L^e_m$ is low dimensional and one is willing to entertain a parametric model for $\bm L^e_m$.

It is also worth noting that $\widehat\theta_\text{IF}$ and $\widetilde\theta_\text{IF}$ do not attain the semiparametric efficiency bound induced by Assumption 4. Nevertheless, studying the form of $R'_\alpha(\widehat P, P) + R'_\beta(\widehat P, P)$, there is some clear appeal to the estimator $\widehat\theta_\text{IF}$. To begin with, consistency only requires that $\eta_1$ and one of the typical ATT nuisance functions are consistent, even if nested nuisance function $\nu$ is inconsistent. This is in contrast to $\widehat\theta_\text{EIF}$ which requires $\eta_1$ and $\eta_0$ to both be consistent if any nested nuisance function is inconsistent, which could be more challenging if there is significant differential missingness by treatment status, and/or interactions between $A$ and $\bm L^*$. Altogether, the optimal choice of estimator may be problem specific based on which assumptions an analyst is most likely to entertain, as well as the complexity of component nuisance functions.

\section{Simulation Study}\label{sec:simulations} 
\subsection{Data Generation}\label{sec:simulations_DGP}

To investigate finite sample performance of $\widehat\theta_\text{EIF}$ and $\widehat\theta_\text{IF}$ we conducted a simulation study tied closely to our motivating bariatric surgery study. In this simulation study, interest lies in the effect of RYGB vs. SG $(A)$ on percent weight change 3 years post surgery $(Y)$ among patients with pre-diabetes or diabetes, defined as an A1C $\geq 5.7\%$ ($\bm L^e_m$). Covariates $\bm L^*$ included health care site, sex, race, baseline BMI, age, smoking status, and eGFR. 

To generate simulated datasets, we began by sampling covariate vectors $\bm L^*$ from the observed distribution in the DURABLE data to preserve the complex correlation structure of $\bm L^*$. Simulated treatment, outcome, and eligibility status (both $\bm L^e_m$ and $R$) were generated using sampled covariates $\bm L^*$ and models for the nuisance functions tied directly to the likelihood factorization in Equation (2) ($\pi, \eta, \lambda, \mu$). In particular, parametric models were fit on a sample of 16,461 surgical patients from the DURABLE database to inform this simulated data generation. This sample is larger than the analytic cohort of 14,809 patients in the data application, because it also includes patients undergoing surgery prior to 2008. Logistic regression was used for $\pi$ and $\eta$, a gamma generalized linear model (GLM) was used for $\lambda$, and linear regression was used for $\mu$. The outcome model $\mu$ was specified with several interactions between $A\times\bm L^e_m, A\times \bm L^*$, and $\bm L^e_m\times \bm L^*$ in order to introduce additional complexity and make $\mu$ difficult to estimate well, perhaps even with more flexible machine learning methods. 

Simulated data were generated following the factorization of the observed data likelihood according to the following steps. In the below steps, we let $\bm\beta_f$ and $\bm X_f$ respectively denote the coefficient vector and design matrix relevant to nuisance function $f(\cdot)$. Exact values of $\bm \beta_f$ are presented in Table \ref{table:sim_coef}.

\begin{enumerate}
    \item Draw $\bm L^*$ from the observed set of non-eligibility defining covariates in the DURABLE database.
    \item Sample surgery type $A~|~\bm L^* \sim \text{Bernoulli}\bigr(\pi(\bm L^*)\bigr)$ where $\text{logit}[\pi(\bm L^*)] = \bm\beta_\pi^T\bm X_\pi$
    \item Sample complete case indicator $R~|~\bm L^*, A \sim \text{Bernoulli}\bigr(\eta(L^*, A)\bigr)$ where $\text{logit}[\eta(\bm L^*, A)] = \bm\beta_\eta^T\bm X_\eta$
    \item Sample A1C according to $(L^e_m-3)~|~\bm L^*, A, R = 1 \sim \text{Gamma}\bigr(\alpha_\lambda, b(\bm L^*, A)\bigr)$ with rate parameter $b(\bm L^*, A) = \frac{\alpha_\lambda}{\mathbb{E}_P[L^e_m~|~\bm L^*, A, R = 1]}$, and we parameterize $\log(\mathbb{E}_P[L^e_m~|~\bm L^*, A, R = 1]) = \bm\beta_\lambda^T \bm X_\lambda$
    \item Sample 3 year weight loss $Y~|~\bm L^*, A, R = 1, L^e_m \sim \mathcal{N}\bigr(\mu_A(\bm L^*,  L^e_m), \sigma_y^2\bigr)$ where $\mu_A(\bm L^*, L^e_m) = \bm \beta_\mu^T\bm X_\mu$
    \item For subjects with $R = 0$, set the value of $L^e_m$ to be missing (\texttt{NA}).
\end{enumerate}

 1,000 simulated datasets were generated for each of $n = 10,000$ and $n = 25,000$ patients. These sample sizes are reasonable in the context many EHR-based settings, including our study of bariatric surgery for T2DM outcomes ($n = 14,809$), the work of McTigue et al. ($n = 9,710$) as well as previous DURABLE studies $(n > 30,000)$ \citep{li2021fiveyear, arterburn2020}. Across simulated datasets, 62\% of patients were treated with RYGB ($A = 1$) and 70\% of patients had ascertainable eligibility ($R = 1$), of whom 57\% were eligible (i.e., $P(E = 1 ~|~R = 1)$).

\subsection{Estimators}
\label{sec:simulations_estimators}

When implementing $\widehat\theta_\text{CC}$, we only estimated $\mu$ with the correct parametric specification, so that any resulting bias was due to selection bias and not model misspecification. When estimating $\widehat\theta_\text{IWOR}$, we considered versions where both $\mu$ and $\eta$ were correctly specified, as well as versions where only one of the two was correctly specified. For $\widehat\theta_\text{IWOR}$ and the two influence function-based estimators, we applied six nonparametric strategies to each for flexible estimation of component nuisance functions using ensemble regression learners from the \texttt{SuperLearner} package in \texttt{R} \citep{van2007super, superlearner}. 

The six strategies arise from all combinations of two distinct sets of libraries included in the SuperLearner (SL) ensemble and three strategies for estimating the outcome regression among SG subjects, $\mu_0$. The first set of SL libraries (SL1) included random forest (RF) learners with several hyperparameter combinations, along with learners for linear model (LM)/GLM, generalized additive model (GAM) (for binary variables), and multivariate adaptive polynomial regression spline (Polymars). The second set (SL2) dropped LM/GLM learners to test estimator behavior when the correct parametric model was not directly among the family of candidate learners. 

\begin{figure}
    \centering
    \includegraphics[width=\textwidth]{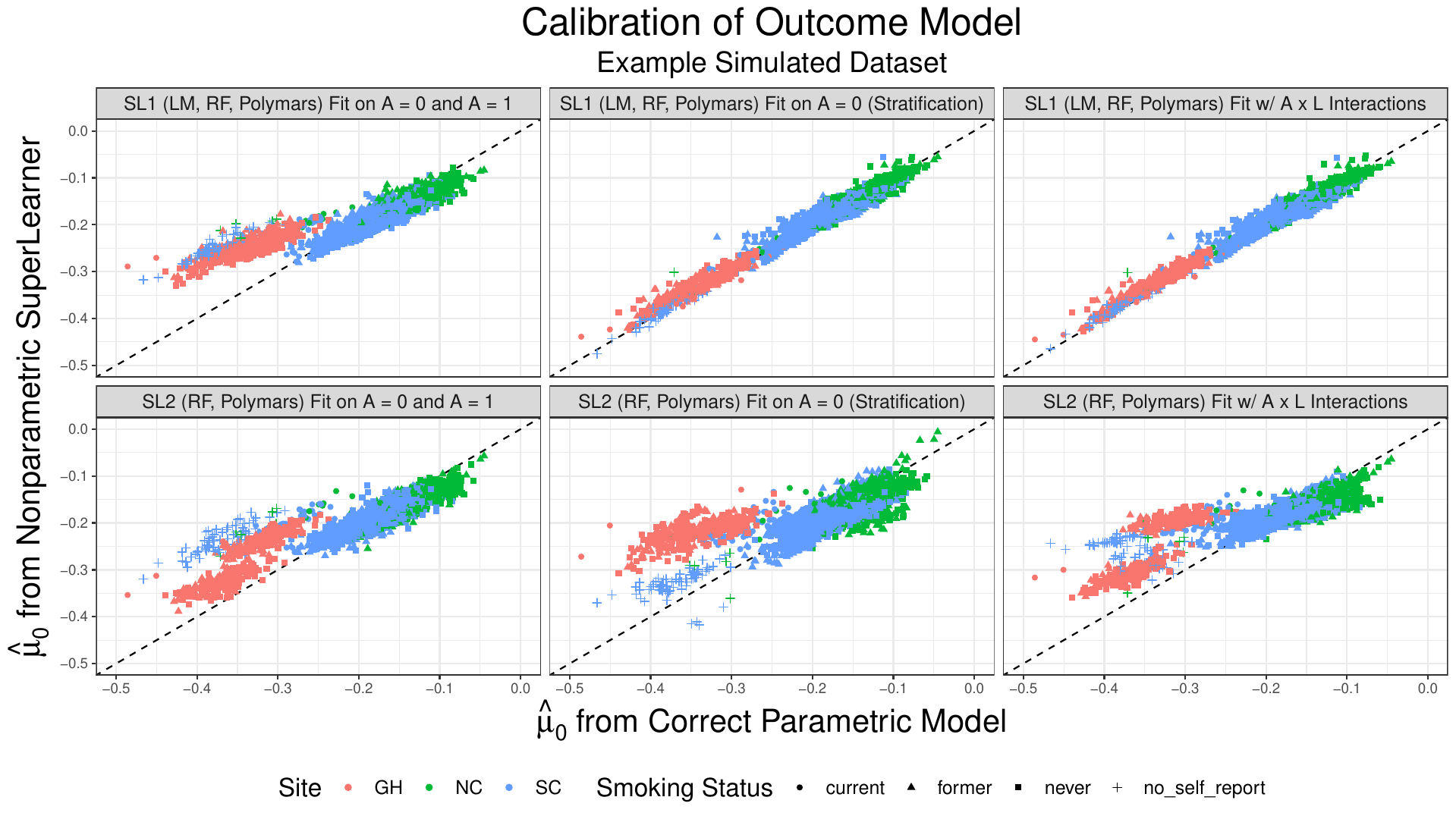}
    \caption{Estimates of $\widehat\mu_0$ on a single simulated dataset across 6 nonparametric strategies compared with estimates of $\widehat\mu_0$ from the correctly specified parametric model. Points are stratified by surgical site and self-reported smoking status, two covariates in $\bm L^*$ which have interactions with surgical procedure $A$ in $\mu$.}
    \label{fig:mu0_calibration}
\end{figure}

Given the complexity of interactions in the outcome model, we considered stratification (e.g., estimating $\mu_0$ only on subjects with $A = 0$) and augmenting the design matrix with all $A\times \bm L$ interactions as alternatives to fitting a single outcome model $\mu$ on all complete cases. The rationale is that fitting a single outcome model might inadequately capture the full scope of interactions in $\mu$, and thus there could be some degree of model misspecification, which may affect various estimators in different ways. 

Figure \ref{fig:mu0_calibration} illustrates estimates of $\widehat\mu_0$ on a single simulated dataset across the six flexible strategies in comparison to estimates of $\widehat\mu_0$ from the correctly specified parametric model. Clearly there is some variability in how well these strategies can estimate the complex form of $\mu$, and thus perhaps the performance of different estimators for $\theta(P)$.

\subsection{Results}
\label{sec:simulations_results}

Table \ref{tab:sim_results} presents relative bias and standard errors for all estimators, as well as coverage of 95\% confidence intervals for the two influence function-based estimators, $\widehat\theta_\text{EIF}$ and $\widehat\theta_\text{IF}$. Unsurprisingly, $\widehat\theta_\text{CC}$ was biased by over 15\%, even with correctly specified outcome model $\mu$, thereby demonstrating that selection bias can be a prominent concern when discarding subjects with incomplete eligibility information. Performance of $\widehat\theta_\text{IWOR}$ was very dependent on model specification. When $\mu$ and $\eta$ were both correctly specified, $\widehat\theta_\text{IWOR}$ was unbiased, but when either nuisance function was misspecified the resulting biases of 17\% and 24\%, respectively, were worse than that of $\widehat\theta_\text{CC}$. Using nonparametric SuperLearner ensembles to estimate $\mu, \eta$ did not guarantee unbiasedness, with plug-in bias ranging from 2-19\%, with bias worse in cases with greater estimation error for $\widehat\mu_0$ (Figure \ref{fig:mu0_calibration}).

\begin{sidewaystable}\scriptsize
\centering
\begin{tabular}[t]{ccccccccccccc}
\toprule
\multicolumn{5}{c}{\textbf{ }} & \multicolumn{4}{c}{\textbf{$n = 10,000$ Patients}} & \multicolumn{4}{c}{\textbf{$n = 25,000$ Patients}} \\
\cmidrule(l{3pt}r{3pt}){6-9} \cmidrule(l{3pt}r{3pt}){10-13}
\textbf{Estimator} & \textbf{Strategy} & \textbf{SL Libs$\textsuperscript{a}$} & \textbf{$\mu_0$ Strategy$\textsuperscript{b}$} & \textbf{True $\mu/\eta$\textsuperscript{c}} & \textbf{\%-Bias} & \textbf{SD} & \textbf{Mean $\widehat{\text{SE}}[\widehat\theta]$} & \textbf{Coverage} & \textbf{\%-Bias} & \textbf{SD} &  \textbf{Mean $\widehat{\text{SE}}[\widehat\theta]$} & \textbf{Coverage}\\
\midrule
$\widehat\theta_\text{CC}$ &  Parametric & --- &  1 &  $\mu$ & 15.4 & 3.72e-03 & --- & --- & 15.7 & 2.39e-03 & --- & ---\\
\cmidrule{1-13}
 &  &  &  &  $\eta/\mu$ & 0.0 & 4.35e-03 & --- & --- & 0.1 & 2.72e-03 & --- & ---\\
\cmidrule{5-13}
 &  &  &  &  $\mu$ & 16.5 & 3.77e-03 & --- & --- & 16.8 & 2.43e-03 & --- & ---\\
\cmidrule{5-13}
 & \multirow{-3}{*}[5pt]{\centering\arraybackslash  Parametric} & \multirow{-3}{*}[5pt]{\centering\arraybackslash ---} & \multirow{-3}{*}[5pt]{\centering\arraybackslash  1} &  $\eta$ & 23.8 & 3.76e-03 & --- & --- & 24.0 & 2.43e-03 & --- & ---\\
\cmidrule{2-13}
 &  &  &  1 &  & 19.1 & 4.30e-03 & --- & --- & 17.9 & 3.18e-03 & --- & ---\\
\cmidrule{4-4}
\cmidrule{6-13}
 &  &  &  2 &  & 3.7 & 4.56e-03 & --- & --- & 1.8 & 2.82e-03 & --- & ---\\
\cmidrule{4-4}
\cmidrule{6-13}
 &  & \multirow{-3}{*}[5pt]{\centering\arraybackslash SL1} &  3 &  & 3.8 & 4.50e-03 & --- & --- & 2.2 & 2.86e-03 & --- & ---\\
\cmidrule{3-4}
\cmidrule{6-13}
 &  &  &  1 &  & 7.9 & 5.99e-03 & --- & --- & 11.0 & 4.53e-03 & --- & ---\\
\cmidrule{4-4}
\cmidrule{6-13}
 &  &  &  2 &  & 17.1 & 4.78e-03 & --- & --- & 11.7 & 3.34e-03 & --- & ---\\
\cmidrule{4-4}
\cmidrule{6-13}
\multirow{-9}{*}[25pt]{\centering\arraybackslash $\widehat\theta_\text{IWOR}$} & \multirow{-6}{*}[15pt]{\centering\arraybackslash  Nonparametric} & \multirow{-3}{*}[5pt]{\centering\arraybackslash SL2} &  3 & \multirow{-6}{*}[15pt]{\centering\arraybackslash  ---} & 17.9 & 5.26e-03 & --- & --- & 13.3 & 4.78e-03 & --- & ---\\
\cmidrule{1-13}
 &  &  &  1 &  & -0.2 & 5.40e-03 & 5.26e-03 & 95.0 & 0.1 & 3.21e-03 & 3.21e-03 & 95.6\\
\cmidrule{4-4}
\cmidrule{6-13}
 &  &  &  2 &  & -0.2 & 4.88e-03 & 4.67e-03 & 93.8 & 0.1 & 2.92e-03 & 2.87e-03 & 94.7\\
\cmidrule{4-4}
\cmidrule{6-13}
 &  & \multirow{-3}{*}[5pt]{\centering\arraybackslash SL1} &  3 &  & -0.1 & 4.92e-03 & 4.66e-03 & 93.3 & 0.1 & 2.91e-03 & 2.87e-03 & 94.3\\
\cmidrule{3-4}
\cmidrule{6-13}
 &  &  &  1 &  & 0.3 & 5.34e-03 & 5.16e-03 & 95.1 & 0.2 & 3.21e-03 & 3.16e-03 & 94.8\\
\cmidrule{4-4}
\cmidrule{6-13}
 &  &  &  2 &  & 0.4 & 5.40e-03 & 5.13e-03 & 93.8 & 0.3 & 3.16e-03 & 3.07e-03 & 93.8\\
\cmidrule{4-4}
\cmidrule{6-13}
\multirow{-6}{*}[15pt]{\centering\arraybackslash $\widehat\theta_\text{IF}$} & \multirow{-6}{*}[15pt]{\centering\arraybackslash  Nonparametric} & \multirow{-3}{*}[5pt]{\centering\arraybackslash SL2} &  3 & \multirow{-6}{*}[15pt]{\centering\arraybackslash  ---} & 0.5 & 5.36e-03 & 5.20e-03 & 94.5 & 0.5 & 3.20e-03 & 3.14e-03 & 94.6\\
\cmidrule{1-13}
 &  &  &  1 &  & -0.4 & 4.89e-03 & 4.53e-03 & 94.2 & -0.2 & 2.83e-03 & 2.71e-03 & 93.8\\
\cmidrule{4-4}
\cmidrule{6-13}
 &  &  &  2 &  & -0.1 & 4.36e-03 & 4.23e-03 & 94.7 & -0.1 & 2.59e-03 & 2.57e-03 & 95.4\\
\cmidrule{4-4}
\cmidrule{6-13}
 &  & \multirow{-3}{*}[5pt]{\centering\arraybackslash SL1} &  3 &  & -0.1 & 4.34e-03 & 4.22e-03 & 94.7 & -0.1 & 2.58e-03 & 2.56e-03 & 95.0\\
\cmidrule{3-4}
\cmidrule{6-13}
 &  &  &  1 &  & 0.0 & 5.05e-03 & 4.67e-03 & 93.6 & -0.1 & 2.92e-03 & 2.78e-03 & 94.8\\
\cmidrule{4-4}
\cmidrule{6-13}
 &  &  &  2 &  & 0.0 & 5.19e-03 & 4.78e-03 & 94.1 & 0.1 & 2.88e-03 & 2.76e-03 & 94.5\\
\cmidrule{4-4}
\cmidrule{6-13}
\multirow{-6}{*}[15pt]{\centering\arraybackslash $\widehat\theta_\text{EIF}$} & \multirow{-6}{*}[15pt]{\centering\arraybackslash  Nonparametric} & \multirow{-3}{*}[5pt]{\centering\arraybackslash SL2} &  3 & \multirow{-6}{*}[15pt]{\centering\arraybackslash  ---} & 0.2 & 5.16e-03 & 4.73e-03 & 94.5 & 0.1 & 2.96e-03 & 2.79e-03 & 93.6\\
\bottomrule\\
\multicolumn{13}{l}{\textsuperscript{a} SL Libs = $\texttt{SuperLearner}$ libraries: S1 = {Random Forest, LM/GLM, GAM, Polymars}; SL2 = {Random Forest, GAM, Polymars}}\\
\multicolumn{13}{l}{\textsuperscript{b} $\mu_0$ Strategy: (1) Fit single $\widehat\mu$ on $A = 0,1$ together (2) Fit $\widehat\mu_0$ on $A = 0$ only (stratification) (3) Specify all $A\times\bm L$ interactions in design matrix for $\widehat\mu$}\\
\multicolumn{13}{l}{\textsuperscript{c} Correctly specified parametric models for $\mu$ and/or $\eta$}\\
\end{tabular}
\caption{Comparison of estimators of $\theta(P)$ in simulation study}\label{tab:sim_results}
\end{sidewaystable}

Both influence function-based estimators were unbiased in all six nonparametric estimation strategies and attained nominal coverage for 95\% confidence intervals, despite a degree of error in estimating $\widehat\mu_0$, and possibly other nuisance functions. Under each strategy, $\widehat\theta_\text{IF}$ had a standard error between 4-14\% larger than the corresponding standard error of $\widehat\theta_\text{EIF}$, demonstrating the additional modeling complexities did indeed increase efficiency. With correctly specified parametric models for $\mu, \eta$, the standard error of $\widehat\theta_\text{IWOR}$ could at best match the smallest standard error among the $\widehat\theta_\text{EIF}$ estimation strategies, and did not beat it: it matched it for $n = 10,000$, and was 5\% larger for $n = 25,000$.

Altogether, these simulations demonstrate that both $\widehat\theta_\text{IF}$ and $\widehat\theta_\text{EIF}$ are implementable with flexible, nonparametric estimation strategies for component nuisance functions, even in the presence of complex nested nuisance functions. The estimators retain good finite sample properties for sample sizes typical of EHR-studies for bariatric surgery. Finally, these simulations demonstrate that $\widehat\theta_\text{EIF}$ attains noticeable efficiency gains if one is willing to estimate slightly more complex nuisance functions (cf. $\widehat\theta_\text{IF}$).

\subsection{Simulation Parameters}\label{supSec:sim_parameters}

\begin{table}[H]
\scriptsize
\centering
\begin{tabular}{|>{}Sc|Sc|Sc|Sc|}
\hline
\multicolumn{3}{|Sc|}{\textbf{Model Information}} & \multicolumn{1}{Sc|}{\multirow{2}{*}{\centering\arraybackslash\textbf{Value}}}  \\
\cline{1-3}
\textbf{Component Model} & \textbf{Coefficient} & \textbf{Covariate} &  \\
\hline
 &  & \texttt{(Intercept)} & 0.96\\
\cline{3-4}
 &  & \texttt{site[NC]} & -0.64\\
\cline{3-4}
 &  & \texttt{site[SC]} & -0.96\\
\cline{3-4}
 &  & \texttt{gender} & $2.7 \times 10^{-2}$\\
\cline{3-4}
 &  & \texttt{race} & 0.35\\
\cline{3-4}
 &  & \texttt{baseline\_bmi} & $1.6 \times 10^{-2}$\\
\cline{3-4}
 &  & \texttt{smoking\_status[former]} & -0.29 \\
\cline{3-4}
 &  & \texttt{smoking\_status[never]} & -0.23 \\
\cline{3-4}
 &  & \texttt{smoking\_status[no\_self\_report]} & -0.32 \\
\cline{3-4}
 &  & \texttt{baseline\_age} & $3 \times 10^{-3}$\\
\cline{3-4}
\multirow{-11}{*}{\centering\arraybackslash Treatment} & \multirow{-11}{*}{\centering\arraybackslash $\bm \beta_\pi$} & \texttt{eGFR} & $-5 \times 10^{-3}$\\
\cline{1-4}
 &  & \texttt{(Intercept)} & 0.38\\
\cline{3-4}
 &  & \texttt{site[NC]} & -0.38\\
\cline{3-4}
 &  & \texttt{site[SC]} & 0.79\\
\cline{3-4}
 &  & \texttt{gender} & -0.15\\
\cline{3-4}
 &  & \texttt{race} & 0.10\\
\cline{3-4}
 &  & \texttt{baseline\_bmi} & $-2 \times 10^{-2}$\\
\cline{3-4}
 &  & \texttt{smoking\_status[former]} & 0.44\\
\cline{3-4}
 &  & \texttt{smoking\_status[never]} & 0.32\\
\cline{3-4}
 &  & \texttt{smoking\_status[no\_self\_report]} & -2.58\\
\cline{3-4}
 &  & \texttt{baseline\_age} & $1.1 \times 10^{-2}$\\
\cline{3-4}
 &  & \texttt{eGFR} & $-1 \times 10^{-4}$\\
\cline{3-4}
\multirow{-12}{*}{\centering\arraybackslash Missingness} & \multirow{-12}{*}{\centering\arraybackslash $\bm \beta_\eta$} & \texttt{bs\_type} & 0.50\\
\cline{1-4}
 &  & \texttt{(Intercept)} & 1.06\\
\cline{3-4}
 &  & \texttt{site[NC]} & 0.23\\
\cline{3-4}
 &  & \texttt{site[SC]} & -0.24\\
\cline{3-4}
 &  & \texttt{gender} & -0.10\\
\cline{3-4}
 &  & \texttt{race} & $-6.9 \times 10^{-2}$\\
\cline{3-4}
 &  & \texttt{baseline\_bmi} & $-7.5 \times 10^{-3}$\\
\cline{3-4}
 &  & \texttt{I(baseline\_bmi$\textrm{\^{}}$2)} & $1 \times 10^{-4}$ \\
\cline{3-4}
 &  & \texttt{smoking\_status[former]} & $-5.7 \times 10^{-2}$\\
\cline{3-4}
 &  & \texttt{smoking\_status[never]} & $-7.6 \times 10^{-2}$\\
\cline{3-4}
 &  & \texttt{smoking\_status[no\_self\_report]} & $-9.4 \times 10^{-2}$\\
\cline{3-4}
 &  & \texttt{baseline\_age} & $9.2 \times 10^{-3}$ \\
\cline{3-4}
 &  & \texttt{eGFR} & $7 \times 10^{-4}$ \\
\cline{3-4}
\multirow{-14}{*}{\centering\arraybackslash Eligibility Defining Covariate} & \multirow{-13}{*}{\centering\arraybackslash $\bm \beta_\lambda$} & \texttt{bs\_type} & 0.10\\
\cline{2-4}
 & $\alpha_\lambda$ & \texttt{---} & 4.83\\
\cline{1-4}
 &  & \texttt{(Intercept)} & -0.24 \\
\cline{3-4}
 &  & \texttt{bs\_type} & $3.3 \times 10^{-2}$ \\
\cline{3-4}
 &  & \texttt{site[NC]} & 0.18 \\
\cline{3-4}
 &  & \texttt{site[SC]} & 0.14\\
\cline{3-4}
 &  & \texttt{gender} & -0.14 \\
\cline{3-4}
 &  & \texttt{race} & $-1.5 \times 10^{-2}$ \\
\cline{3-4}
 &  & \texttt{baseline\_bmi} & $-3.8 \times 10^{-3}$ \\
\cline{3-4}
 &  & \texttt{smoking\_status[former]} & $3.8 \times 10^{-2}$ \\
\cline{3-4}
 &  & \texttt{smoking\_status[never]} & $4.9 \times 10^{-2}$ \\
\cline{3-4}
 &  & \texttt{smoking\_status[no\_self\_report]} & -0.15 \\
\cline{3-4}
 &  & \texttt{baseline\_age} & $9.7 \times 10^{-4}$ \\
\cline{3-4}
 &  & \texttt{eGFR} & $1.4 \times 10^{-4}$ \\
\cline{3-4}
 &  & \texttt{baseline\_A1C} & $2.2 \times 10^{-4}$\\
\cline{3-4}
 &  & \texttt{bs\_type:baseline\_A1C} & $3.8 \times 10^{-3}$\\
\cline{3-4}
 &  & \texttt{gender:baseline\_A1C} & $4.8 \times 10^{-3}$ \\
\cline{3-4}
 &  & \texttt{gender:baseline\_bmi} & $2 \times 10^{-3}$ \\
\cline{3-4}
 &  & \texttt{smoking\_status[no\_self\_report]:bs\_type}  & 0.17\\
\cline{3-4}
 &  & \texttt{smoking\_status[never]:bs\_type} & $-2.4 \times 10^{-2}$ \\
\cline{3-4}
 &  & \texttt{smoking\_status[former]:bs\_type} & $-2.4 \times 10^{-2}$ \\
\cline{3-4}
 &  & \texttt{site[NC]:bs\_type} & -0.12\\
\cline{3-4}
 & \multirow{-21}{*}{\centering\arraybackslash $\bm \beta_\mu$} & \texttt{site[SC]:bs\_type} & -0.10\\
\cline{2-4}
\multirow{-22}{*}{\centering\arraybackslash Outcome} & $\sigma^2_{\text{y}}$ & \texttt{---} & $1 \times 10^{-2}$\\
\hline
\end{tabular}\caption{Coefficients values used to generate simulated datasets}
\label{table:sim_coef}
\end{table}

\section{Additional Data Application Information}\label{supSec:data_app}
\subsection{Methodological Details}\label{supSec:data_app_meth}
In this section, we provide some additional details on the data application which were omitted from the main text due to space constraints. \\

\noindent\underline{\textbf{Model Specifications}}
\begin{itemize}
    \item $\bm L^*$: surgery site, race, sex, age, estimated glomerular filtration rate (eGFR), self reported smoking status, hypertension, dyslipidemia, and calendar year of surgery
    \item $\bm L^e_m$: baseline BMI, baseline A1C, T2DM medication usage, DiaRem score (remission outcome only)
    \item $A$: RYGB ($A =1$) vs. SG ($A = 0$)
    \item $Y$: $\%$ weight change at 3 years post surgery (continuous), remission of T2DM at any point within 3 years post surgery (binary)
\end{itemize}

\noindent As mentioned in the main text, T2DM status is a discrete covariate which has no variance among the study eligible population, and thus it is not included in modeling any component nuisance functions, even though it may reasonably be considered in $\bm L^e_m$ if desired. While several medication types were considered for establishment of T2DM status, when modeling nuisance functions, we only considered an indicator of insulin usage in the corresponding lookback window, as insulin is typically utilized when other common T2DM medications fail to achieve desired level of glycemic control \citep{Thota2023}.

$\widehat\theta_\text{EIF}$ and $\widehat\theta_\text{IF}$ were estimated with the SL1 set of SuperLearner libraries. For nuisance functions with continuous modeling targets $(\mu, \xi, \gamma, \chi, \nu)$, these libraries included random forest (\texttt{SL.ranger}), linear models (\texttt{SL.lm}), and multivariate adaptive polynomial regression spline (\texttt{SL.polymars}). For binary modeling targets $(\eta, u, \varepsilon, \omega)$, these libraries included random forest (\texttt{SL.ranger}), generalized linear models (\texttt{SL.glm}), and generalized additive models (\texttt{SL.gam}). Rather than just use a single candidate random forest learner, all nuisance models used 27 candidate random forest learners over a grid of three important hyperparameters. 

\begin{itemize}
    \item \texttt{mtry} (number of predictors that will be randomly sampled at each split): $\{0.5, 1, 2\} \times \lfloor \sqrt{p}\rfloor$, where $p$ denotes the rank of the corresponding design matrix
    \item \texttt{num.trees} (number of trees used for prediction): $\{250, 500, 1000\}$
    \item \texttt{min.node.size} (minimal node size to split at): $\{5, 30, 50\}$
\end{itemize}

Finally, both outcome model $\mu$ and propensity score $u$ were trained only using eligible subjects. Given that the contributions of these nuisance functions to $\beta(P)$ are only non-zero for eligible subjects, this is strictly a decision about whether one believes they can model these contributions better by including ineligible subjects, and thus increasing sample size, or by getting a smaller but more narrow training set for the model which better reflects the patient who ultimately contribute.

$\widehat\theta_\text{CC}$ and $\widehat\theta_\text{IWOR}$ used a linear model for the outcome regression, also fit among ascertainably eligible subjects by definition. The following $A \times \bm L$ interactions were specified, motivated by existing bariatric surgery literature \citep{arterburn2020, McTigue2020}.

\begin{itemize}
    \item \% Weight Change: Sex, Age, Race, Baseline BMI
    \item Diabetes Remission: Sex, Age, Baseline A1C, DiaRem
\end{itemize}

\noindent\underline{\textbf{Additional Missing Data Considerations}}\\
The focus of this work was on missingness in eligibility defining covariates $\bm L^e$, and thus implicit to this work was the assumption that there was no missingness in other variables. In practice, in EHR-based observational studies, there is likely to be missingness elsewhere, which was the case in our data application. While the missingness was not nearly as pervasive as in eligibility defining covariates, we describe additional missing data challenges and our approach in dealing with such challenges in the data application. 

In our application on bariatric surgery, there was no missingness in treatment $A$, which may not be the case in all applications. Only a single covariate in $\bm L^*$, baseline eGFR, was missing for a small number of subjects. Estimated glomerular filtration rate is a function of serum creatinine, age, and sex \citep{ckd-epi2021}, so whenever eGFR was missing it was because a patient's serum creatinine value was not available. Similar to previous work by \citet{benz2024}, we imputed missing serum creatinine values (prior to operationalizing the eligibility criteria) using a gamma GLM and used the imputed value to compute eGFR.

Recall that covariates $\bm L^e_m$ are used in two nuisance functions, $\mu_a(\bm L^*, \bm L^e_m)$ and $u(\bm L^*, \bm L^e_m)$. In our application, we estimated these models using eligible complete cases $(E = 1, R = 1)$ rather than all complete cases $(R = 1)$ given that their respective influence function contributions are multiplied by $E$ and thus do not contribute if $E = 0$. Nevertheless, because T2DM status could be established without a baseline A1C measure for a small subset of patients, baseline A1C was not available for use as an important confounder. For study eligible patients without available A1C, we imputed A1C values using a gamma GLM as in previous work \citep{benz2024}. Imputation was not used to retroactively determine study eligibility (eg., to determine T2DM status or DiaRem score). 

Finally, weight change at 3 years was computed following the method of \citet{thaweethai2021robust}, and could be missing if patients were missing a baseline BMI and/or were missing a weight measure within $\pm$ 6 months of 3 years post-surgery. Given that outcomes $Y$ are in the conditioning set for several nuisance functions used by $\widehat\theta_\text{EIF}$, we decided to impute weight outcomes when missing so that we could illustrate the use of $\widehat\theta_\text{EIF}$. Table~\ref{tab:imputation_counts} reports the number of imputed values for each variable across the 40 operationalizations of the eligibility criteria. Baseline eGFR was imputed for 108 patients (0.7\% of the full analytic cohort). Across operationalizations of eligibility status, baseline A1c was imputed, among eligible complete cases ($E = 1$, $R = 1$), for 33 to 2,189 patients depending on the A1c lookback window; expressed as a percentage of the full analytic cohort for each operationalization, this corresponds to 0.2\% to 14.8\%. Three-year percent weight change was imputed for 3,519 - 5,446 patients (23.8\%-36.8\% of the full cohort), and 619 - 948 patients (20.4\%-22.6\% of eligible complete cases; e.g., where the outcome has greater impact in the influence function for $\beta(P)$). 

In all cases, the fraction of imputed values was far smaller than the fraction of patients with missing eligibility information (Section 2.2). In other applications, outcome missingness might be more pervasive, which might motivate analysts to consider using $\widehat\theta_\text{IF}$ instead of $\widehat\theta_\text{EIF}$, as the former does not use $Y$ in the conditioning set of any component nuisance function. We used a two tiered approach towards imputing outcomes: if baseline BMI was available, we imputed $\%$ weight change at 3 years via linear regression using $A, \bm L^*$, and baseline BMI as predictors; if baseline BMI was not available  we imputed $\%$ weight change at 3 years via linear regression using $A, \bm L^*$.

\begin{table}[!h]

\centering
\resizebox{\linewidth}{!}{
\begin{tabular}[t]{cccccccc}
\toprule
\multicolumn{4}{c}{Scenario ID} & \multicolumn{4}{c}{n (\%)} \\
\cmidrule(l{3pt}r{3pt}){1-4} \cmidrule(l{3pt}r{3pt}){5-8}
Scenario & BMI lookback & Diabetes lookback & Rx lookback & \% Weight Change & \% Weight Change, complete-case eligible & HbA1c imputed & eGFR imputed\\
\midrule
1 & 1 & 1 & 0 & 5,446 (36.8) & 619 (22.6) & 1,328 (9.0) & 108 (0.7)\\
2 & 1 & 1 & 12 & 5,446 (36.8) & 727 (22.4) & 1,771 (12.0) & 108 (0.7)\\
3 & 1 & 3 & 0 & 5,446 (36.8) & 675 (22.2) & 711 (4.8) & 108 (0.7)\\
4 & 1 & 3 & 12 & 5,446 (36.8) & 764 (22.2) & 966 (6.5) & 108 (0.7)\\
5 & 1 & 6 & 0 & 5,446 (36.8) & 737 (22.2) & 151 (1.0) & 108 (0.7)\\
\addlinespace
6 & 1 & 6 & 12 & 5,446 (36.8) & 807 (22.3) & 216 (1.5) & 108 (0.7)\\
7 & 1 & 12 & 0 & 5,446 (36.8) & 767 (22.3) & 46 (0.3) & 108 (0.7)\\
8 & 1 & 12 & 12 & 5,446 (36.8) & 825 (22.4) & 67 (0.5) & 108 (0.7)\\
9 & 1 & 24 & 0 & 5,446 (36.8) & 776 (22.4) & 33 (0.2) & 108 (0.7)\\
10 & 1 & 24 & 12 & 5,446 (36.8) & 829 (22.5) & 49 (0.3) & 108 (0.7)\\
\addlinespace
11 & 3 & 1 & 0 & 4,285 (28.9) & 672 (21.6) & 1,552 (10.5) & 108 (0.7)\\
12 & 3 & 1 & 12 & 4,285 (28.9) & 791 (21.5) & 2,046 (13.8) & 108 (0.7)\\
13 & 3 & 3 & 0 & 4,285 (28.9) & 736 (21.3) & 829 (5.6) & 108 (0.7)\\
14 & 3 & 3 & 12 & 4,285 (28.9) & 836 (21.4) & 1,111 (7.5) & 108 (0.7)\\
15 & 3 & 6 & 0 & 4,285 (28.9) & 806 (21.4) & 161 (1.1) & 108 (0.7)\\
\addlinespace
16 & 3 & 6 & 12 & 4,285 (28.9) & 887 (21.6) & 229 (1.5) & 108 (0.7)\\
17 & 3 & 12 & 0 & 4,285 (28.9) & 841 (21.5) & 50 (0.3) & 108 (0.7)\\
18 & 3 & 12 & 12 & 4,285 (28.9) & 908 (21.7) & 72 (0.5) & 108 (0.7)\\
19 & 3 & 24 & 0 & 4,285 (28.9) & 850 (21.6) & 35 (0.2) & 108 (0.7)\\
20 & 3 & 24 & 12 & 4,285 (28.9) & 911 (21.7) & 52 (0.4) & 108 (0.7)\\
\addlinespace
21 & 6 & 1 & 0 & 3,653 (24.7) & 691 (20.9) & 1,652 (11.2) & 108 (0.7)\\
22 & 6 & 1 & 12 & 3,653 (24.7) & 816 (20.9) & 2,170 (14.7) & 108 (0.7)\\
23 & 6 & 3 & 0 & 3,653 (24.7) & 757 (20.6) & 894 (6.0) & 108 (0.7)\\
24 & 6 & 3 & 12 & 3,653 (24.7) & 863 (20.8) & 1,195 (8.1) & 108 (0.7)\\
25 & 6 & 6 & 0 & 3,653 (24.7) & 832 (20.7) & 172 (1.2) & 108 (0.7)\\
\addlinespace
26 & 6 & 6 & 12 & 3,653 (24.7) & 917 (21.0) & 244 (1.6) & 108 (0.7)\\
27 & 6 & 12 & 0 & 3,653 (24.7) & 872 (20.9) & 52 (0.4) & 108 (0.7)\\
28 & 6 & 12 & 12 & 3,653 (24.7) & 941 (21.1) & 74 (0.5) & 108 (0.7)\\
29 & 6 & 24 & 0 & 3,653 (24.7) & 882 (21.0) & 35 (0.2) & 108 (0.7)\\
30 & 6 & 24 & 12 & 3,653 (24.7) & 945 (21.1) & 52 (0.4) & 108 (0.7)\\
\addlinespace
31 & 12 & 1 & 0 & 3,519 (23.8) & 694 (20.7) & 1,667 (11.3) & 108 (0.7)\\
32 & 12 & 1 & 12 & 3,519 (23.8) & 819 (20.7) & 2,189 (14.8) & 108 (0.7)\\
33 & 12 & 3 & 0 & 3,519 (23.8) & 760 (20.4) & 903 (6.1) & 108 (0.7)\\
34 & 12 & 3 & 12 & 3,519 (23.8) & 866 (20.6) & 1,207 (8.2) & 108 (0.7)\\
35 & 12 & 6 & 0 & 3,519 (23.8) & 835 (20.5) & 176 (1.2) & 108 (0.7)\\
\addlinespace
36 & 12 & 6 & 12 & 3,519 (23.8) & 920 (20.8) & 250 (1.7) & 108 (0.7)\\
37 & 12 & 12 & 0 & 3,519 (23.8) & 875 (20.7) & 53 (0.4) & 108 (0.7)\\
38 & 12 & 12 & 12 & 3,519 (23.8) & 944 (20.9) & 75 (0.5) & 108 (0.7)\\
39 & 12 & 24 & 0 & 3,519 (23.8) & 885 (20.8) & 36 (0.2) & 108 (0.7)\\
40 & 12 & 24 & 12 & 3,519 (23.8) & 948 (20.9) & 53 (0.4) & 108 (0.7)\\
\bottomrule
\end{tabular}}
\caption{\label{tab:imputation-scenarios}Imputation counts by lookback scenario, n (\%).}
\label{tab:imputation_counts}
\end{table}

That our (eligibility) complete case analysis closely matches the results reported by \citet{McTigue2020} suggests that the small amount of imputation did not substantively drive any observed results. Moreover, $\widehat\theta_\text{IF}$ does not condition on $Y$ in any nuisance function, and so does not depend on imputed outcomes through the nested functions $\varepsilon_1$, $\xi$, $\gamma$ and $\chi$. The clinical conclusions from $\widehat\theta_\text{IF}$ and $\widehat\theta_\text{EIF}$ are consistent (Figure 5), which further suggests that outcome imputation did not drive the results. We did not propagate imputation uncertainty into the reported confidence intervals, as missing outcomes and non-eligibility covariates are not the focus of this work and a bootstrap over the entire estimation pipeline is computationally prohibitive.

\clearpage
\subsection{Additional Figures and Tables}\label{supSec:data_app_figs}
\begin{figure}[H]
    \centering
    \includegraphics[width=\textwidth]{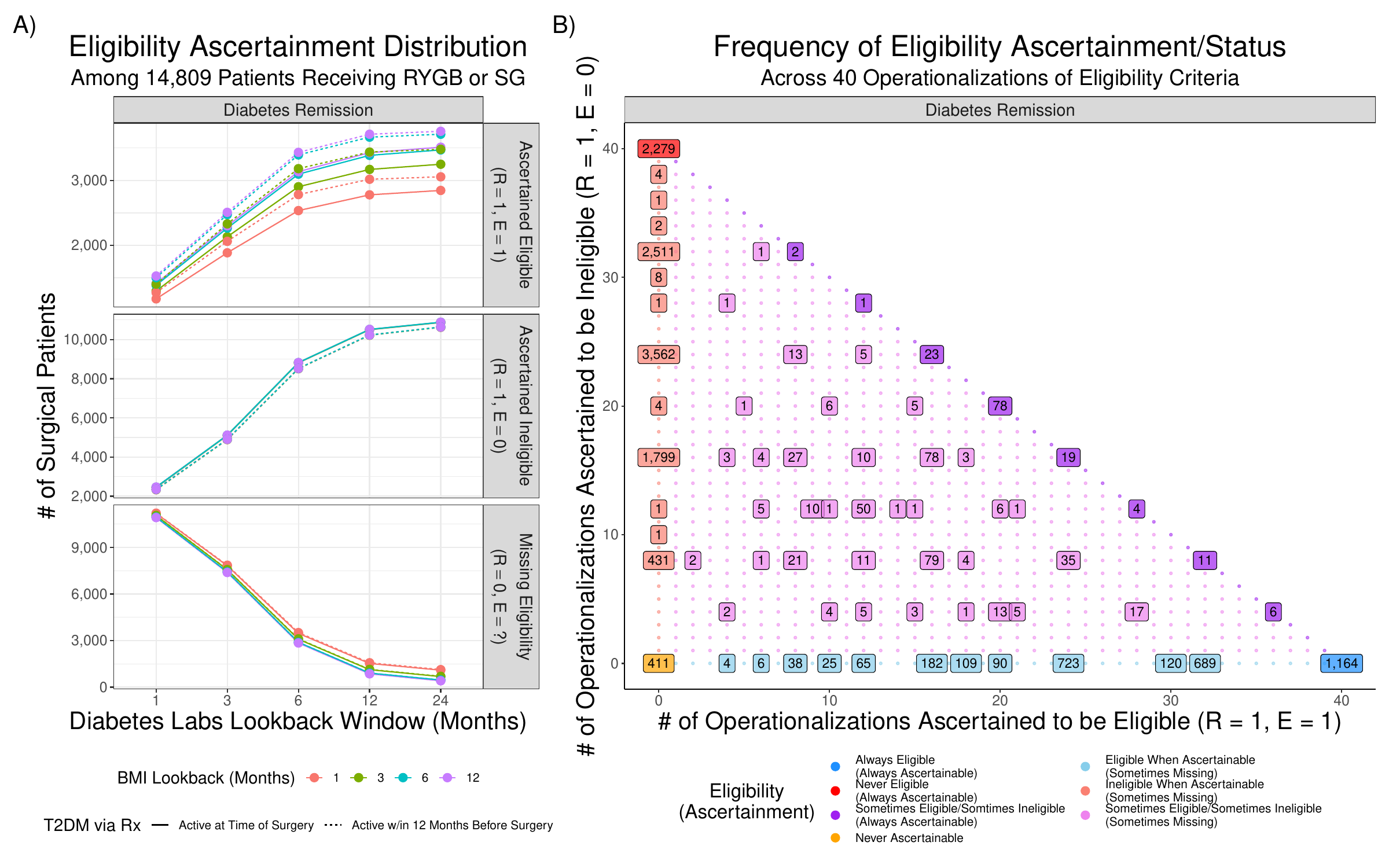}
    \caption{Distribution of $(n_{11}, n_{10})$ for diabetes remission outcome, where $n_{re}$ denotes the number of ways of operationalizing the study eligibility criteria that a subject has $R = r, E = e$. This figure is analogous to Figure 2 in the main text, which shows the same distribution for the relative weight change outcome.}
    \label{fig:S2}
\end{figure}

\begin{figure}[H]
    \centering
    \includegraphics[width=\textwidth]{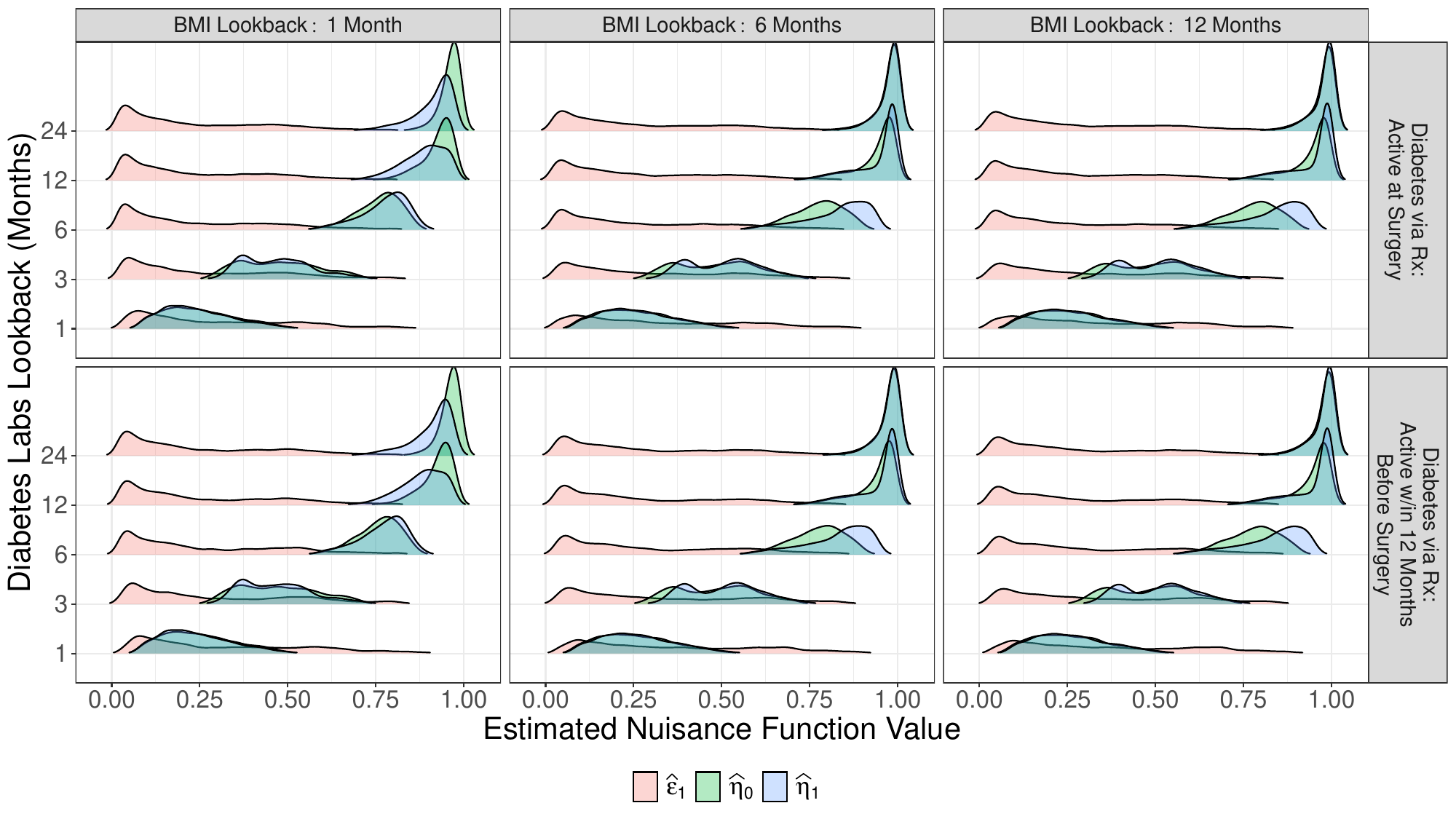}
    \caption{Distributions of select nuisance function estimates from $\widehat\theta_\text{EIF}$ related to ascertainment ($\widehat\eta$) and eligibility ($\widehat\varepsilon$) for diabetes remission outcome.}
    \label{fig:S3}
\end{figure}

\begin{figure}[H]
    \centering
    \includegraphics[width=\textwidth]{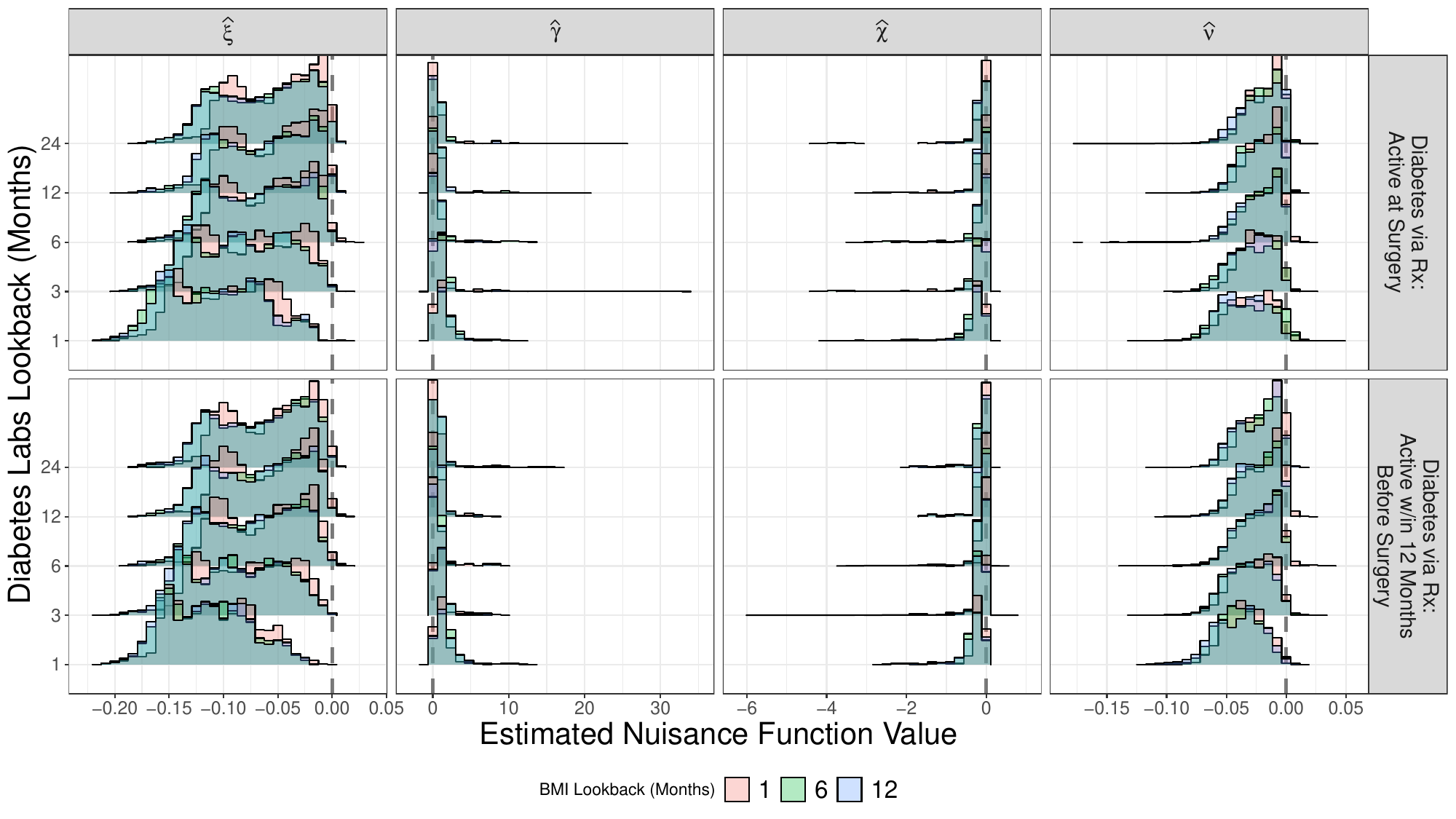}
    \caption{Distributions of nested nuisance function estimates from $\widehat\theta_\text{EIF}$ ($\widehat\xi$, $\widehat\gamma$, $\widehat\chi$) and $\widehat\theta_\text{IF}$ ($\widehat\nu)$ for relative weight change outcome.}
    \label{fig:S4}
\end{figure}

\begin{sidewaystable}[H]
\centering
\resizebox{\linewidth}{!}{
\begin{tabular}[t]{cccccccccccc}
\toprule
\multicolumn{4}{c}{ } & \multicolumn{2}{c}{$\widehat\theta_\text{CC}$} & \multicolumn{2}{c}{$\widehat\theta_\text{IWOR}$} & \multicolumn{2}{c}{$\widehat\theta_\text{IF}$} & \multicolumn{2}{c}{$\widehat\theta_\text{EIF}$} \\
\cmidrule(l{3pt}r{3pt}){5-6} \cmidrule(l{3pt}r{3pt}){7-8} \cmidrule(l{3pt}r{3pt}){9-10} \cmidrule(l{3pt}r{3pt}){11-12}
Scenario & \shortstack{BMI \\ Lookback} & \shortstack{Diabetes \\ Lookback} & \shortstack{Rx \\ Lookback} & ATT (95\% CI) & E-value (CI) & ATT (95\% CI) & E-value (CI) & ATT (95\% CI) & E-value (CI) & ATT (95\% CI) & E-value (CI)\\
\midrule
1 & 1 & 1 & 0 & -0.0623 (-0.0708, -0.0539) & 2.84 (2.58) & -0.0615 (-0.0690, -0.0539) & 2.81 (2.58) & -0.0488 (-0.0647, -0.0329) & 2.43 (2.00) & -0.0517 (-0.0714, -0.0320) & 2.52 (1.97)\\
2 & 1 & 1 & 12 & -0.0625 (-0.0713, -0.0537) & 2.82 (2.56) & -0.0606 (-0.0678, -0.0535) & 2.76 (2.55) & -0.0426 (-0.0655, -0.0197) & 2.24 (1.65) & -0.0465 (-0.0684, -0.0245) & 2.35 (1.77)\\
3 & 1 & 3 & 0 & -0.0616 (-0.0695, -0.0536) & 2.80 (2.56) & -0.0619 (-0.0681, -0.0556) & 2.81 (2.62) & -0.0479 (-0.0661, -0.0298) & 2.40 (1.91) & -0.0495 (-0.0673, -0.0316) & 2.44 (1.96)\\
4 & 1 & 3 & 12 & -0.0640 (-0.0723, -0.0558) & 2.87 (2.61) & -0.0636 (-0.0697, -0.0576) & 2.85 (2.67) & -0.0506 (-0.0682, -0.0330) & 2.47 (1.99) & -0.0497 (-0.0668, -0.0326) & 2.44 (1.98)\\
5 & 1 & 6 & 0 & -0.0658 (-0.0740, -0.0576) & 2.90 (2.65) & -0.0657 (-0.0711, -0.0602) & 2.90 (2.73) & -0.0499 (-0.0668, -0.0330) & 2.43 (1.98) & -0.0472 (-0.0671, -0.0273) & 2.36 (1.84)\\
\addlinespace
6 & 1 & 6 & 12 & -0.0694 (-0.0769, -0.0618) & 3.03 (2.79) & -0.0693 (-0.0745, -0.0641) & 3.02 (2.86) & -0.0548 (-0.0748, -0.0347) & 2.58 (2.03) & -0.0575 (-0.0719, -0.0430) & 2.66 (2.25)\\
7 & 1 & 12 & 0 & -0.0629 (-0.0709, -0.0549) & 2.82 (2.58) & -0.0628 (-0.0679, -0.0577) & 2.82 (2.66) & -0.0453 (-0.0649, -0.0256) & 2.31 (1.80) & -0.0462 (-0.0658, -0.0266) & 2.33 (1.82)\\
8 & 1 & 12 & 12 & -0.0646 (-0.0720, -0.0571) & 2.86 (2.64) & -0.0642 (-0.0694, -0.0591) & 2.85 (2.70) & -0.0509 (-0.0689, -0.0330) & 2.46 (1.98) & -0.0516 (-0.0694, -0.0337) & 2.48 (2.00)\\
9 & 1 & 24 & 0 & -0.0649 (-0.0728, -0.0570) & 2.89 (2.65) & -0.0646 (-0.0699, -0.0593) & 2.88 (2.71) & -0.0522 (-0.0674, -0.0369) & 2.51 (2.09) & -0.0543 (-0.0693, -0.0392) & 2.57 (2.15)\\
10 & 1 & 24 & 12 & -0.0681 (-0.0759, -0.0603) & 2.97 (2.73) & -0.0678 (-0.0729, -0.0628) & 2.97 (2.81) & -0.0508 (-0.0704, -0.0312) & 2.46 (1.94) & -0.0525 (-0.0697, -0.0354) & 2.51 (2.04)\\
\addlinespace
11 & 3 & 1 & 0 & -0.0625 (-0.0709, -0.0541) & 2.80 (2.55) & -0.0610 (-0.0683, -0.0537) & 2.75 (2.54) & -0.0429 (-0.0623, -0.0235) & 2.24 (1.74) & -0.0504 (-0.0671, -0.0336) & 2.44 (2.00)\\
12 & 3 & 1 & 12 & -0.0685 (-0.0767, -0.0604) & 2.99 (2.73) & -0.0674 (-0.0745, -0.0604) & 2.95 (2.73) & -0.0523 (-0.0732, -0.0313) & 2.50 (1.94) & -0.0552 (-0.0754, -0.0350) & 2.58 (2.03)\\
13 & 3 & 3 & 0 & -0.0666 (-0.0748, -0.0584) & 2.95 (2.69) & -0.0666 (-0.0725, -0.0607) & 2.95 (2.76) & -0.0532 (-0.0705, -0.0358) & 2.54 (2.06) & -0.0570 (-0.0738, -0.0402) & 2.65 (2.18)\\
14 & 3 & 3 & 12 & -0.0668 (-0.0744, -0.0591) & 2.94 (2.71) & -0.0665 (-0.0722, -0.0607) & 2.93 (2.75) & -0.0511 (-0.0700, -0.0322) & 2.47 (1.96) & -0.0527 (-0.0717, -0.0337) & 2.52 (2.00)\\
15 & 3 & 6 & 0 & -0.0673 (-0.0749, -0.0597) & 2.96 (2.73) & -0.0671 (-0.0722, -0.0620) & 2.96 (2.80) & -0.0549 (-0.0716, -0.0382) & 2.58 (2.12) & -0.0521 (-0.0694, -0.0347) & 2.50 (2.03)\\
\addlinespace
16 & 3 & 6 & 12 & -0.0685 (-0.0760, -0.0610) & 2.98 (2.75) & -0.0684 (-0.0735, -0.0633) & 2.98 (2.82) & -0.0537 (-0.0728, -0.0345) & 2.54 (2.02) & -0.0540 (-0.0716, -0.0364) & 2.55 (2.07)\\
17 & 3 & 12 & 0 & -0.0680 (-0.0759, -0.0601) & 2.97 (2.73) & -0.0678 (-0.0727, -0.0630) & 2.97 (2.82) & -0.0562 (-0.0748, -0.0377) & 2.62 (2.10) & -0.0569 (-0.0747, -0.0391) & 2.64 (2.14)\\
18 & 3 & 12 & 12 & -0.0699 (-0.0771, -0.0626) & 3.02 (2.79) & -0.0697 (-0.0745, -0.0648) & 3.01 (2.86) & -0.0532 (-0.0723, -0.0341) & 2.52 (2.01) & -0.0532 (-0.0725, -0.0338) & 2.52 (2.00)\\
19 & 3 & 24 & 0 & -0.0645 (-0.0719, -0.0571) & 2.87 (2.65) & -0.0645 (-0.0692, -0.0598) & 2.87 (2.73) & -0.0530 (-0.0692, -0.0368) & 2.53 (2.08) & -0.0551 (-0.0712, -0.0389) & 2.59 (2.14)\\
20 & 3 & 24 & 12 & -0.0711 (-0.0785, -0.0637) & 3.06 (2.83) & -0.0710 (-0.0757, -0.0663) & 3.06 (2.91) & -0.0545 (-0.0740, -0.0350) & 2.56 (2.03) & -0.0550 (-0.0751, -0.0350) & 2.57 (2.03)\\
\addlinespace
21 & 6 & 1 & 0 & -0.0659 (-0.0743, -0.0575) & 2.92 (2.67) & -0.0644 (-0.0716, -0.0572) & 2.88 (2.66) & -0.0399 (-0.0650, -0.0149) & 2.17 (1.53) & -0.0449 (-0.0690, -0.0209) & 2.31 (1.68)\\
22 & 6 & 1 & 12 & -0.0689 (-0.0766, -0.0611) & 3.00 (2.76) & -0.0674 (-0.0741, -0.0607) & 2.96 (2.75) & -0.0552 (-0.0704, -0.0400) & 2.59 (2.16) & -0.0588 (-0.0727, -0.0448) & 2.69 (2.29)\\
23 & 6 & 3 & 0 & -0.0667 (-0.0745, -0.0589) & 2.95 (2.71) & -0.0667 (-0.0727, -0.0607) & 2.95 (2.76) & -0.0499 (-0.0684, -0.0314) & 2.44 (1.95) & -0.0531 (-0.0708, -0.0353) & 2.53 (2.05)\\
24 & 6 & 3 & 12 & -0.0684 (-0.0761, -0.0607) & 2.98 (2.74) & -0.0683 (-0.0741, -0.0626) & 2.98 (2.80) & -0.0648 (-0.0810, -0.0486) & 2.87 (2.39) & -0.0657 (-0.0806, -0.0507) & 2.90 (2.45)\\
25 & 6 & 6 & 0 & -0.0643 (-0.0719, -0.0566) & 2.85 (2.62) & -0.0642 (-0.0693, -0.0591) & 2.85 (2.70) & -0.0524 (-0.0642, -0.0406) & 2.50 (2.18) & -0.0478 (-0.0637, -0.0319) & 2.37 (1.95)\\
\addlinespace
26 & 6 & 6 & 12 & -0.0677 (-0.0751, -0.0603) & 2.97 (2.74) & -0.0676 (-0.0725, -0.0626) & 2.96 (2.81) & -0.0591 (-0.0747, -0.0435) & 2.70 (2.26) & -0.0577 (-0.0731, -0.0423) & 2.66 (2.23)\\
27 & 6 & 12 & 0 & -0.0678 (-0.0754, -0.0602) & 2.96 (2.72) & -0.0678 (-0.0726, -0.0630) & 2.96 (2.81) & -0.0592 (-0.0806, -0.0378) & 2.70 (2.10) & -0.0576 (-0.0787, -0.0365) & 2.65 (2.07)\\
28 & 6 & 12 & 12 & -0.0688 (-0.0761, -0.0615) & 2.99 (2.76) & -0.0687 (-0.0734, -0.0640) & 2.99 (2.84) & -0.0593 (-0.0739, -0.0447) & 2.70 (2.28) & -0.0605 (-0.0752, -0.0458) & 2.73 (2.31)\\
29 & 6 & 24 & 0 & -0.0667 (-0.0740, -0.0594) & 2.95 (2.72) & -0.0667 (-0.0714, -0.0620) & 2.95 (2.81) & -0.0608 (-0.0756, -0.0460) & 2.77 (2.34) & -0.0594 (-0.0736, -0.0452) & 2.73 (2.31)\\
30 & 6 & 24 & 12 & -0.0714 (-0.0788, -0.0640) & 3.07 (2.84) & -0.0713 (-0.0760, -0.0667) & 3.07 (2.92) & -0.0537 (-0.0709, -0.0366) & 2.54 (2.07) & -0.0557 (-0.0723, -0.0391) & 2.59 (2.14)\\
\addlinespace
31 & 12 & 1 & 0 & -0.0666 (-0.0754, -0.0579) & 2.95 (2.68) & -0.0653 (-0.0724, -0.0583) & 2.91 (2.69) & -0.0495 (-0.0712, -0.0277) & 2.43 (1.85) & -0.0507 (-0.0715, -0.0300) & 2.47 (1.91)\\
32 & 12 & 1 & 12 & -0.0686 (-0.0768, -0.0605) & 2.98 (2.73) & -0.0673 (-0.0739, -0.0607) & 2.94 (2.74) & -0.0551 (-0.0695, -0.0406) & 2.57 (2.18) & -0.0574 (-0.0707, -0.0441) & 2.64 (2.27)\\
33 & 12 & 3 & 0 & -0.0654 (-0.0732, -0.0575) & 2.91 (2.67) & -0.0649 (-0.0709, -0.0589) & 2.89 (2.71) & -0.0487 (-0.0664, -0.0309) & 2.41 (1.93) & -0.0519 (-0.0690, -0.0347) & 2.50 (2.03)\\
34 & 12 & 3 & 12 & -0.0681 (-0.0756, -0.0607) & 2.98 (2.75) & -0.0680 (-0.0737, -0.0622) & 2.98 (2.80) & -0.0504 (-0.0711, -0.0296) & 2.45 (1.90) & -0.0521 (-0.0719, -0.0323) & 2.50 (1.96)\\
35 & 12 & 6 & 0 & -0.0673 (-0.0747, -0.0598) & 2.95 (2.72) & -0.0672 (-0.0723, -0.0622) & 2.95 (2.79) & -0.0520 (-0.0672, -0.0367) & 2.49 (2.08) & -0.0516 (-0.0668, -0.0363) & 2.48 (2.07)\\
\addlinespace
36 & 12 & 6 & 12 & -0.0657 (-0.0734, -0.0581) & 2.89 (2.66) & -0.0657 (-0.0706, -0.0607) & 2.89 (2.74) & -0.0464 (-0.0673, -0.0256) & 2.33 (1.79) & -0.0464 (-0.0672, -0.0256) & 2.33 (1.79)\\
37 & 12 & 12 & 0 & -0.0689 (-0.0763, -0.0614) & 2.96 (2.74) & -0.0688 (-0.0737, -0.0639) & 2.96 (2.81) & -0.0552 (-0.0716, -0.0388) & 2.56 (2.12) & -0.0576 (-0.0728, -0.0424) & 2.63 (2.21)\\
38 & 12 & 12 & 12 & -0.0705 (-0.0777, -0.0632) & 3.03 (2.80) & -0.0704 (-0.0751, -0.0656) & 3.02 (2.88) & -0.0532 (-0.0740, -0.0324) & 2.51 (1.96) & -0.0544 (-0.0747, -0.0341) & 2.55 (2.00)\\
39 & 12 & 24 & 0 & -0.0683 (-0.0756, -0.0609) & 2.98 (2.76) & -0.0683 (-0.0729, -0.0636) & 2.98 (2.84) & -0.0619 (-0.0765, -0.0472) & 2.78 (2.36) & -0.0628 (-0.0776, -0.0480) & 2.81 (2.38)\\
40 & 12 & 24 & 12 & -0.0697 (-0.0770, -0.0624) & 3.01 (2.79) & -0.0697 (-0.0744, -0.0651) & 3.02 (2.87) & -0.0628 (-0.0749, -0.0507) & 2.80 (2.45) & -0.0599 (-0.0747, -0.0452) & 2.71 (2.29)\\
\bottomrule
\end{tabular}}

\caption{Comparison of estimators of $\theta(P)$ on 3-year weight loss outcomes.}
\label{tbl:wt_results}
\end{sidewaystable}

\begin{table}[H]
\centering
\resizebox{\linewidth}{!}{
\begin{tabular}[t]{cccccccc}
\toprule
\multicolumn{4}{c}{ } & \multicolumn{1}{c}{$\widehat\theta_\text{CC}$} & \multicolumn{1}{c}{$\widehat\theta_\text{IWOR}$} & \multicolumn{1}{c}{$\widehat\theta_\text{IF}$} & \multicolumn{1}{c}{$\widehat\theta_\text{EIF}$} \\
\cmidrule(l{3pt}r{3pt}){5-5} \cmidrule(l{3pt}r{3pt}){6-6} \cmidrule(l{3pt}r{3pt}){7-7} \cmidrule(l{3pt}r{3pt}){8-8}
Scenario & \shortstack{BMI \\ Lookback} & \shortstack{Diabetes \\ Lookback} & \shortstack{Rx \\ Lookback} & ATT (95\% CI) & ATT (95\% CI) & ATT (95\% CI) & ATT (95\% CI)\\
\midrule
1 & 1 & 1 & 0 & 0.0365 (-0.0256, 0.0986) & 0.0518 (-0.0093, 0.1129) & 0.0453 (-0.0151, 0.1058) & 0.0066 (-0.0605, 0.0736)\\
2 & 1 & 1 & 12 & 0.0561 (-0.0016, 0.1138) & 0.0693 (0.0113, 0.1273) & 0.0657 (0.0078, 0.1235) & 0.0401 (-0.0208, 0.1009)\\
3 & 1 & 3 & 0 & 0.0424 (-0.0031, 0.0880) & 0.0585 (0.0150, 0.1021) & 0.0239 (-0.0965, 0.1443) & 0.0227 (-0.0791, 0.1245)\\
4 & 1 & 3 & 12 & 0.0588 (0.0143, 0.1032) & 0.0800 (0.0379, 0.1221) & 0.0492 (-0.0420, 0.1404) & 0.0354 (-0.0236, 0.0943)\\
5 & 1 & 6 & 0 & 0.0491 (0.0091, 0.0892) & 0.0671 (0.0306, 0.1036) & 0.0358 (-0.0308, 0.1024) & 0.0209 (-0.0356, 0.0774)\\
\addlinespace
6 & 1 & 6 & 12 & 0.0660 (0.0261, 0.1060) & 0.0900 (0.0550, 0.1251) & 0.0545 (-0.0075, 0.1164) & 0.0423 (-0.0116, 0.0961)\\
7 & 1 & 12 & 0 & 0.0442 (0.0055, 0.0829) & 0.0583 (0.0238, 0.0928) & 0.0393 (-0.0275, 0.1061) & 0.0374 (-0.0465, 0.1214)\\
8 & 1 & 12 & 12 & 0.0616 (0.0264, 0.0968) & 0.0787 (0.0453, 0.1121) & 0.0600 (-0.0084, 0.1285) & 0.0561 (-0.0027, 0.1149)\\
9 & 1 & 24 & 0 & 0.0497 (0.0134, 0.0859) & 0.0631 (0.0287, 0.0974) & 0.0575 (-0.0203, 0.1353) & 0.0430 (-0.0160, 0.1019)\\
10 & 1 & 24 & 12 & 0.0633 (0.0265, 0.1001) & 0.0805 (0.0475, 0.1136) & 0.0680 (-0.0187, 0.1547) & 0.0679 (-0.0082, 0.1440)\\
\addlinespace
11 & 3 & 1 & 0 & 0.0521 (-0.0068, 0.1110) & 0.0671 (0.0082, 0.1260) & 0.0687 (-0.0073, 0.1447) & 0.0058 (-0.0628, 0.0743)\\
12 & 3 & 1 & 12 & 0.0665 (0.0098, 0.1233) & 0.0797 (0.0248, 0.1346) & 0.0713 (0.0099, 0.1328) & 0.0422 (-0.0158, 0.1001)\\
13 & 3 & 3 & 0 & 0.0628 (0.0183, 0.1072) & 0.0776 (0.0355, 0.1197) & 0.0549 (-0.0321, 0.1420) & 0.0239 (-0.0564, 0.1042)\\
14 & 3 & 3 & 12 & 0.0726 (0.0315, 0.1136) & 0.0922 (0.0520, 0.1324) & 0.0576 (-0.0386, 0.1538) & 0.0439 (-0.0464, 0.1342)\\
15 & 3 & 6 & 0 & 0.0677 (0.0272, 0.1082) & 0.0810 (0.0457, 0.1162) & 0.0605 (-0.0254, 0.1465) & 0.0359 (-0.0421, 0.1139)\\
\addlinespace
16 & 3 & 6 & 12 & 0.0803 (0.0425, 0.1180) & 0.1003 (0.0668, 0.1338) & 0.0714 (0.0108, 0.1320) & 0.0599 (0.0050, 0.1149)\\
17 & 3 & 12 & 0 & 0.0627 (0.0255, 0.0999) & 0.0741 (0.0406, 0.1076) & 0.0600 (-0.0107, 0.1307) & 0.0486 (-0.0189, 0.1160)\\
18 & 3 & 12 & 12 & 0.0756 (0.0391, 0.1121) & 0.0905 (0.0584, 0.1227) & 0.0727 (-0.0017, 0.1471) & 0.0705 (0.0124, 0.1287)\\
19 & 3 & 24 & 0 & 0.0669 (0.0309, 0.1030) & 0.0789 (0.0458, 0.1119) & 0.0662 (-0.0097, 0.1422) & 0.0548 (-0.0091, 0.1187)\\
20 & 3 & 24 & 12 & 0.0757 (0.0393, 0.1121) & 0.0918 (0.0601, 0.1235) & 0.0705 (-0.0157, 0.1566) & 0.0736 (0.0163, 0.1309)\\
\addlinespace
21 & 6 & 1 & 0 & 0.0550 (-0.0023, 0.1122) & 0.0717 (0.0141, 0.1293) & 0.0595 (-0.0070, 0.1260) & 0.0387 (-0.0289, 0.1062)\\
22 & 6 & 1 & 12 & 0.0676 (0.0139, 0.1212) & 0.0798 (0.0252, 0.1345) & 0.0730 (-0.0019, 0.1479) & 0.0389 (-0.0317, 0.1095)\\
23 & 6 & 3 & 0 & 0.0669 (0.0222, 0.1116) & 0.0827 (0.0412, 0.1242) & 0.0502 (-0.0580, 0.1584) & 0.0377 (-0.0691, 0.1445)\\
24 & 6 & 3 & 12 & 0.0762 (0.0315, 0.1208) & 0.0946 (0.0543, 0.1348) & 0.0537 (-0.0236, 0.1309) & 0.0342 (-0.0711, 0.1395)\\
25 & 6 & 6 & 0 & 0.0702 (0.0318, 0.1086) & 0.0858 (0.0507, 0.1208) & 0.0464 (-0.0254, 0.1182) & 0.0353 (-0.0239, 0.0945)\\
\addlinespace
26 & 6 & 6 & 12 & 0.0819 (0.0448, 0.1190) & 0.1034 (0.0699, 0.1370) & 0.0675 (0.0097, 0.1254) & 0.0540 (-0.0043, 0.1123)\\
27 & 6 & 12 & 0 & 0.0648 (0.0277, 0.1020) & 0.0786 (0.0457, 0.1116) & 0.0630 (0.0014, 0.1246) & 0.0411 (-0.0155, 0.0976)\\
28 & 6 & 12 & 12 & 0.0774 (0.0422, 0.1126) & 0.0945 (0.0625, 0.1264) & 0.0830 (0.0162, 0.1499) & 0.0722 (0.0059, 0.1384)\\
29 & 6 & 24 & 0 & 0.0690 (0.0334, 0.1046) & 0.0833 (0.0509, 0.1156) & 0.0621 (0.0011, 0.1232) & 0.0628 (-0.0054, 0.1309)\\
30 & 6 & 24 & 12 & 0.0772 (0.0420, 0.1124) & 0.0959 (0.0645, 0.1273) & 0.0720 (0.0146, 0.1294) & 0.0708 (0.0180, 0.1237)\\
\addlinespace
31 & 12 & 1 & 0 & 0.0549 (-0.0015, 0.1113) & 0.0708 (0.0138, 0.1278) & 0.0663 (-0.0050, 0.1376) & 0.0136 (-0.0484, 0.0757)\\
32 & 12 & 1 & 12 & 0.0684 (0.0146, 0.1223) & 0.0798 (0.0254, 0.1341) & 0.0673 (0.0070, 0.1277) & 0.0273 (-0.0341, 0.0887)\\
33 & 12 & 3 & 0 & 0.0652 (0.0217, 0.1087) & 0.0811 (0.0399, 0.1222) & 0.0671 (-0.0426, 0.1768) & 0.0274 (-0.0629, 0.1177)\\
34 & 12 & 3 & 12 & 0.0749 (0.0333, 0.1165) & 0.0934 (0.0535, 0.1334) & 0.0535 (-0.0387, 0.1458) & 0.0483 (-0.0332, 0.1299)\\
35 & 12 & 6 & 0 & 0.0683 (0.0291, 0.1074) & 0.0845 (0.0497, 0.1193) & 0.0511 (-0.0055, 0.1076) & 0.0326 (-0.0281, 0.0932)\\
\addlinespace
36 & 12 & 6 & 12 & 0.0804 (0.0434, 0.1175) & 0.1026 (0.0692, 0.1360) & 0.0683 (0.0143, 0.1222) & 0.0550 (-0.0015, 0.1116)\\
37 & 12 & 12 & 0 & 0.0628 (0.0270, 0.0987) & 0.0777 (0.0449, 0.1104) & 0.0571 (-0.0069, 0.1211) & 0.0410 (-0.0225, 0.1045)\\
38 & 12 & 12 & 12 & 0.0757 (0.0402, 0.1112) & 0.0939 (0.0621, 0.1257) & 0.0686 (0.0061, 0.1311) & 0.0665 (0.0131, 0.1200)\\
39 & 12 & 24 & 0 & 0.0669 (0.0302, 0.1036) & 0.0823 (0.0501, 0.1144) & 0.0601 (0.0057, 0.1144) & 0.0581 (-0.0018, 0.1179)\\
40 & 12 & 24 & 12 & 0.0754 (0.0420, 0.1087) & 0.0953 (0.0640, 0.1265) & 0.0829 (0.0106, 0.1552) & 0.0710 (0.0027, 0.1392)\\
\bottomrule
\end{tabular}}
\caption{Comparison of estimators of $\theta(P)$ on 3-year diabetes remission.}
\label{tbl:t2dm_results}
\end{table}

Tables \ref{tbl:wt_results} and \ref{tbl:t2dm_results} present full point estimates 95\% confidence intervals for 3-year weight change and T2DM remission differences, respectively. For the weight change outcome, we also present E-values \citep{haneuse2019evalue, vanderweele2017sensitivity}, which has been used in DURABLE studies \citep{Harrington2024, fisher2018association, coleman2022bariatric, obrien2018microvascular} as way to quantify the strength of unmeasured confounding that would be required to overturn observed results. Because the E-value was derived in the context of risk ratio, and we estimate an ATT-type contrast for a continuous outcome, we apply the approximation outlined in \citep{vanderweele2017sensitivity},$RR \approx \exp(0.91 \times \text{SMD})$ where SMD denotes the standardized mean difference. We take SMD $= \widehat\theta / \widehat\sigma_Y$, where $\widehat\sigma_Y$ outcome standard deviation among eligible complete cases. Given that all E-values exceed 2, an unmeasured confounder would need to be associated with both the exposure (bariatric surgery) and 3-year weight change by a risk ratio of at least 2-fold each, above and beyond the measured covariates $\bm L^*$ and $\bm L^e_m$, to fully explain away the observed treatment effect. This magnitude of confounding exceeds that observed for any of our measured covariates and is not thought to characterize unmeasured factors in other DURABLE studies \citep{Harrington2024, fisher2018association, coleman2022bariatric, obrien2018microvascular}, lending support to the robustness of our estimates to plausible unmeasured confounding.

Table \ref{tab:cohort_characteristics} reports baseline demographic, comorbidity, and clinical characteristics of the study cohort by bariatric procedure type.

\begin{table}[H]
\centering
\scriptsize
\renewcommand{\arraystretch}{0.85}
\begin{tabular}[t]{lcc}
\toprule
 & RYGB (N=9,614) & Sleeve (N=5,195) \\
\midrule
\textbf{Baseline Age} & & \\
\hspace{1em}Mean (SD) & 46.3 (11.0) & 45.8 (11.1) \\
\addlinespace[2pt]
\textbf{Sex} & & \\
\hspace{1em}F & 7,819 (81.3\%) & 4,220 (81.2\%) \\
\hspace{1em}M & 1,795 (18.7\%) & 975 (18.8\%) \\
\addlinespace[2pt]
\textbf{Race} & & \\
\hspace{1em}Non-white & 3,233 (33.6\%) & 2,140 (41.2\%) \\
\hspace{1em}White & 6,381 (66.4\%) & 3,055 (58.8\%) \\
\addlinespace[2pt]
\textbf{Site of Surgery} & & \\
\hspace{1em}KPNC & 3,023 (31.4\%) & 981 (18.9\%) \\
\hspace{1em}KPSC & 5,778 (60.1\%) & 4,200 (80.8\%) \\
\hspace{1em}KPWA & 813 (8.5\%) & 14 (0.3\%) \\
\addlinespace[2pt]
\textbf{Hypertension} & & \\
\hspace{1em}Yes & 5,786 (60.2\%) & 2,767 (53.3\%) \\
\hspace{1em}No & 3,828 (39.8\%) & 2,428 (46.7\%) \\
\addlinespace[2pt]
\textbf{Dyslipidemia} & & \\
\hspace{1em}Yes & 4,573 (47.6\%) & 2,310 (44.5\%) \\
\hspace{1em}No & 5,041 (52.4\%) & 2,885 (55.5\%) \\
\addlinespace[2pt]
\textbf{Self-Reported Smoking Status} & & \\
\hspace{1em}Current & 162 (1.7\%) & 72 (1.4\%) \\
\hspace{1em}Former & 3,360 (34.9\%) & 1,775 (34.2\%) \\
\hspace{1em}Never & 5,927 (61.6\%) & 3,323 (64.0\%) \\
\hspace{1em}Not Self-Reported & 165 (1.7\%) & 25 (0.5\%) \\
\addlinespace[2pt]
\textbf{eGFR} & & \\
\hspace{1em}Mean (SD) & 95.6 (20.5) & 96.9 (20.4) \\
\addlinespace[2pt]
\textbf{Calendar Year of Surgery} & & \\
\hspace{1em}2008 & 2,371 (24.7\%) & 185 (3.6\%) \\
\hspace{1em}2009 & 2,348 (24.4\%) & 557 (10.7\%) \\
\hspace{1em}2010 & 2,445 (25.4\%) & 1,427 (27.5\%) \\
\hspace{1em}2011 & 2,450 (25.5\%) & 3,026 (58.2\%) \\
\addlinespace[2pt]
\textbf{Baseline BMI (1-month Lookback)} & & \\
\hspace{1em}Mean (SD) & 43.7 (7.02) & 43.2 (6.64) \\
\hspace{1em}Missing & 2,010 (20.9\%) & 386 (7.4\%) \\
\addlinespace[2pt]
\textbf{Baseline BMI (3-month Lookback)} & & \\
\hspace{1em}Mean (SD) & 44.2 (7.20) & 43.3 (6.71) \\
\hspace{1em}Missing & 793 (8.2\%) & 174 (3.3\%) \\
\addlinespace[2pt]
\textbf{Baseline BMI (6-month Lookback)} & & \\
\hspace{1em}Mean (SD) & 44.5 (7.31) & 43.4 (6.77) \\
\hspace{1em}Missing & 189 (2.0\%) & 32 (0.6\%) \\
\addlinespace[2pt]
\textbf{Baseline BMI (12-month Lookback)} & & \\
\hspace{1em}Mean (SD) & 44.6 (7.37) & 43.4 (6.79) \\
\hspace{1em}Missing & 64 (0.7\%) & 5 (0.1\%) \\
\addlinespace[2pt]
\textbf{Baseline HbA1c (1-month Lookback)} & & \\
\hspace{1em}Mean (SD) & 6.56 (1.39) & 6.20 (1.25) \\
\hspace{1em}Missing & 6,360 (66.2\%) & 3,785 (72.9\%) \\
\addlinespace[2pt]
\textbf{Baseline HbA1c (3-month Lookback)} & & \\
\hspace{1em}Mean (SD) & 6.13 (1.31) & 5.79 (1.13) \\
\hspace{1em}Missing & 4,291 (44.6\%) & 2,723 (52.4\%) \\
\addlinespace[2pt]
\textbf{Baseline HbA1c (6-month Lookback)} & & \\
\hspace{1em}Mean (SD) & 5.90 (1.25) & 5.56 (1.02) \\
\hspace{1em}Missing & 1,676 (17.4\%) & 1,182 (22.8\%) \\
\addlinespace[2pt]
\textbf{Baseline HbA1c (12-month Lookback)} & & \\
\hspace{1em}Mean (SD) & 5.91 (1.25) & 5.53 (1.02) \\
\hspace{1em}Missing & 663 (6.9\%) & 228 (4.4\%) \\
\addlinespace[2pt]
\textbf{Baseline HbA1c (24-month Lookback)} & & \\
\hspace{1em}Mean (SD) & 5.92 (1.26) & 5.56 (1.03) \\
\hspace{1em}Missing & 334 (3.5\%) & 83 (1.6\%) \\
\addlinespace[2pt]
\textbf{Active Insulin Rx at Surgery} & & \\
\hspace{1em}Yes & 836 (8.7\%) & 187 (3.6\%) \\
\hspace{1em}No & 8,778 (91.3\%) & 5,008 (96.4\%) \\
\addlinespace[2pt]
\textbf{Insulin Rx in 12-months Before Surgery} & & \\
\hspace{1em}Yes & 1,328 (13.8\%) & 314 (6.0\%) \\
\hspace{1em}No & 8,286 (86.2\%) & 4,881 (94.0\%) \\
\bottomrule
\end{tabular}
\caption{Baseline demographic, comorbidity, and clinical characteristics of the study cohort, by bariatric procedure type. Values are n (\%) unless otherwise noted.}
\label{tab:cohort_characteristics}
\end{table}

\section{Sensitivity Analysis}
\subsection{Sensitivity Analysis for Assumption 4 (MAR)}
\label{supSec:sensitivity}

In order to explore the impact of certain violations of Assumption 4, we conduct a comprehensive sensitivity analysis. In particular, we follow the general ideas of \citet{scharfstein1999adjusting}.

First, observe that Assumption 4, which states that $R \indep (\bm L^e_m, Y)~|~\bm L^*, A$, implies that
$$
P(R = 1~|~\bm L^*, A, Y, \bm L^e_m) = P(R = 1~|~\bm L^*, A)
$$

\noindent To formally explore violations of Assumption 4 to certain forms of MNAR,  we suppose that 

\begin{equation}\label{eq:sens_model}
\text{logit}\bigr[P(R = 1~|~\bm L^*, A, Y, \bm L^e_m)\bigr] =  b(\bm L^*, A) + h(\bm L^e_m, Y; \bm \alpha)    
\end{equation}

\noindent for some functions $b(\cdot)$ $h(\cdot)$ and sensitivity parameter(s) $\bm \alpha$, where $h(\, \cdot\, ,\cdot \, ; \boldsymbol{0}) \equiv 0$. Note that Assumption 4 holds when $\bm \alpha = \boldsymbol{0}$. (In this section only, $\bm\alpha$ denotes the vector of sensitivity parameters, and should not be confused with the functional $\alpha(P)$ that appears in the denominator of $\theta(P) = \beta(P)/\alpha(P)$.)

Given that both clinical outcomes examined reflect changes from baseline (\% weight change from study baseline or remission of baseline diabetes status), we work just with $\bm L^e_m$. In particular, we take $h(\bm L_m^e;\bm \alpha)$ to be

$$
h(\bm L_m^e;\bm \alpha) = \alpha_\text{A1C}(\text{Baseline A1C} - 6.5\%) + \alpha_\text{BMI}(\text{Baseline BMI} - 35\text{kg/m}^2)
$$

\noindent where $\alpha_\text{A1C}$ and $\alpha_\text{BMI}$ can interpreted as the log-odds of having completely observed eligibility information, per unit change in A1C/BMI, relative to their respective eligibility-defining thresholds.

Because $h$ depends only on $\bm L^e_m$ and not on $Y$, this sensitivity model perturbs the ascertainment probability only through the eligibility-defining covariates $\bm L^e_m$, holding $\bm L^*$ and $A$ fixed. As such, it addresses departures from Assumption 4 in which ascertainment depends on $\bm L^e_m$ given $\bm L^*$ and $A$; it does not address missingness that depends on the outcome $Y$ itself, given $\bm L^*$, $\bm L^e_m$, and $A$.

Under the assumed sensitivity model in Equation \eqref{eq:sens_model}, and the fact that $h(\bm L_m^e;\bm \alpha)$ does not depend on $Y$, we have

$$
\text{logit}\bigr[P(R = 1~|~\bm L^*, A, \bm L^e_m)\bigr] =  b(\bm L^*, A) + h(\bm L^e_m; \bm \alpha) 
$$

\noindent Now, notice that by the law of iterated expectation,

$$
\mathbb{E}\Bigr[P(R = 1~|~\bm L^*, A, \bm L^e_m)~|~\bm L^*, A\Bigr] = P(R = 1 ~|~ \bm L^*, A) = \eta(\bm L^*, A)
$$

\noindent and 

$$
\begin{aligned}
\mathbb{E}\Bigr[\frac{R}{P(R = 1~|~\bm L^*, A, \bm L^e_m)}~\Bigr|~\bm L^*, A\Bigr] &= \mathbb{E}\Bigr[\frac{P(R = 1~|~\bm L^*, A, \bm L^e_m)}{P(R = 1~|~\bm L^*, A, \bm L^e_m)}~\Bigr|~\bm L^*, A\Bigr]  \implies \\
\mathbb{E}\Bigr[\frac{R}{P(R = 1~|~\bm L^*, A, \bm L^e_m)}~\Bigr|~\bm L^*, A\Bigr] &= 1  \implies \\
\eta(\bm L^*, A) \mathbb{E}\Bigr[\frac{1}{P(R = 1~|~\bm L^*, A, \bm L^e_m)}~\Bigr|~\bm L^*, A, R = 1\Bigr] &= 1 ~~~(\text{Lemma~\ref{lemma:bvt}}) \\
\end{aligned}
$$

\noindent Solving for $b(\bm L^*, A)$ for a given sensitivity parameter $\bm \alpha$, we have that 
$$
\begin{aligned}
P(R = 1~|~\bm L^*, A, \bm L^e_m) &= \frac{1}{1 + \exp[-b(\bm L^*, A) - h(\bm L^e_m; \bm \alpha)]} \implies \\
\frac{1}{P(R = 1~|~\bm L^*, A, \bm L^e_m)} &= 1 + \exp[-b(\bm L^*, A) - h(\bm L^e_m; \bm \alpha)] \implies \\
1 &= \eta(\bm L^*, A) \times \mathbb{E}\Bigr[ 1 + \exp[-b(\bm L^*, A) - h(\bm L^e_m; \bm \alpha)]~|~\bm L^*, A,  R = 1\Bigr] \implies \\
1 &= \eta(\bm L^*, A)\Biggr(1 + \exp[-b(\bm L^*, A)] \times \mathbb{E}\Bigr[\exp\bigr(-h(\bm L^e_m; \bm \alpha)\bigr)~|~\bm L^*, A,  R = 1\Bigr]  \Biggr) \implies \\
\exp[-b(\bm L^*, A)] &=  \frac{1 -\eta(\bm L^*, A)}{\eta(\bm L^*, A)\times\mathbb{E}\Bigr[\exp\bigr(-h(\bm L^e_m; \bm \alpha)\bigr)~|~\bm L^*, A,  R = 1\Bigr] } 
\end{aligned}
$$

\noindent Thus, under the assumed sensitivity model, that 

\begin{equation}
P(R = 1~|~\bm L^*, A, \bm L^e_m) = \Biggr[1 + \frac{1 - \eta(\bm L^*, A)}{\eta(\bm L^*, A)} \times \frac{\exp[-h(\bm L^e_m; \bm \alpha)]}{\mathbb{E}\bigr[\exp\bigr(-h(\bm L^e_m; \bm \alpha)\bigr)~|~\bm L^*, A,  R = 1\bigr]} \Biggr]^{-1}
\end{equation}\\~\\

The idea for the sensitivity analysis is to replace $P(R = 1~|~\bm L^*, A)$ by $P(R = 1~|~\bm L^*, A, \bm L^e_m)$ in the expression to estimate $\widehat\theta_\text{EIF}$. First, note that to estimate $P(R = 1~|~\bm L^*, A, \bm L^e_m)$ under this sensitivity model, we need to estimate one additional nuisance function, $\mathbb{E}\bigr[\exp\bigr(-h(\bm L^e_m; \bm \alpha)\bigr)~|~\bm L^*, A,  R = 1\bigr]$, which can be estimated from the observed data. Once $P(R = 1~|~\bm L^*, A, \bm L^e_m)$ is computed for every complete case, it can replace $\eta(\bm L^*, A)$ in \textit{nearly} every location in Equations (3) and (4). This is due to the fact the majority of terms where $\eta(\bm L^*, A)$ appears, it is pre-multiplied by the complete case indicator $R$, and thus only contributes to the influence function contribution if the subject has $\bm L^e_m$ observed. 

However, the term $\frac{\eta(\bm{L}^*, 0)}{\eta(\bm{L}^*, 1)}$ would at first appear to be problematic to replace by $\frac{\widetilde{\eta}(\bm{L}^*, 0, \bm{L}_m^e; \bm{\alpha})}{\widetilde{\eta}(\bm{L}^*, 1, \bm{L}_m^e; \bm{\alpha})}$,given that this term is contributed by all subjects, not just complete cases, and the ratio depends on $\bm{L}_m^e$. Instead, we replace the two augmentation terms in
Equation (4) involving $\gamma(\bm{L}^*, Y)Y - \chi(\bm{L}^*, Y)$, namely

$$
(1 - A)\left[\frac{R}{\eta(\bm{L}^*, 1)} - \frac{\eta(\bm{L}^*, 0)}{\eta(\bm{L}^*, 1)}\right]
\Big(\gamma(\bm{L}^*, Y)Y - \chi(\bm{L}^*, Y)\Big),
$$

by the single term

\begin{equation}
(1 - A)\,\frac{\eta(\bm{L}^*, 0)}{\eta(\bm{L}^*, 1)}
\left(\frac{R}{\widetilde{\eta}(\bm{L}^*, 0, \bm{L}_m^e; \bm{\alpha})} - 1\right)
\Big(\gamma(\bm{L}^*, Y)Y - \chi(\bm{L}^*, Y)\Big)
\label{eq:S4}
\end{equation}

When $\bm{\alpha} = \bm{0}$,  Equation\eqref{eq:S4} reduces exactly to the original two terms. Moreover, Equation \eqref{eq:S4} has mean zero under the sensitivity model, since $R \perp\!\!\!\perp Y \mid \bm{L}^*, A, \bm{L}_m^e$ implies
\[
\mathbb{E}\left[\frac{R}{\widetilde{\eta}(\bm{L}^*, 0, \bm{L}_m^e; \bm{\alpha})} \;\middle|\; \bm{L}^*, A = 0, Y\right]
= \mathbb{E}\left[\frac{\widetilde{\eta}(\bm{L}^*, 0, \bm{L}_m^e; \bm{\alpha})}{\widetilde{\eta}(\bm{L}^*, 0, \bm{L}_m^e; \bm{\alpha})} \;\middle|\; \bm{L}^*, A = 0, Y\right] = 1.
\]
This swap may affect efficiency under violations of MAR, but not unbiasedness when the
sensitivity model is correct. Critically, the only component of Equation \eqref{eq:S4} that depends on $\bm{L}_m^e$ is premultiplied by $R$; for subjects with $R = 0$, Equation \eqref{eq:S4} depends only on $\eta(\bm{L}^*, 0)$ and $\eta(\bm{L}^*, 1)$. Formal justification that $\widehat{\theta}_{\text{EIF}}$ remains unbiased under the procedure below is provided in Section \ref{supSec:sens_proof}.

Instead, we replace the augmentation term in Equation (4), $(1-A)\frac{ \eta(\bm L^*, 0)}{ \eta(\bm L^*, 1)} \Bigr(\gamma(\bm L^*, Y)Y - \chi(\bm L^*, Y)\Bigr)$ by 

$$
(1-A)\frac{ \eta(\bm L^*, 0)}{ \eta(\bm L^*, 1)} \Biggr(\frac{R}{P(R = 1~|~\bm L^*, 0, \bm L^e_m)} - 1{}\Biggr)\Bigr(\gamma(\bm L^*, Y)Y - \chi(\bm L^*, Y)\Bigr)
$$

\noindent This swap keeps this term mean 0, which may impact the efficiency under violations of MAR, but not the unbiasedness (if the sensitivity model is correct). The unbiasedness of $\widehat\theta_\text{EIF}$ under the sensitivity model is demonstrated in Section \ref{supSec:sens_proof}. Critically, this term as the component dependent on $\bm L^e_m$ is premultipled by $R$.\\~\\

To query the sensitivity of $\widehat{\theta}_{\text{EIF}}$, we take the following steps:
\begin{enumerate}
\item Estimate $\widehat{\theta}_{\text{EIF}}$ following Algorithm 1 retaining the cross-fitted nuisance estimates $\widehat{\eta}$, $\widehat{u}$, $\widehat{\mu}_0$, $\widehat{\varepsilon}_1$, $\widehat{\xi}$, $\widehat{\gamma}$, and $\widehat{\chi}$ for each subject.
 
\item For a given pair $(\alpha_{\text{A1C}}, \alpha_{\text{BMI}})$, compute $h(\bm{L}_m^e; \bm{\alpha})$ for all patients with completely observed eligibility information ($R = 1$).
 
\item Within each training fold of Algorithm 1, estimate the regression $\kappa(\bm{L}^*, a; \bm{\alpha}) = \mathbb{E}\big[\exp\big(-h(\bm{L}_m^e; \bm{\alpha})\big) \mid \bm{L}^*, A = a, R = 1\big]$, and obtain predictions $\widehat{\kappa}(\bm{L}^*, 0; \bm{\alpha})$ and $\widehat{\kappa}(\bm{L}^*, 1; \bm{\alpha})$ for all subjects in the corresponding held-out fold.
 
\item For patients with completely observed eligibility information ($R = 1$), compute

\begin{align*}
\widetilde{\eta}(\bm{L}^*, 1, \bm{L}_m^e; \bm{\alpha})
&= \left[1 + \frac{1 - \widehat{\eta}(\bm{L}^*, 1)}{\widehat{\eta}(\bm{L}^*, 1)}
\times \frac{\exp\big[-h(\bm{L}_m^e; \bm{\alpha})\big]}{\widehat{\kappa}(\bm{L}^*, 1; \bm{\alpha})}\right]^{-1}, \\
\widetilde{\eta}(\bm{L}^*, 0, \bm{L}_m^e; \bm{\alpha})
&= \left[1 + \frac{1 - \widehat{\eta}(\bm{L}^*, 0)}{\widehat{\eta}(\bm{L}^*, 0)}
\times \frac{\exp\big[-h(\bm{L}_m^e; \bm{\alpha})\big]}{\widehat{\kappa}(\bm{L}^*, 0; \bm{\alpha})}\right]^{-1}.
\end{align*}
 
\item In Equations (3) and (4), replace $\widehat{\eta}(\bm{L}^*, 1)$ by $\widetilde{\eta}(\bm{L}^*, 1, \bm{L}_m^e; \bm{\alpha})$ in every term premultiplied by $R$, including the terms premultiplied by $(1 - A)R$, and replace the two $\big(\widehat{\gamma}Y - \widehat{\chi}\big)$ augmentation terms by Equation \eqref{eq:S4}. All other nuisance estimates
($\widehat{u}$, $\widehat{\mu}_0$, $\widehat{\varepsilon}_1$, $\widehat{\xi}$, $\widehat{\gamma}$, $\widehat{\chi}$) are unchanged. Recompute $\widehat{\theta}_{\text{EIF}}$ following steps 5 and 6 of Algorithm 1.
 
\item Repeat steps 2--5 over a grid of $(\alpha_{\text{A1C}}, \alpha_{\text{BMI}})$.
\end{enumerate}

At $\bm{\alpha} = \bm{0}$, $\widehat{\kappa} \equiv 1$ and $\widetilde{\eta}(\bm{L}^*, A, \bm{L}_m^e; \bm{0}) = \widehat{\eta}(\bm{L}^*, A)$, so the procedure reproduces $\widehat{\theta}_{\text{EIF}}$ from Algorithm 1 exactly. Finally, We note that other types of MNAR violations of Assumption 4 are possible, but such violations are not easy to explore, particularly with interpretable sensitivity parameters, which play an important role in describing the robustness of clinical conclusions to MNAR violations of this form. Formal justification that $\widehat\theta_\text{EIF}$ remains unbiased when under the above procedure is provided in Section \ref{supSec:sens_proof}. \\~\\

\begin{figure}[H]
    \centering
    \includegraphics[width=\textwidth]{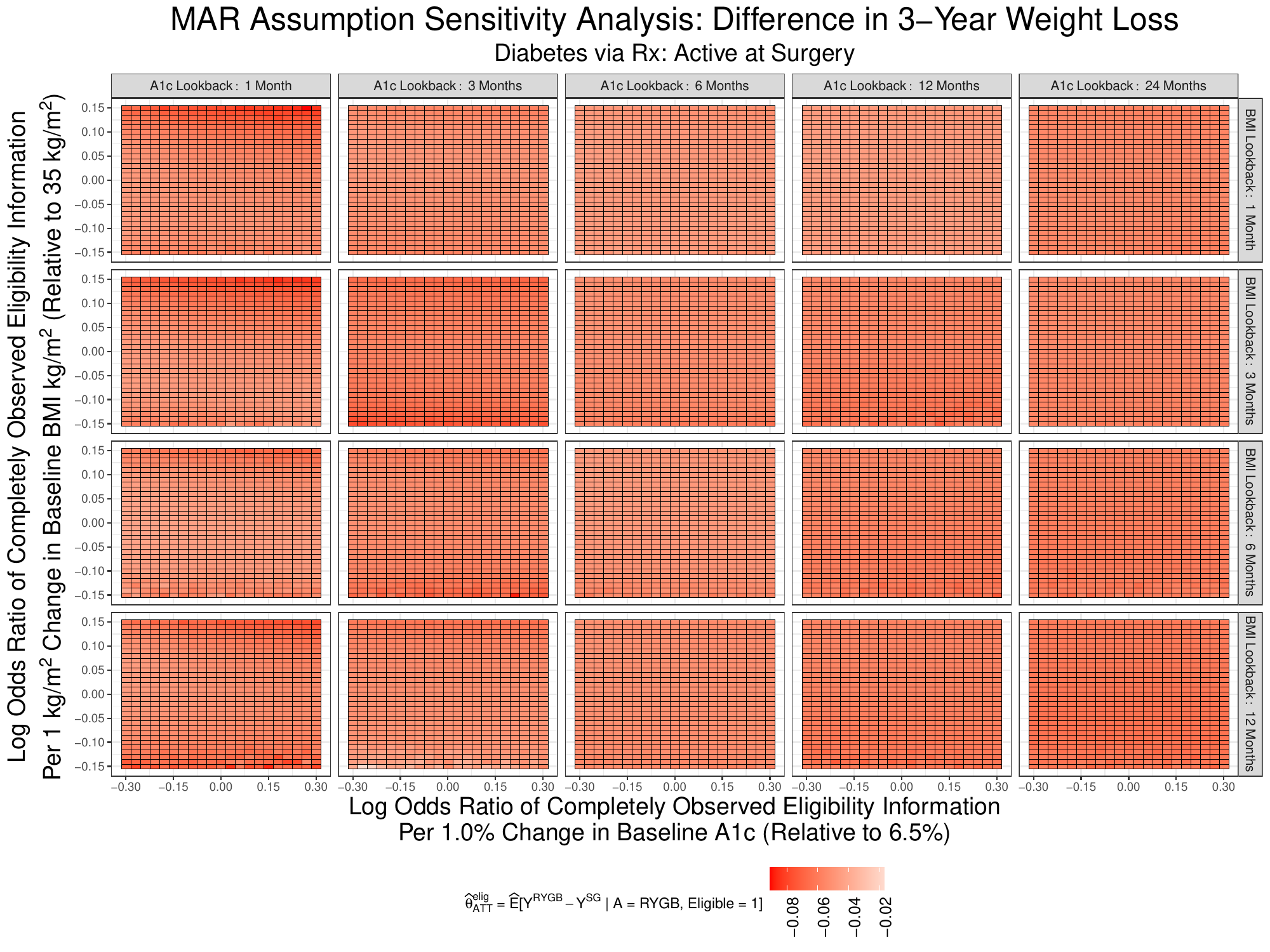}
    \caption{Sensitivity analysis for Assumption 4 (MAR): 3-Year Weight Change (T2DM ascertainment via prescriptions requires active prescription at the time of surgery)}
    \label{fig:sens_weight_0}
\end{figure}

\begin{figure}[H]
    \centering
    \includegraphics[width=\textwidth]{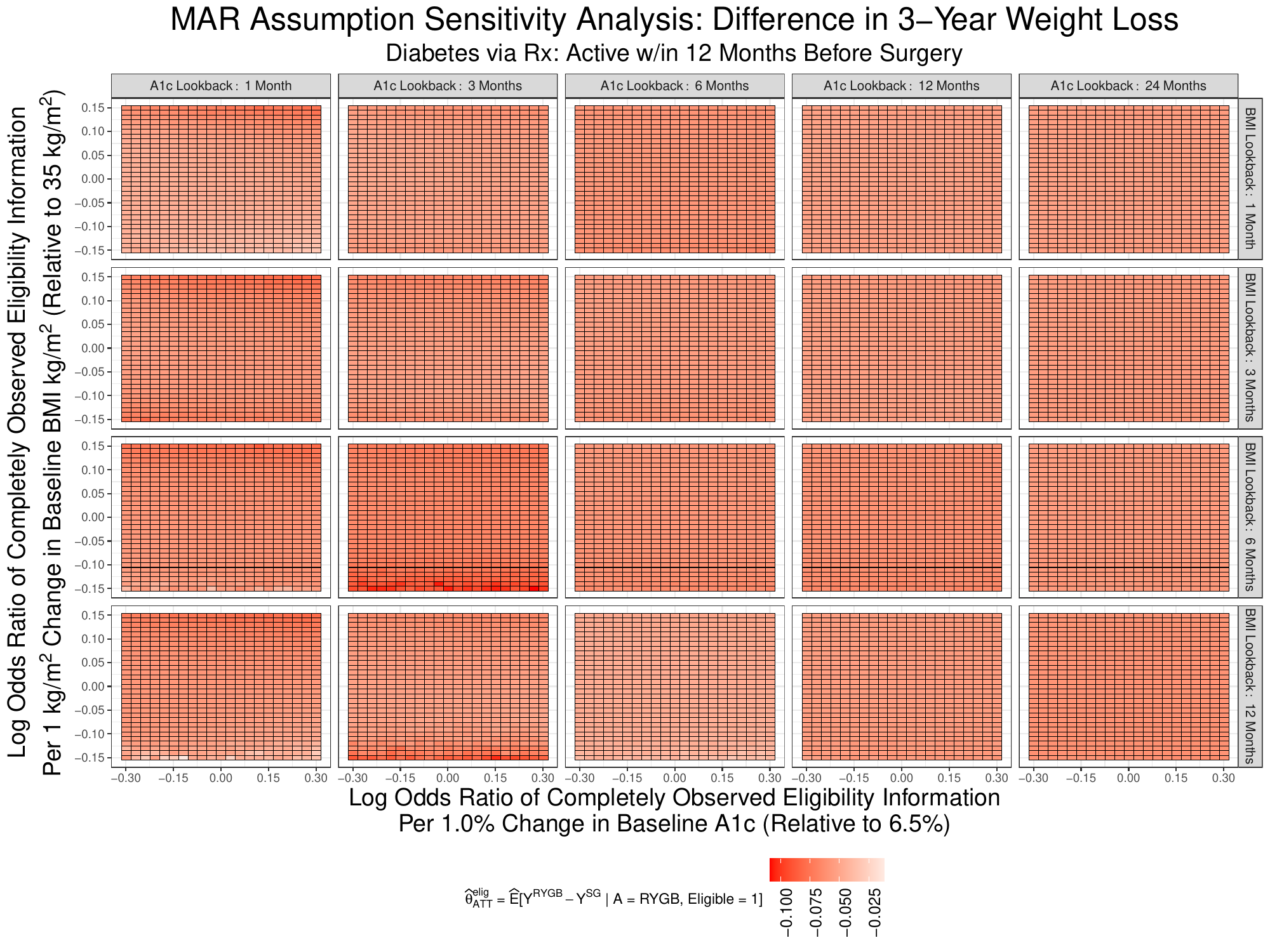}
    \caption{Sensitivity analysis for Assumption 4 (MAR): 3-Year Weight Change (T2DM ascertainment via prescriptions requires active prescription at any time within the year prior to surgery)}
    \label{fig:sens_weight_12}
\end{figure}

\begin{figure}[H]
    \centering
    \includegraphics[width=\textwidth]{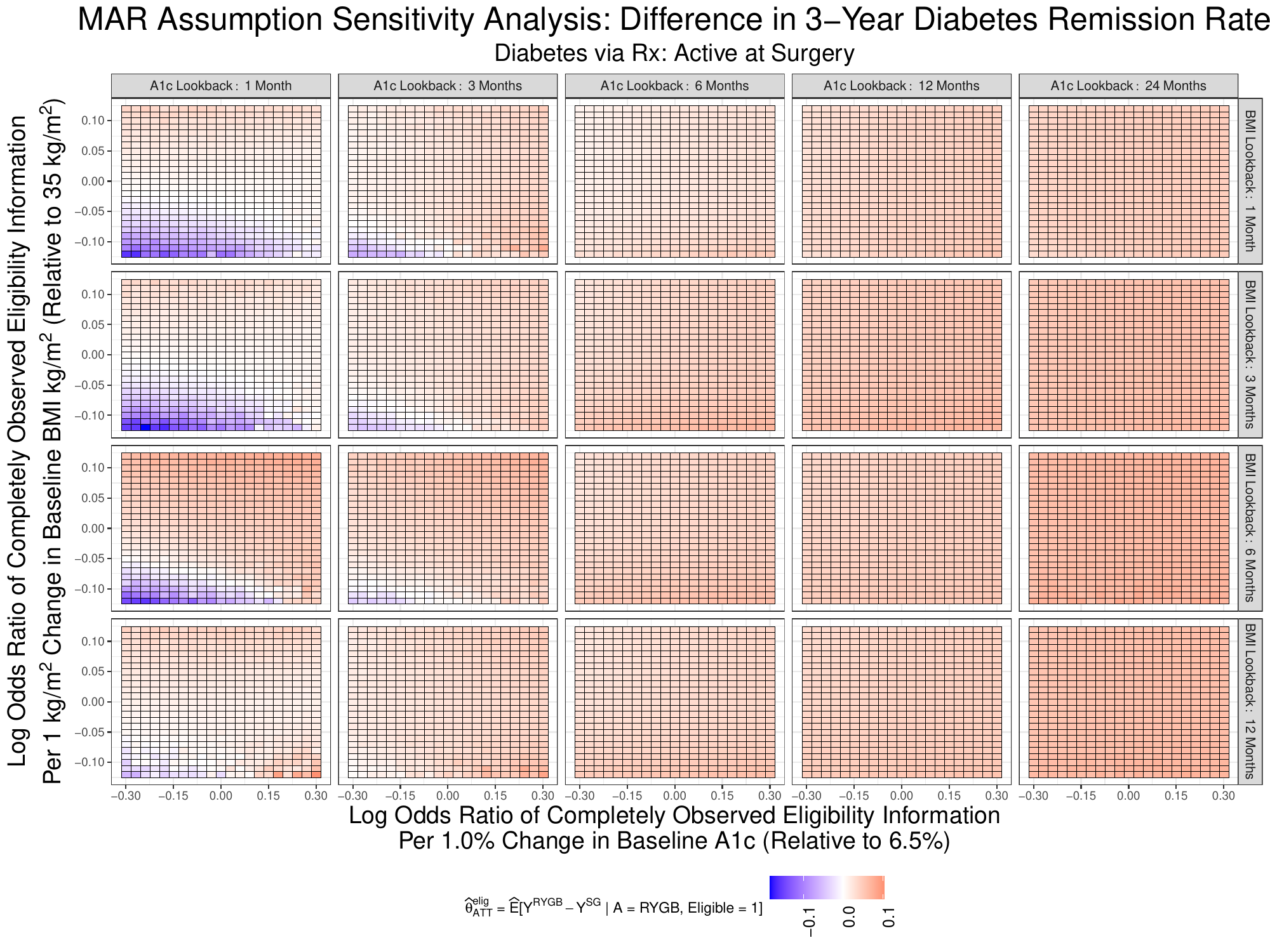}
    \caption{Sensitivity analysis for Assumption 4 (MAR): 3-Year T2DM Remission Rate (T2DM ascertainment via prescriptions requires active prescription at the time of surgery)}
    \label{fig:sens_t2dm_0}
\end{figure}

\begin{figure}[H]
    \centering
    \includegraphics[width=\textwidth]{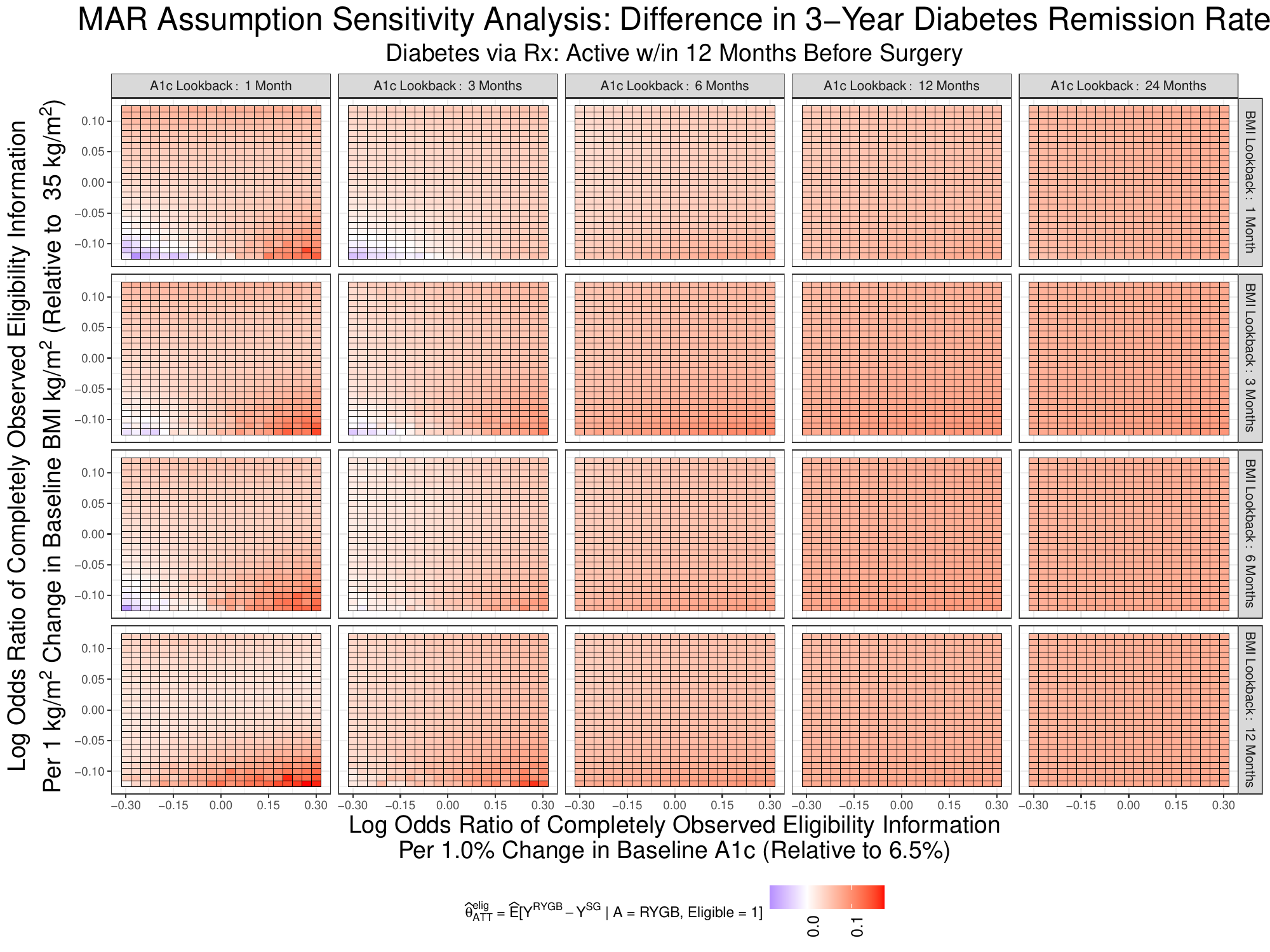}
    \caption{Sensitivity analysis for Assumption 4 (MAR): T2DM Remission Rate (T2DM ascertainment via prescriptions requires active prescription at any time within the year prior to surgery)}
    \label{fig:sens_t2dm_12}
\end{figure}

Results over a grid of $(\alpha_\text{A1C}, \alpha_\text{BMI})$ are presented in Figures \ref{fig:sens_weight_0}-\ref{fig:sens_t2dm_12}. In particular, each cell plots \textbf{point estimates} for $\widehat \theta_\text{EIF}$ under each respective pair of sensitivity parameters. Cells in red favor RYGB, cells in blue favor SG, and cells in white reflect that point estimates very close to 0. Thus, the white regions separating red and blue cells can be viewed as the a ``tipping point'' region for the point estimate of $\widehat\theta_\text{EIF}$

For the weight change outcome, in the most stringent of BMI/A1C lookback windows, no combination of $(\alpha_\text{A1C}, \alpha_\text{BMI})$ changed the point estimate in favor of SG. More extreme values of of both sensitivity parameters were considered but such exponential tilt led to extreme positivity violations which made $\widehat\theta_\text{EIF}$ impossible to compute. The confidence interval crossed 0 for values of $\alpha_\text{BMI} < -0.14$. Such an extreme sensitivity value implies patients with baseline BMI of $60$ kg/m$^2$ (a threshold exceeded by 313 patients just 1 month prior to surgery) would have an odds ratio of < 0.05. That is, in order to obtain a confidence interval consistent with no weight loss difference between the two surgical procedures, patients with the most severe obesity would need to have at least 1/20$^\text{th}$ the odds of ascertainable eligibility compared to otherwise identical patients with BMI of  $35$ kg/m$^2$. 

By contrast, for the T2DM remission outcome, the difference in T2DM remission rate can be made to favor SG with much smaller values of $\alpha_\text{BMI}$ across the entire spectrum of $\alpha_\text{A1C}$, at least for shorter A1C lookback windows. For example with $\alpha_\text{A1C} = 0$, the point estimate can be made to favor SG under $\alpha_\text{BMI} = -0.03$, which corresponds to an odds ratio of 0.97 per 1 kg/m$^2$ of having ascertainable eligibility status. This is not surprising given that $\theta_\text{EIF}$ for the T2DM outcome is closer to the null and the confidence interval already contains 0. Such results substantiate our interpretation that remission of T2DM should be an auxiliary consideration when deciding between SG and RYGB, at least relative to desired weight loss, along with surgical safety.

\subsection{Proof of Unbiasedness Under Sensitivity Model}
\label{supSec:sens_proof}
In this section, we offer justification that the procedure outlined in Section \ref{supSec:sensitivity} preserves unbiased estimation of $\theta_\text{ATT}^\text{elig}$ under correctly specified sensitivity model (which we assume and query under a large number of sensitivity models).

In the remainder of this section we let 
$$
\eta(\,\cdot, \bm \alpha) \equiv \mathbb{P}[R = 1~|~\bm L^*, A, \bm L^e_m, Y]
$$

In Section \ref{supSec:sensitivity}, we conducted sensitivity analysis only with respect $\bm L^e_m$, given that both clinical outcomes examined reflect changes from baseline (\% weight change from study baseline or remission of baseline diabetes status). Implicitly, this invokes the assumption that $R \indep Y ~|~ \bm L^*, \bm L^e_m, A$.
Thus, $\eta(\,\cdot, \bm \alpha)$ does not depend on $Y$ and $\mathbb{P}[R = 1~|~\bm L^*, A, \bm L^e_m, Y] = \mathbb{P}[R = 1~|~\bm L^*, A, \bm L^e_m]$. We then show that replacing $\eta(\bm L^*, A)$ by our sensitivity model $\eta(\,\cdot, \bm \alpha)$ in uncentered influence functions $\dot\alpha_P(O)$ (not to be confused with sensitivity parameter $\bm \alpha$) and $\dot\beta_P(O)$ preserve 
$$
\begin{aligned}
\mathbb{E}[\dot\alpha_P(O)] &= \mathbb{E}[EA]\\
\mathbb{E}[\dot\beta_P(O)] &= \mathbb{E}[EA]\times \theta_\text{ATT}^\text{elig}
\end{aligned}
$$

\noindent\underline{\textbf{Expectation of $\dot\alpha_P(O)$ Under Ascertainment Probability Substitution}}
$$
\begin{aligned}
\mathbb{E}[\dot \alpha_P(O)] &= \mathbb{E}\biggr[A\Bigr(1 - \frac{R}{\eta(\bm L^*, A, \bm L^e_m; \bm \alpha)}\Bigr)\varepsilon_1(\bm L^*, Y) + \frac{ARE}{\eta(\bm L^*, A, \bm L^e_m; \bm \alpha)}\biggr] \\
&= \mathbb{E}\biggr[A\Bigr(1 - \frac{\eta(\bm L^*, A, \bm L^e_m; \bm \alpha)}{\eta(\bm L^*, A, \bm L^e_m; \bm \alpha)}\Bigr)\varepsilon_1(\bm L^*, Y) + \frac{AE\times\eta(\bm L^*, A, \bm L^e_m; \bm \alpha)}{\eta(\bm L^*, A, \bm L^e_m; \bm \alpha)}\biggr]    
~~~\text{(It. Exp. $\bm L^*, \bm L^e_m, A, Y$)} \\
&= \mathbb{E}[EA]
\end{aligned}
$$

\noindent\underline{\textbf{Expectation of $\dot\beta_P(O)$ Under Ascertainment Probability Substitution}}

$$
\begin{aligned}
\mathbb{E}[\dot \beta_P(O)] &= \mathbb{E}\Biggr\{\frac{AR}{\eta(\bm L^*, 1, \bm L^e_m;\bm \alpha)}\biggr[\Bigr(E - \varepsilon_1(\bm L^*, Y)\Bigr)Y - 
   \Bigr(E\mu_0(\bm L^*, \bm L^e_m) - \xi(\bm L^*, Y)\Bigr)\biggr]\\  
   &~~~~~~~~~~+ 
   A\Bigr( \varepsilon_1(\bm L^*, Y)Y - \xi(\bm L^*, Y)\Bigr)  \\
   &~~~~~~~~~~- \frac{(1-A)R}{\eta(\bm L^*, 1, \bm L^e_m;\bm \alpha)}\biggr[E\frac{ u(\bm L^*, \bm L^e_m)}{1- u(\bm L^*, \bm L^e_m)}\Bigr(Y -  \mu_0(\bm L^*, \bm L^e_m)\Bigr) \biggr] \\
&~~~~~~~~~~+(1-A)\frac{ \eta(\bm L^*, 0)}{ \eta(\bm L^*, 1)} \Biggr(\frac{R}{\eta(\bm L^*, 0, \bm L^e_m;\bm \alpha)} - 1{}\Biggr)\Bigr(\gamma(\bm L^*, Y)Y - \chi(\bm L^*, Y)\Bigr)\Biggr\}
\end{aligned}
$$

\noindent After applying iterated expectation conditioning on $\bm L^*, A, \bm L^e_m, Y$, the first 2 terms simplify to 
$$
\mathbb{E}\Bigr[EA[Y - \mu_0(\bm L^*, \bm L^e_m)]\Bigr]
$$
\noindent and 

$$
\mathbb{E}\biggr\{\frac{(1-A)R}{\eta(\bm L^*, 1, \bm L^e_m;\bm \alpha)}\biggr[E\frac{ u(\bm L^*, \bm L^e_m)}{1- u(\bm L^*, \bm L^e_m)}\Bigr(Y -  \mu_0(\bm L^*, \bm L^e_m)\Bigr)\biggr]\biggr\} = 0
$$

Thus, we are left with

$$
\begin{aligned}
\mathbb{E}[\dot \beta_P(O)] &= \mathbb{E}\Biggr\{EA[Y - \mu_0(\bm L^*, \bm L^e_m)] \\
&~~~~~~~~~~-(1-A)\frac{ \eta(\bm L^*, 0)}{ \eta(\bm L^*, 1)} \Biggr(\frac{R}{\eta(\bm L^*, 0, \bm L^e_m;\bm \alpha)} - 1{}\Biggr)\Bigr(\gamma(\bm L^*, Y)Y - \chi(\bm L^*, Y)\Bigr)\Biggr\}
\end{aligned}
$$

Finally, we see that
$$
\begin{aligned}
   &\mathbb{E}\biggr[(1 - A)\,\frac{\eta(\bm{L}^*, 0)}{\eta(\bm{L}^*, 1)}
\left(\frac{R}{\eta(\bm{L}^*, 0, \bm{L}_m^e; \bm{\alpha})} - 1\right)
\Big(\gamma(\bm{L}^*, Y)Y - \chi(\bm{L}^*, Y)\Big) \biggr] = \\
&\mathbb{E}\biggr[ (1 - A)\,\frac{\eta(\bm{L}^*, 0)}{\eta(\bm{L}^*, 1)}
\left(\frac{P(R = 1 ~|~\bm L^*, A = 0, \bm L^e_m, Y)}{\eta(\bm{L}^*, 0, \bm{L}_m^e; \bm{\alpha})} - 1\right)
\Big(\gamma(\bm{L}^*, Y)Y - \chi(\bm{L}^*, Y)\Big) \biggr] = \\
&\mathbb{E}\biggr[ (1 - A)\,\frac{\eta(\bm{L}^*, 0)}{\eta(\bm{L}^*, 1)}
\left(\frac{\eta(\bm{L}^*, 0, \bm{L}_m^e; \bm{\alpha})}{\eta(\bm{L}^*, 0, \bm{L}_m^e; \bm{\alpha})} - 1\right) 
\Big(\gamma(\bm{L}^*, Y)Y - \chi(\bm{L}^*, Y)\Big) \biggr] = 0
\end{aligned}
$$

The last line follows as $Y\indep R ~|~ \bm L^*, \bm L^e_m, A$ under the working sensitivity model.

$$
\begin{aligned}
\mathbb{E}_P[Y(0)~|~A = 1, E = 1 ] &= \mathbb{E}_P[\mathbb{E}_P(Y(0)~|~A = 1, E = 1, \bm L^*, \bm L^e_m)~|~A = 1, E = 1 ] \\ 
&= \mathbb{E}_P[\mathbb{E}_P(Y(0)~|~A = 0, E = 1, \bm L^*, \bm L^e_m)~|~A = 1, E = 1] ~~~\text{(A2, A3)} \\ 
&= \mathbb{E}_P[\mathbb{E}_P(Y~|~A = 0, E = 1, \bm L^*, \bm L^e_m)~|~A = 1, E = 1] ~~~\text{(A1)} \\ 
&= \mathbb{E}_P[\mathbb{E}_P(Y~|~A = 0, E, \bm L^*, \bm L^e_m)~|~A = 1, E = 1] \\
&= \mathbb{E}_P[\mathbb{E}_P(Y~|~A = 0, \bm L^*, \bm L^e_m)~|~A = 1, E = 1]~~~\text{($E = g(\bm L^e)$ fixed)} \\
&= \mathbb{E}_P[\mu_0(\bm L^*, \bm L^e_m)~|~A = 1, E = 1]~~~Y\indep R ~|~ \bm L^*, \bm L^e_m, A~\text{(sensitivity model)} \\
&= \frac{\mathbb{E}_P[E  \mu_0(\bm L^*, \bm L^e_m)~|~A = 1]}{P(E = 1~|~A = 1)}~~~\text{(Lemma \ref{lemma:bvt})} 
\end{aligned}
$$

Crucially, that we can apply $Y\indep R ~|~ \bm L^*, \bm L^e_m, A$ is because of the assumed sensitivity model rather than Assumption 4. Thus, $\mathbb{E}[\dot\beta_P(O)] = \mathbb{E}[EA]\times \theta_\text{ATT}^\text{elig}$, and so the sensitivity procedure in Section \ref{supSec:sensitivity} is unbiased under the assumed sensitivity model.

\section{Practitioner Guidance}
\label{supSec:guidance}

The proposed analysis framework centered $\widehat\theta_\text{EIF}$ has many moving parts, by design. Given such moving parts, it might be difficult for practitioners to evaluate estimator stability and diagnose potential causes for concern. In this section, we offer some general guidance and considerations for implementing $\widehat\theta_\text{EIF}$. In particular, we list several diagnostic checks an analyst might consider when implementing $\widehat\theta_\text{EIF}$. While we cannot offer hard and fast rules, we offer some general commentary on possible causes for concern. We conclude by conducting the checks in our analysis of 3-year weight change following surgery.

\subsection{Diagnostic Checklist}

\begin{itemize}
    \item \textbf{Check}: Distribution of $\widehat\eta(\bm L^*, 1)$ 
    \begin{itemize}
        \item \textbf{Relevant Assumption(s)}: Assumption 5 (Complete Case Positivity)
        \item \textbf{Reason}: Division by $\eta(\bm L^*, 1)$ in component influence functions $\dot\alpha_P(O)$ and $\dot\beta_P(O)$
        \item \textbf{What to look for}: $\eta(\bm L^*, 1)$ can be interpreted as an inverse probability of ascertainment weight, so analysts should examine $\min \eta(\bm L^*, 1)$. Causes for concern include extremely small weights (e.g., $\min \widehat \eta(\bm L^*, 1) < 0.01$), or a large concentration of small weights (e.g. at least 25\% of $\widehat \eta(\bm L^*, 1) \leq 0.1)$.
    \end{itemize}

    \item \textbf{Check}: Distribution of $\widehat\eta(\bm L^*, 1), \widehat\eta(\bm L^*, 0)$ together
    \begin{itemize}
        \item \textbf{Relevant Assumption(s)}: ---
        \item \textbf{Reason}: Differential missingness by treatment status
        \item \textbf{What to look for}: If there are large areas where the distributions of $\widehat\eta(\bm L^*, 1)$ and $\widehat\eta(\bm L^*, 0)$ do not overlap, this suggests that missingness in eligibility-defining criteria is differential by treatment status, even after accounting for completely observed covariates. While differential missingness does not necessarily imply violations of Assumption 4 (MAR), it would be worth noting and may give analysts reason to select alternative lookback windows where missingness is not differential by treatment status (after accounting for $\bm L^*$). A reasonable diagnostic to check and report is $\mathbb{P}_n[\widehat\eta(\bm L^*, 1) - \widehat\eta(\bm L^*, 0)]$. 
    \end{itemize}

    \item \textbf{Check}: Effective Sample Size (ESS) 
    \begin{itemize}
        \item \textbf{Relevant Assumption(s)}: Assumption 5 (Complete Case Positivity)
        \item \textbf{Reason}: Calibration of $\widehat{\eta}(\bm L^*, 1)$
        \item \textbf{What to look for}: Following the ideas of \citet{kish1992weighting} and \citet{MatchIt}, an approximate ESS based on the inverse probability of ascertainment weights is given as follows, 
        $$
        ESS  = \frac{\Bigr[\sum_{i = 1}^n R_iE_i \times 1/\widehat\eta(\bm L^*_i, 1)\Bigr]^2}{\sum_{i = 1}^n R_iE_i \times 1/\widehat\eta(\bm L^*_i, 1)^2}
        $$

    When weights are ``well behaved'' (e.g., not concentrated at the boundaries, Assumption 2 holds, missingness non-differential by treatment status), the effective sample size will be similar to the number of complete cases who are eligible for the study.
        
    \end{itemize}

    \item \textbf{Check}: Distribution of $\widehat u(\bm L^*, \bm L^e_m)$
    \begin{itemize}
        \item \textbf{Relevant Assumption(s)}: Assumption 2 (Treatment Positivity Among Eligible)
        \item \textbf{Reason}: Division by $u(\bm L^*, \bm L^e_m)$ in influence function $\dot\beta_P(O)$
        \item \textbf{What to look for}: $\widehat u(\bm L^*, \bm L^e_m)$ only contributes to $\widehat \theta_\text{EIF}$ among eligible complete cases for whom $A = 0$, through the ratio $\frac{u(\bm L^*, \bm L^e_m)}{1-u(\bm L^*, \bm L^e_m)}$. Thus, it is useful to summarize $\min_{A = 0, E = 1, R = 1} u(L^*, L^e_m)$. Similar to the advice regarding $\eta(\bm L^*, 1)$, extreme weights (e.g., $\geq 0.99$), or concentration of small weights are cause for concern. In the presence of extreme weights, analysts might consider weight trimming; we refer readers to \citet{cole2008} and \citet{chernozhukov2018} for more discussion on extreme inverse probability weights.
    \end{itemize}

        \item \textbf{Check}: Distribution of $\widehat\varepsilon_1(\bm L^*, Y)$ across lookback windows
        \begin{itemize}
            \item \textbf{Relevant Assumption(s)}: Assumption 4 (Eligibility MAR)
            \item \textbf{Reason}: Choice of lookback windows
            \item \textbf{What to look for}: Note that in contrast to nuisance function like $u$ and $\eta$, $\varepsilon$ is not used as a form of inverse probability weight. Inherently, values $\varepsilon_1(\bm L^*, Y)$ close to 0 are not inherently problematic as there are always patients in the EHR with completely observed information who just do not meet the study eligibility criteria. It is far more valuable to examine the distribution of $\widehat\varepsilon_1(\bm L^*, Y)$ across lookback windows. Shifting of the distribution towards 0 suggests that relaxing eligibility criteria by extending the length of time over which eligibility may be ascertained may be capturing lower quality of information. Altogether, this may speak to greater plausibility that the MAR assumption holds close to study baseline. Alternatively, if the distribution of $\widehat\varepsilon_1(\bm L^*, Y)$ is fairly stable across lookback windows, it would suggest that longer lookback windows may be more reasonable.
    \end{itemize}

        \item \textbf{Check}: Tipping point analysis for $\widehat\theta_\text{EIF}$ 
        \begin{itemize}
            \item \textbf{Relevant Assumption(s)}: Assumption 4 (Eligibility MAR)
            \item \textbf{Reason}: Sensitivity analysis to violations of Assumption 4
            \item \textbf{What to look for}: Degree of robustness of clinical conclusions to certain MNAR violations of assumption 4. See Section \ref{supSec:sensitivity} for details.
        \end{itemize}

        \item \textbf{Check}: Distribution of nested nuisance functions $(\widehat \xi, \widehat\gamma, \widehat \chi)$
        \begin{itemize}
            \item \textbf{Relevant Assumption(s)}: ---
            \item \textbf{Reason}: Variability of Estimator
            \item \textbf{What to look for}: Note that the scales of nested nuisance functions $(\xi, \gamma, \chi)$ do not have easily interpretable meaning, and are likely to be very context specific, particularly $\xi$ and $\chi$ which contain a nested outcome regression ($\mu_0$). Across lookback windows, analysts might look for signs of zero-inflation, multimodality, and skewness, particularly when such phenomena differ across scenarios. In isolation, none of these problems is itself problematic, but when patterns change over lookback windows, it may speak to problems in underlying components of these nested nuisance functions. For example extreme propensity scores $u$, or high concentration of ineligible patients. 
    \end{itemize}

\end{itemize}

\subsection{Diagnostic Checks for Weight Change Outcomes}

\begin{enumerate}
    \item \textbf{Ascertainment probability distribution}
    \begin{itemize}
        \item The distribution of $\widehat\eta(\bm L^*, 1)$ across lookback windows is shown in Figure 4 in the main manuscript.
        \item $\min \widehat\eta(\bm L^*, 1) = 0.069$ across all scenarios considered.
        \item The maximal value of $\mathbb{P}_n[\widehat \eta(\bm L^*, 1) \leq 0.1]$ across all lookback windows is 1.8\%
        \item \textbf{Conclusion:} Small inverse probability of ascertainment weights are not of great concern
    \end{itemize}

    \item \textbf{Differential missingness by treatment status}
    \begin{itemize}
        \item The distribution of $\widehat\eta(\bm L^*, 1)$ and $\widehat\eta(\bm L^*, 0)$ across lookback windows is shown in Figure 4 in the main manuscript.
        \item Covariate-adjusted average difference in observing complete eligibility information between RYGB and SG, $\mathbb{P}_n[\widehat \eta_1(\bm L^*) - \widehat \eta_0(\bm L^*)]$, ranged from -6.1\% to 7.6\%. 
        \item Differential missingness by procedure was most prominent when the length of time for BMI and T2DM lookback windows differed. 
        \item \textbf{Conclusion:} There are reasonable degrees of overlap between the two distributions, but given that differential missingness appears to be associated with lookback windows that differed in length, we would recommend selecting similar lookback window lengths for BMI and A1C. In light of Figure 3, this might be BMI lookback of 1 month and A1C Lookback $\leq$ 6 months. For such settings, $\mathbb{P}_n[\widehat \eta_1(\bm L^*) - \widehat \eta_0(\bm L^*)]$ ranged from 0.5\% to 2.5\%. 
    \end{itemize}

    \item \textbf{Effective Sample Size}
    \begin{itemize}
        \item Across all lookback windows the ratio $\frac{ESS}{n_{cc}}$, where $n_{cc}$ denotes the number of ascertainably eligible patients, ranged from 0.913 to 0.998, with median 0.993.
        \item \textbf{Conclusion:} Ascertainment model $\eta$ is well-behaved.
    \end{itemize}

    \item \textbf{Distribution of propensity scores}
    \begin{itemize}
        \item Summary statistics from the distribution of $\widehat u(\bm L^*, \bm L^e_m)$ across the most reasonable scenarios from above (BMI lookback = 1 month, A1C lookback $\leq 6$ months):
        \begin{itemize}
            \item Minimum: 0.275 
            \item 1\% quantile: 0.333
            \item Median: 0.610  
            \item Mean: 0.600 
            \item 99\% quantile: 0.924
            \item Max: 0.993 
            \item Proportion $> 0.99$: 0.092\%
        \end{itemize}
        \item \textbf{Conclusion:} The weights fall in a reasonable range, with fewer than 0.1\% of $\widehat u(\bm L^*, \bm L^e_m)$ exceeding 0.99.  Given how few propensity scores are near the boundary, we elect not to clip weights based on $\frac{\widehat u(\bm L^*, \bm L^e_m)}{1-\widehat u(\bm L^*, \bm L^e_m)}$, as it can violate the required rate conditions for $\sqrt{n}$-consistency \citep{chernozhukov2018}.
    \end{itemize}

    \item \textbf{Distribution of $\widehat \varepsilon_1(\bm L^*, Y)$}
    \begin{itemize}
         \item The distribution of $\widehat \varepsilon_1(\bm L^*, Y)$ across lookback windows is shown in Figure 4 in the main manuscript. 
         \item Average values of $\varepsilon_1$ ranged from 0.613 in the most stringent settings with to 0.300 in the most relaxed, indicating that a higher proportion of subjects with complete information were judged to be eligible in lookback windows with narrower scope. 
         \item \textbf{Conclusion}: While this observation cannot directly assess the validity of Assumption 4 across operationalizations, it speaks to the higher quality of information available closer to the time of surgery, and thus in our view, greater plausibility that the MAR assumption holds. 
    \end{itemize}

    \item \textbf{Tipping Point Analysis}
    \begin{itemize} 
        \item See Section \ref{supSec:sensitivity}
        \item See Figures \ref{fig:sens_weight_0} and \ref{fig:sens_weight_12} 
        \item \textbf{Conclusion:} Clinical conclusions for each given lookback window are robust to violations of Assumption 4, at least violations of the form described in Section \ref{supSec:sensitivity}.
    \end{itemize}
    
    \item \textbf{Distribution of Nested Nuisance Functions}
    \begin{itemize} 
        \item See Figure \ref{fig:S4}
        \item That the distribution of $\widehat\xi$ is pulled towards 0 with longer diabetes lab lookbacks, which speaks to a lower eligibility rate with longer lookback times.
        \item Long tails in 3 month diabetes lab look back window in $\widehat \gamma, \widehat\chi$ $\to$ possibly higher variance in $\widehat \theta_\text{EIF}$
        \item \textbf{Conclusion:} Thinking through the clinical rationale for MAR plausibility within 6 months and estimator stability, a reasonable choice for an investigator seeking guidance is a BMI lookback of 1 month and an A1C lookback of 6 months. We do not designate this as a primary operationalization in the main text, since conclusions (particularly for T2DM remission) could differ across operationalizations, and instead report results across all 40. The conclusions we report apply broadly across the range of lookback windows for which MAR is most plausible.
    \end{itemize}

\end{enumerate}

\section{Alternative Covariate Partitioning}\label{supSec:alt_covariate}
\subsection{Notation and Assumptions}\label{supSec:alt_notation}

In this section, we explore an alternative way of partitioning baseline covariates $\bm L$. In particular, we show that though this partitioning may facilitate assumptions which are slightly more parsimonious than those in the main paper, identification of $\theta_\text{ATT}^\text{elig}$ is much more complex. In this exploration, we partition $\bm L$ as follows:

\begin{itemize}
    \item $\bm L^{e, \widebar c}_m$: Eligibility defining covariates with some degree of missingness, which are not confounders
    \item $\bm L^{e, c}_m$: Eligibility defining covariates with some degree of missingness, which are confounders
    \item $\bm L^{*, c}$: Completely observed covariates which are confounders
    \item $\bm L^{*, \widebar c}$: Completely observed covariates which are not confounders (but may for example be necessary to predict missingness in $\bm L^e$)
\end{itemize}

The connection to this covariate partitioning and the one utilized in the majority of the work is given by $\bm L^e_m = (\bm L^{e, \widebar c}_m, \bm L^{e, c}_m)$ and $\bm L^* = (\bm L^{*, c}, \bm L^{*, \widebar c})$ Assumptions are as follows:\\

\noindent\textbf{Assumption 1}: $Y(A) = Y~|~E = 1$\\
\noindent\textbf{Assumption 2$'$}: $\exists ~\epsilon > 0$ such that $0 < \epsilon \leq P(A = 1 ~|~\bm  L^{*, c}, \bm L^{e, c}_m, E = 1) \leq 1 - \epsilon < 1$, almost surely \\
\noindent\textbf{Assumption 3$'$}: $Y(a) \indep A~|~\bm L^{*, c}, \bm L^{e, c}_m, E = 1$ for $a \in \{0, 1\}$ \\
\noindent\textbf{Assumption 4}: $R \indep (Y, \bm L^e_m)~|~\bm L^*, A$ \\
\noindent\textbf{Assumption 5}: $\exists ~\epsilon > 0$ such that $0 < \epsilon \leq P(R = 1 ~|~\bm L^*, A) $, almost surely \\

\noindent While Assumptions 1, 4 and 5 remain unchanged, Assumptions 2$'$ and 3$'$ only require confounders $\bm L^{e, c}_m$ and $\bm L^{*, c}$ rather than all of $\bm L$.\\~\\
We introduce an additional nuisance function for identification of $\theta_\text{ATT}^\text{elig}$ in the following section. Not all nusiance functions are used for identification but are useful in intermediate derivations. Henceforth, let 

$$
\begin{aligned}
\Lambda_a(\bm L^{e,c}_m;\bm L^{*,c}) &= \int_{\mathcal{L}^{*, \widebar c}} p(E = 1, \bm L^{e,c}_m~|~A = a, \bm L^{*, c}, \bm \ell^{*, \widebar c}, R = 1)\, p(\bm \ell^{*, \widebar c}~|~A = a, \bm L^{*, c})\, d\bm\ell^{*, \widebar c} \\
\delta_a(\bm L^{e,c}_m;\bm L^{*,c}) &= \int_{\mathcal{L}^{*, \widebar c}} p(\bm L^{e,c}_m~|~A = a, \bm L^{*, c}, \bm \ell^{*, \widebar c}, R = 1)\, p(\bm \ell^{*, \widebar c}~|~A = a, \bm L^{*, c})\, d\bm\ell^{*, \widebar c}\\
\kappa(\bm L^{*,c}) &= P(A = 1~|~\bm L^{*,c})
\end{aligned}
$$

\subsection{Identification of \texorpdfstring{$\theta_{\text{ATT}}^{\text{elig}}$}{ATTE}  Under Alternative Covariate Partition}\label{supSec:alt_id}

\begin{theorem}
Under Assumptions 1, 2$'$, 3$'$, 4, and 5, $\theta_\text{ATT}^\text{elig}$ is identified by $\frac{\zeta(P)}{\alpha(P)}$ where 
$$
\begin{aligned}
\zeta(P) &= \mathbb{E}_P\Biggr[REY\biggr\{\frac{A}{\eta(\bm L^*, 1)} - \frac{(1 - A)}{\eta(\bm L^*,0)}\cdot\frac{\kappa(\bm L^{*, c})\Lambda_1(\bm L^{e,c}_m; \bm L^{*, c})}{\bigr(1 - \kappa(\bm L^{*, c})\bigr)\Lambda_0(\bm L^{e,c}_m; \bm L^{*, c})}\biggr\}\Biggr] \\
\alpha(P) &= \mathbb{E}_P\Biggr[\frac{ARE}{\eta(\bm L^*, 1)}\Biggr]
\end{aligned}
$$
\end{theorem}

\noindent\underline{\textbf{Proof}:}
\noindent We first note that the following result holds unchanged, given that it only relied on consistency (A1) and Lemma \ref{lemma:bvt}.

$$
\mathbb{E}_P[Y(1)~|~A = 1, E = 1] = \mathbb{E}_P[Y~|~A = 1, E = 1] = \frac{\mathbb{E}_P[E  Y~|~A = 1]}{P(E = 1~|~A = 1)}
$$

\noindent Moreover, the identification of $\frac{\mathbb{E}_P[E  Y~|~A = 1]}{P(E = 1~|~A = 1)}$ remains unchanged. In the notation of this updated covariate partitioning, we have

$$
\begin{aligned}
P(E = 1~|~A = 1) &= \mathbb{E}_P[\mathbb{E}_P(E~|~\bm L^*, A = 1)~|~ A = 1)] \\
&= \mathbb{E}_P[\mathbb{E}_P(E~|~\bm L^*, A = 1, R = 1)~|~ A = 1)]~~~(\text{A4, A5}) \\
&= \mathbb{E}_P\biggr[\mathbb{E}_P\biggr(\frac{RE}{\eta(\bm L^*, 1)}~\Bigr|~\bm L^*, A = 1\biggr)~\Bigr|~ A = 1\biggr]~~~(\text{Lemma \ref{lemma:bvt}, Defn. of $\eta$}) \\
&= \mathbb{E}_P\biggr[\frac{RE}{\eta(\bm L^*, 1)}~\Bigr|~ A = 1\biggr]\\
&= \frac{\mathbb{E}_P\biggr[\frac{ARE}{\eta(\bm L^*, 1)}\biggr]}{P(A = 1)}~~~(\text{Lemma \ref{lemma:bvt}})\\
&= \frac{\alpha(P)}{P(A = 1)}\\
\end{aligned}
$$

$$
\begin{aligned}
\mathbb{E}_P[E Y~|~A = 1] &= 
\mathbb{E}_P[\mathbb{E}_P(E Y~|~\bm L^*, \bm L^e_m, A = 1)~|~ A = 1)] \\
&= \mathbb{E}_P[E\mathbb{E}_P(Y~|~\bm L^*, \bm L^e_m, A = 1)~|~ A = 1)]~~~\text{($E = g(\bm L^e)$, fixed function of $\bm L^e$)} \\
&= \mathbb{E}_P[E\mathbb{E}_P(Y~|~\bm L^*, \bm L^e_m, A = 1, R = 1)~|~ A = 1)]~~~\text{(A4, A5)} \\
&= \mathbb{E}_P\biggr[E\mathbb{E}_P\biggr(\frac{RY}{P(R = 1~|~\bm L^*, \bm L^e_m, A = 1)}~\Bigr|~\bm L^*, \bm L^e_m, A = 1\biggr)~\Bigr|~ A = 1\biggr]~~~(\text{Lemma \ref{lemma:bvt}}) \\
&= \mathbb{E}_P\biggr[\frac{E}{\eta(\bm L^*, 1)}\mathbb{E}_P[RY~|~\bm L^*, \bm L^e_m, A = 1]~\Bigr|~A = 1 \biggr] ~~~(\text{A4, A5}) \\
&= \mathbb{E}_P\biggr[\frac{RE}{\eta(\bm L^*, 1)}Y~\Bigr|~ A = 1\biggr]\\
&= \frac{\mathbb{E}_P\Bigr[\frac{ARE}{\eta(\bm L^*, 1)}Y\Bigr]}{P(A = 1)}~~~(\text{Lemma \ref{lemma:bvt}}) \\
\end{aligned}
$$

\noindent Thus we have that 

$$
\mathbb{E}_P[Y(1)~|~A = 1, E = 1] = \frac{\mathbb{E}_P\Bigr[\frac{ARE}{\eta(\bm L^*, 1)}Y\Bigr]}{\alpha(P)}
$$

\noindent Things become more complicated for $\mathbb{E}_P[Y(0)~|~A = 1, E = 1]$  because the set of covariates to satisfy the assumption of no unmeasured confounding ($\bm L^{e, c}_m, \bm L^{*, c})$ is no longer sufficient to define eligibility or satisfy MAR (A4). In particular

$$
\begin{aligned}
&~~~~\mathbb{E}_P[Y(0)~|~A = 1, E = 1 ]\\ 
&= \mathbb{E}_P[\mathbb{E}_P(Y(0)~|~A = 1, E = 1, \bm L^{*, c}, \bm L^{e, c}_m)~|~A = 1, E = 1 ] \\ 
&= \mathbb{E}_P[\mathbb{E}_P(Y(0)~|~A = 0, E = 1, \bm L^{*, c}, \bm L^{e, c}_m)~|~A = 1, E = 1 ] ~~~\text{(A2$'$, A3$'$)} \\
&= \mathbb{E}_P[\mathbb{E}_P(Y~|~A = 0, E = 1, \bm L^{*, c}, \bm L^{e, c}_m)~|~A = 1, E = 1 ]~~~\text{(A1)} \\
&= \mathbb{E}_P\Bigr[\mathbb{E}_P\Bigr(\mathbb{E}_P[Y~|~A = 0, E = 1, \bm L^*, \bm L^e_m]~\bigr|~A = 0, E = 1, \bm L^{*, c}, \bm L^{e, c}_m\Bigr)~\Bigr|~A = 1, E = 1 \Bigr] \\
&= \mathbb{E}_P\Bigr[\mathbb{E}_P\Bigr(\mathbb{E}_P[Y~|~A = 0, \bm L^*, \bm L^e_m]~\bigr|~A = 0, E = 1, \bm L^{*, c}, \bm L^{e, c}_m\Bigr)~\Bigr|~A = 1, E = 1 \Bigr]~~~\text{($E = g(\bm L^e)$, fixed)} \\
&= \mathbb{E}_P\Bigr[\mathbb{E}_P\Bigr(\mathbb{E}_P[Y~|~A = 0, \bm L^*, \bm L^e_m, R = 1]~\bigr|~A = 0, E = 1, \bm L^{*, c}, \bm L^{e, c}_m\Bigr)~\Bigr|~A = 1, E = 1 \Bigr]~~~\text{(A4, A5)} \\
&= \mathbb{E}_P\Bigr[\mathbb{E}_P\Bigr(\mu_0(\bm L^*, \bm L^e_m) ~\bigr|~A = 0, E = 1, \bm L^{*, c}, \bm L^{e, c}_m\Bigr)~\Bigr|~A = 1, E = 1 \Bigr]\\
\end{aligned}
$$

\noindent Repeated application of Lemma \ref{lemma:bvt} yields

$$
\begin{aligned}
&~~~~\mathbb{E}_P\Bigr[\mathbb{E}_P\Bigr(\mu_0(\bm L^*, \bm L^e_m) ~\bigr|~A = 0, E = 1, \bm L^{*, c}, \bm L^{e, c}_m\Bigr)~\Bigr|~A = 1, E = 1 \Bigr]\\
&= \mathbb{E}_P\Biggr[\frac{E}{P(E = 1~|~A =1)}\mathbb{E}_P\Bigr(\mu_0(\bm L^*, \bm L^e_m) ~\bigr|~A = 0, E = 1, \bm L^{*, c}, \bm L^{e, c}_m\Bigr)~\Bigr|~A = 1 \Biggr]~~~\text{(Lemma \ref{lemma:bvt})} \\
&= \mathbb{E}_P\Biggr[\frac{EA}{\alpha(P)}\mathbb{E}_P\biggr(\frac{E\mu_0(\bm L^*, \bm L^e_m)}{P(E = 1~|~A = 0, \bm L^{*, c}, \bm L^{e, c}_m)} ~\Bigr|~A = 0, \bm L^{*, c}, \bm L^{e, c}_m\Biggr)\Biggr]~~~\text{(Lemma \ref{lemma:bvt})} \\
&= \mathbb{E}_P\Biggr[\frac{EA}{\alpha(P)}\mathbb{E}_P\biggr(\frac{(1-A)}{P(A = 0~|~\bm L^{*, c}, \bm L^{e, c}_m)}\frac{E\mu_0(\bm L^*, \bm L^e_m)}{P(E = 1~|~ A = 0, \bm L^{*, c}, \bm L^{e, c}_m)} ~\Bigr|~\bm L^{*, c}, \bm L^{e, c}_m\Biggr)\Biggr]~~~\text{(Lemma \ref{lemma:bvt})} \\
&= \mathbb{E}_P\Biggr[\frac{\mathbb{E}_P[EA~|~\bm L^{*, c}, \bm L^{e, c}_m]}{\alpha(P)}\frac{(1-A)}{P(A = 0~|~\bm L^{*, c}, \bm L^{e, c}_m)}\frac{E\mu_0(\bm L^*, \bm L^e_m)}{P(E = 1~|~ A = 0, \bm L^{*, c}, \bm L^{e, c}_m)} \Biggr] \\
&= \mathbb{E}_P\Biggr[\frac{1 - A}{\alpha(P)}\cdot\frac{P(A = 1~|~\bm L^{*, c}, \bm L^{e, c}_m)P(E = 1~|~ A = 1, \bm L^{*, c}, \bm L^{e, c}_m)}{P(A = 0~|~\bm L^{*, c}, \bm L^{e, c}_m)P(E = 1~|~ A = 0, \bm L^{*, c}, \bm L^{e, c}_m)}E\mu_0(\bm L^*, \bm L^e_m) \Biggr]
\end{aligned}
$$

\noindent Next we see that 
$$
\begin{aligned}
&~~~~P(E = 1~|~A = a, \bm L^{*,c}, \bm L^{e,c}_m) \\
&= \frac{p(E = 1, \bm L^{e,c}_m~|~A = a, \bm L^{*,c})}{p(\bm L^{e,c}_m~|~A = a, \bm L^{*,c})} \quad \text{(Bayes Rule)} \\
&= \frac{\int_{\mathcal{L}^{*, \widebar c}} p(E = 1, \bm L^{e,c}_m~|~A = a, \bm L^{*, c}, \bm \ell^{*, \widebar c})\, p(\bm \ell^{*, \widebar c}~|~A = a, \bm L^{*, c})\, d\bm\ell^{*, \widebar c}}{\int_{\mathcal{L}^{*, \widebar c}} p(\bm L^{e,c}_m~|~A = a, \bm L^{*, c}, \bm \ell^{*, \widebar c})\, p(\bm \ell^{*, \widebar c}~|~A = a, \bm L^{*, c})\, d\bm\ell^{*, \widebar c}} \\
&= \frac{\int_{\mathcal{L}^{*, \widebar c}} p(E = 1, \bm L^{e,c}_m~|~A = a, \bm L^{*, c}, \bm \ell^{*, \widebar c}, R = 1)\, p(\bm \ell^{*, \widebar c}~|~A = a, \bm L^{*, c})\, d\bm\ell^{*, \widebar c}}{\int_{\mathcal{L}^{*, \widebar c}} p(\bm L^{e,c}_m~|~A = a, \bm L^{*, c}, \bm \ell^{*, \widebar c}, R = 1)\, p(\bm \ell^{*, \widebar c}~|~A = a, \bm L^{*, c})\, d\bm\ell^{*, \widebar c}} \quad \text{(A4, A5)} \\
&= \frac{\Lambda_a(\bm L^{e,c}_m;\bm L^{*,c})}{\delta_a(\bm L^{e,c}_m;\bm L^{*,c})}
\end{aligned}
$$

The third equality applies A4 inside each integral, where the conditioning set includes all of $\bm L^* = (\bm L^{*,c}, \bm L^{*,\widebar c})$; since $E$ is a function of $\bm L^e_m$ and $\bm L^*$, $R \perp\!\!\!\perp (E, \bm L^{e,c}_m)~|~\bm L^*, A$ follows from A4. Similarly,

$$
\begin{aligned}
P(A = a~|~\bm L^{*,c}, \bm L^{e,c}_m) &= \frac{p(\bm L^{e,c}_m~|~A = a, \bm L^{*,c})P(A = a~|~\bm L^{*,c})}{\sum_{a' = 0}^1 p(\bm L^{e,c}_m~|~A = a', \bm L^{*,c})P(A = a'~|~\bm L^{*,c})} \quad \text{(Bayes Rule)} \\
&= \frac{P(A = a~|~\bm L^{*,c})\,\delta_a(\bm L^{e,c}_m;\bm L^{*,c})}{\sum_{a' = 0}^1 P(A = a'~|~\bm L^{*,c})\,\delta_{a'}(\bm L^{e,c}_m;\bm L^{*,c})} \quad \text{(A4, A5)}
\end{aligned}
$$

\noindent Combining the two results, $\delta_a$ cancels:
$$
\begin{aligned}
\frac{P(A = 1~|~\bm L^{*,c}, \bm L^{e,c}_m)P(E = 1~|~A = 1, \bm L^{*,c}, \bm L^{e,c}_m)}{P(A = 0~|~\bm L^{*,c}, \bm L^{e,c}_m)P(E = 1~|~A = 0, \bm L^{*,c}, \bm L^{e,c}_m)} = \frac{\kappa(\bm L^{*,c})\Lambda_1(\bm L^{e,c}_m;\bm L^{*,c})}{\big(1 - \kappa(\bm L^{*,c})\big)\Lambda_0(\bm L^{e,c}_m;\bm L^{*,c})}
\end{aligned}
$$

\noindent Thus our expression simplifies to 
$$
\begin{aligned}
&~~~~\mathbb{E}_P\Biggr[\frac{1 - A}{\alpha(P)}\cdot\frac{P(A = 1~|~\bm L^{*, c}, \bm L^{e, c}_m)P(E = 1~|~ A = 1, \bm L^{*, c}, \bm L^{e, c}_m)}{P(A = 0~|~\bm L^{*, c}, \bm L^{e, c}_m)P(E = 1~|~ A = 0, \bm L^{*, c}, \bm L^{e, c}_m)}E\mu_0(\bm L^*, \bm L^e_m) \Biggr] \\
&= \mathbb{E}_P\Biggr[\frac{1 - A}{\alpha(P)}\cdot\frac{\kappa(\bm L^{*,c}) \Lambda_1(\bm L^{e,c}_m; \bm L^{*, c})}{\bigr(1-\kappa(\bm L^{*,c}) \bigr)\Lambda_0(\bm L^{e,c}_m; \bm L^{*, c})}E\mu_0(\bm L^*, \bm L^e_m) \Biggr] \\
&= \mathbb{E}_P\Biggr[\frac{1 - A}{\alpha(P)}\cdot\frac{\kappa(\bm L^{*,c}) \Lambda_1(\bm L^{e,c}_m; \bm L^{*, c})}{\bigr(1-\kappa(\bm L^{*,c}) \bigr)\Lambda_0(\bm L^{e,c}_m; \bm L^{*, c})}E\mathbb{E}_P\biggr(\frac{R}{\eta(\bm L^*, 0)}Y~\biggr|~A = 0, \bm L^*, \bm L^e_m\biggr) \Biggr]~~~\text{(Lemma \ref{lemma:bvt}, A4, A5)} \\
&=\mathbb{E}_P\Biggr[\frac{(1 - A)REY}{\eta(\bm L^*,0)}\cdot\frac{\kappa(\bm L^{*,c}) \Lambda_1(\bm L^{e,c}_m; \bm L^{*, c})}{\bigr(1-\kappa(\bm L^{*,c}) \bigr)\Lambda_0(\bm L^{e,c}_m; \bm L^{*, c})}\Biggr]\Biggr/\alpha(P)~~~\text{(tower property)}
\end{aligned}
$$

\noindent Combining pieces, we obtain the desired result, that

$$
\begin{aligned}
\theta_\text{ATT}^\text{elig} &= \mathbb{E}_P[Y(1)~|~A = 1, E = 1 ] - \mathbb{E}_P[Y(0)~|~A = 1, E = 1 ] \\
&= \frac{1}{\alpha(P)}\Biggr\{\mathbb{E}_P\Biggr[\frac{ARE}{\eta(\bm L^*, 1)}Y\Biggr] - \mathbb{E}_P\Biggr[\frac{(1 - A)REY}{\eta(\bm L^*,0)}\cdot \frac{\kappa(\bm L^{*,c})\Lambda_1(\bm L^{e,c}_m;\bm L^{*,c})}{\big(1 - \kappa(\bm L^{*,c})\big)\Lambda_0(\bm L^{e,c}_m;\bm L^{*,c})}\Biggr]\Biggr\} \\
&= \frac{\zeta(P)}{\alpha(P)}
\end{aligned}
$$

Though $\theta_\text{ATT}^\text{elig}$ can still be identified under this alternative covariate partitioning, the derivation of any influence function would likely entail something similar to $\Lambda_a$ or $\delta_a$, possibly in a nested form. As such, further pursuing (efficient) influence function-based estimation techniques may yield estimators which are computationally very difficult to work with in practice.

\bibliographystyle{plainnat}
\clearpage
\bibliography{references}